\documentclass[prb,twocolumn,superscriptaddress,amsmath,amssymb,floatfix]{revtex4-2}

\usepackage{amssymb,amsmath,epsfig}
\usepackage{color}

\usepackage{ragged2e}
\usepackage{caption}

\usepackage{subcaption}
\usepackage{float}

 \usepackage[colorlinks=true, pdfstartview=FitV, linkcolor=blue, citecolor=red, urlcolor=magenta, breaklinks=true]{hyperref}

\usepackage{graphicx}
\usepackage{latexsym}
\usepackage{amsmath}
\usepackage{mathrsfs}
\usepackage{amsfonts}
\usepackage{amssymb}
\usepackage{booktabs}
\usepackage{multirow}

\usepackage{verbatim}
\usepackage{orcidlink}
\graphicspath{{Graf/}}
\usepackage{array}
\usepackage{float} 

\begin{document}

\title{Geodesic dynamics and tidal forces in a quantum-regularized black hole from string T-duality}

\author{G. Alencar \orcidlink{0000-0002-3020-4501}}\email{geova@fisica.ufc.br}
\affiliation{Departamento de F\'{\i}sica, Universidade Federal do Ceará\\
Caixa Postal 6030, Campus do Pici, 60455-760 Fortaleza, Ceará, Brazil}

\author{Micael S. Oliveira \orcidlink{0009-0009-2742-2311}}\email{micaeloliveira@fisica.ufc.br}
\affiliation{Departamento de F\'{\i}sica, Universidade Federal do Ceará\\
Caixa Postal 6030, Campus do Pici, 60455-760 Fortaleza, Ceará, Brazil}

\author{Diego S\'aez-Chill\'on G\'omez
        \orcidlink{0000-0002-0714-2680}}
\email{diego.saez@uva.es} 
\affiliation{Department of Theoretical Physics, Atomic and Optics, and Laboratory for Disruptive Interdisciplinary Science (LaDIS), Campus Miguel Delibes, \\ University of Valladolid UVA, Paseo Bel\'en, 7,
47011 - Valladolid, Spain}
\affiliation{Departamento de F\'{\i}sica, Universidade Federal do Ceará\\
Caixa Postal 6030, Campus do Pici, 60455-760 Fortaleza, Ceará, Brazil}

\author{Marcos V. de S. Silva \orcidlink{0000-0002-8080-9277}}
\email{marcos.sousa@uva.es}
\affiliation{Department of Theoretical Physics, Atomic and Optics, and Laboratory for Disruptive Interdisciplinary Science (LaDIS), Campus Miguel Delibes, \\ University of Valladolid UVA, Paseo Bel\'en, 7,
47011 - Valladolid, Spain}

\date{\today}

\begin{abstract}

In this work, we investigate the geodesic dynamics and tidal force effects in neutral and electrically charged regular spacetimes inspired by T-duality. We first analyze the geometric properties of these solutions, including their regular central structure and horizon configurations, and show how the interplay between the T-duality scale $\ell_0$ and the electric charge $Q$ determines the transition between black hole and horizonless geometries. We then study the motion of massive and massless particles through the corresponding effective potentials, examining the existence and stability of circular orbits and the behavior of their characteristic radii. We further investigate the tidal forces experienced by observers in radial free fall, focusing on the behavior of the radial and angular components, their extrema and sign changes, and the location of these characteristic points relative to the horizon structure of the geometry. We show that the T-duality regularization renders the tidal forces finite in the central region and gives rise to different stretching and compression regimes, including a region of simultaneous compression in both the radial and angular directions, while the electric charge modifies the strength and location of these tidal features. Finally, we analyze the evolution of the geodesic deviation vector for two sets of initial conditions. We find that both the T-duality scale and the electric charge tend to suppress the accumulated radial stretching, whereas the angular deformation is considerably less sensitive to the charge. These results show how the quantum regularization induced by T-duality and the presence of electric charge modify the orbital dynamics and tidal response of these geometries in the strong-field regime. Furthermore, we demonstrate that the corresponding classical limits are recovered appropriately.
\end{abstract}
\pacs{11.15.-q, 11.10.Kk} \maketitle

\section{INTRODUCTION}\label{sec_I}

Over the past decades, observational evidence has consolidated the role of black holes as some of the most fundamental astrophysical objects. The detection of gravitational waves from the coalescence of binary black hole systems by the LIGO and Virgo collaborations opened a new observational window for investigating the dynamics of these objects and testing the predictions of General Relativity (GR)~\cite{LIGOScientific:2016aoc, LIGOScientific:2016sjg, LIGOScientific:2017ycc, LIGOScientific:2020zkf, LIGOScientific:2021qlt, LIGOScientific:2025rsn}. In parallel, observations performed by the Event Horizon Telescope (EHT) have provided images of the emission on scales close to the horizons of the supermassive compact objects M87* and Sgr~A*, offering information about the structure of spacetime in strong-gravity regions~\cite{EventHorizonTelescope:2019dse, EventHorizonTelescope:2019ggy, EventHorizonTelescope:2022wkp}. These observational advances reinforce the importance of theoretically investigating the properties of black hole geometries and the phenomena associated with strong-field regions, where possible deviations from classical GR solutions may become more relevant.

Black holes constitute one of the most remarkable predictions of GR and provide a privileged setting for investigating gravity in the strong-field regime. Despite the success of the Schwarzschild solution in describing the exterior gravitational field of a spherically symmetric object, its extension to the central region exhibits a curvature singularity at $r=0$, where geometric invariants diverge and the classical description of spacetime breaks down~\cite{Penrose:1964wq, Penrose:1969pc, Hawking:1970zqf, Gubser:2000nd}. The presence of such singularities is often interpreted as an indication of the limitations of the classical theory, motivating the search for a quantum description of gravity and for models capable of regularizing the geometry at sufficiently small scales. In this context, regular black holes constitute a particularly interesting class of solutions in which the central singularity is removed while the Schwarzschild behavior is recovered at large distances~\cite{Hayward:2005gi, Frolov:2016pav, Ayon-Beato:1998hmi}.

Since the pioneering proposal by Bardeen~\cite{Bardeen:1968}, which was later interpreted as a solution of the Einstein equations sourced by nonlinear electrodynamics~\cite{Ayon:Beato:2000mjt,Rodrigues:2018bdc}, several models of regular black holes have been developed and investigated in the literature~\cite{Ayon-Beato:1999kuh, Ayon-Beato:1999qin, Balart:2014cga, Lan:2023cvz, Hennigar:2025yqm, Bueno:2025dqk, Estrada:2024uuu}, providing a framework for exploring mechanisms that regularize the central region and their consequences for the causal structure and gravitational phenomena in strong-field regimes. Among the approaches considered, particular attention has been given to those involving the introduction of a fundamental length scale. In particular, within the framework of string theory, T-duality relates physical descriptions at large and small length scales, providing motivation for the existence of a minimum length~\cite{Padmanabhan:1996ap, Padmanabhan:1998yya}.

Recently, a regular black hole solution incorporating corrections inspired by T-duality was obtained by Nicolini, Spallucci, and Wondrak~\cite{Nicolini:2019irw}. In this construction, the connection between T-duality and duality in the path-integral formalism naturally introduces an ultraviolet cutoff associated with the so-called zero-point length, which modifies the behavior of the gravitational potential at short distances~\cite{Padmanabhan:1996ap,Nicolini:2022rlz}. Starting from this regularized potential, the authors obtained a spherically symmetric and electrically neutral geometry in which the length scale $\ell_0$, associated with T-duality, controls the modifications in the central region of the spacetime. As a consequence, the singularity is replaced by a regular region with finite curvature, while the classical Schwarzschild behavior is recovered at large distances or in the limit $\ell_0\rightarrow0$~\cite{Nicolini:2019irw}. This approach was subsequently extended to include regular configurations in the presence of electric charge and angular momentum~\cite{Gaete:2022une,Gaete:2022ukm}.

Beyond the construction of these solutions, considerable attention has been devoted to investigating their physical properties and possible differences from classical black holes. In this context, geodesic dynamics constitutes one of the main tools for probing the strong-field structure of these geometries, allowing the motion of particles and photons to be characterized and possible modifications introduced by the regularization to be identified~\cite{Stuchlik:2014qja,Berry:2020ntz,Chiba:2017nml,Hu:2018old}. These studies have been complemented by investigations of different phenomena associated with regular spacetimes, including black strings, quasinormal modes, wormholes, grey-body factors, and other possible observational signatures~\cite{Jusufi:2022rbt,Konoplya:2023ahd,Lobo:2025nng,Lutfuoglu:2026etg,Javed:2024wbc,Lutfuoglu:2026rqe}. Such analyses are particularly relevant in view of the increasing precision of observations in strong-field regimes, as they may provide means of distinguishing regular geometries from their classical counterparts.

Among the properties that can be used to probe the local structure of the gravitational field, tidal forces play a particularly important role, as they describe the relative acceleration between neighboring geodesics and, consequently, the deformation experienced by a freely falling body~\cite{10.1093/oso/9780198862024.001.0001,hobson2006}. In the Schwarzschild spacetime, these forces produce stretching in the radial direction and compression in the angular directions, with amplitudes that increase as the observer approaches the central singularity, resulting in the characteristic deformation known as spaghettification~\cite{hobson2006}. In the Reissner--Nordstr\"om spacetime, as well as in other charged black hole geometries, the tidal-force components may undergo sign changes at certain radii, giving rise to transitions between stretching and compression regimes~\cite{Crispino:2016pnv,Lima:2020wcb,Sharif:2018gzj,Crispim:2025cql,Cordeiro:2025cfo,Uniyal:2025sdr,Sharif:2018gaj, Junior:2026upy}.

More complex tidal structures have also been found in other black hole spacetimes with a central singularity. In the Kottler spacetime, for instance, the presence of the cosmological constant qualitatively modifies the behavior of the tidal forces, allowing the radial and angular components to change sign at certain radii~\cite{Vandeev:2021yan}. Transitions between stretching and compression regimes have also been observed in Kiselev black holes, where the tidal components are influenced by the parameters associated with the matter fields surrounding the object~\cite{Shahzad:2017vwi}. Effects associated with the presence of quintessence and a cloud of strings have also been investigated in extensions of the Schwarzschild spacetime, revealing modifications in the profiles of the radial and angular tidal-force components~\cite{Uniyal:2022ouc}. More recently, in the Letelier--Alencar black hole, it was shown that the cloud of strings modifies the tidal-force structure and can induce a reversal between stretching and compression regimes during radial infall~\cite{Silva:2025hwl}.

On the other hand, regular geometries and nonsingular compact objects may exhibit a qualitatively distinct tidal structure due to modifications of the geometry at small scales. In several of these models, the radial and angular components of the tidal forces may exhibit extrema and sign changes, thereby defining different regions of stretching and compression. In particular, studies of black-bounce geometries have shown that the tidal forces can remain finite and exhibit different stretching and compression regimes along radial infall~\cite{Crispim:2025cql}. The regularization of tidal forces in the central region has also been observed in different regular geometries~\cite{Sharif:2018gaj,Lima:2020wcb,Huo:2026jjn,LimaJunior:2022gko,Arora:2023ltv,Macedo:2026ehd, Idrissov:2026kzg, Silva:2026mlo}. These results, however, do not indicate a universal behavior, since the tidal-force structure depends on the specific properties of each geometry. Moreover, there exist spacetimes in which divergent tidal effects can arise even at the event horizon~\cite{Nandi:2000gt}.

Some aspects of particle dynamics and photon propagation in T-duality-inspired spacetimes have already been explored in the literature. In particular, the motion and circular orbits of spinning particles around charged T-dual black holes were investigated in~\cite{Rakhimova:2024ind}, while null geodesic dynamics and the properties of photon orbits were analyzed for the charged T-dual solution in~\cite{Javed:2024ocf}. Other phenomena related to light propagation, such as gravitational lensing, as well as the optical properties of rotating and charged extensions of the T-dual geometry, were considered in~\cite{Orzuev:2026szf,Jha:2023jde}. More recently, T-duality-inspired black-bounce geometries have been investigated from different perspectives, including their thermodynamic properties, energy conditions, shadows, and possible observational signatures~\cite{Alencar:2026qeb}. The geodesic completeness of the neutral regular T-dual solution, in turn, was studied through the Raychaudhuri equation and the evolution of geodesic congruences~\cite{Jusufi:2024dtr}. However, a combined characterization of the geodesic dynamics and tidal effects in both the neutral and electrically charged T-duality-inspired regular geometries remains largely unexplored. In particular, it is relevant to understand how the interplay between the minimum length scale $\ell_0$ and the electric charge $Q$ modifies the causal structure, the existence and stability of circular orbits, the tidal force profiles, and the accumulated deformation of neighboring geodesics during radial infall.

The paper is organized as follows. In Sec.~\ref{sec_II}, we introduce the neutral and electrically charged T-duality-inspired regular geometries and analyze their regularity, asymptotic behavior, and horizon structure, including the transition between black hole and horizonless configurations. In Sec.~\ref{sec_III}, we investigate the geodesic dynamics of massive and massless particles, analyzing their effective potentials, circular orbits, and corresponding characteristic radii. In Sec.~\ref{sec_IV}, we study the tidal forces experienced by radially freely falling observers, focusing on the radial and angular components, their characteristic extrema and sign changes, and their relation to the causal structure of the spacetime. In Sec.~\ref{sec_V}, we analyze the geodesic deviation equations and the evolution of the radial and angular components of the separation vector for two sets of initial conditions, considering both the neutral and charged geometries. Finally, in Sec.~\ref{sec_VI}, we present our conclusions.

Throughout this work, we adopt the metric signature $(-,+,+,+)$ and use geometrized units, in which $G=\hbar=c=1$.

\section{T-duality-inspired regular spacetime}
\label{sec_II}

One of the expected consequences of a quantum description of gravity is the emergence of a fundamental length scale below which the classical description of spacetime should be modified. Within the path-integral formulation for a relativistic particle, Padmanabhan proposed the invariance of the propagation amplitude under the transformation~\cite{Padmanabhan:1996ap,Padmanabhan:1998yya}
\begin{equation}
    ds \longrightarrow \frac{\ell_0^2}{ds},
    \label{eq:pathduality}
\end{equation}
where $\ell_0$ represents a minimum length scale. This duality suppresses contributions associated with arbitrarily small proper lengths and can be interpreted in terms of a zero-point length, effectively leading to the modification $ds^2\rightarrow ds^2+\ell_0^2$. This modification preserves the structure of the theory at large distances while altering its behavior in the ultraviolet regime, thereby providing an effective mechanism for regularizing divergences associated with field propagation~\cite{Padmanabhan:1996ap,Padmanabhan:1998yya}.

In momentum space, this regularization modifies the scalar propagator, which for a particle of mass $m$ takes the form
\begin{equation}
    G(k)=
    -\frac{\ell_0}{\sqrt{k^2+m^2}}
    K_1\!\left(\ell_0\sqrt{k^2+m^2}\right),
    \label{eq:propagator}
\end{equation}
where $K_1$ is the modified Bessel function of the second kind. The modified propagator can be used to obtain the effective gravitational potential associated with a point source of mass $M$. Writing the potential as the Fourier transform of the propagator in the static limit, we obtain the regularized potential~\cite{Nicolini:2019irw},
\begin{align}
V(r) &=-M\int\frac{d^3k}{(2\pi)^3}\,G(k)e^{i\mathbf{k}\cdot\mathbf{r}} \\
&=-\frac{M}{\sqrt{r^2+\ell_0^2}},
\label{eq:potentialfourier}
\end{align} 
which remains finite at the origin and recovers the Newtonian behavior for $r\gg\ell_0$. Thus, the scale $\ell_0$ replaces the point source with an effective regular matter distribution whose density, obtained from the Poisson equation, takes the form
\begin{equation}
    \rho(r)
    =
    \frac{1}{4\pi}\nabla^2V(r)
    =
    \frac{3M\ell_0^2}
    {4\pi\left(r^2+\ell_0^2\right)^{5/2}}.
    \label{eq:density}
\end{equation}
This distribution is finite at the origin, falls off as $r^{-5}$ at large distances, and preserves the total mass of the configuration~\cite{Nicolini:2019irw,Nicolini:2022rlz}. The corresponding mass function is
\begin{equation}
    m(r)
    =
    4\pi\int_0^r
    \rho(\tilde r)\tilde r^{\,2}d\tilde r
    =
    \frac{Mr^3}
    {(r^2+\ell_0^2)^{3/2}},
    \label{eq:massfunction}
\end{equation}
with $m(r)\rightarrow M$ as $r\rightarrow\infty$ and $m(r)\rightarrow Mr^3/\ell_0^3$ as $r\rightarrow 0$.

\subsection{Neutral T-duality black hole}
\label{subsec:neutral}

The corresponding gravitational solution is obtained by solving the Einstein equations for a spherically symmetric matter distribution. In Schwarzschild-like coordinates, the line element is given by
\begin{equation}
    ds^2
    =
    -f(r)dt^2
    +\frac{dr^2}{f(r)}
    +r^2\left(d\theta^2+\sin^2\theta\,d\phi^2\right),
    \label{eq-et}
\end{equation}
with
\begin{equation}
    f(r)
    =
    1-\frac{2m(r)}{r}
    =
    1-\frac{2Mr^2}
    {\left(r^2+\ell_0^2\right)^{3/2}},
    \label{eq-f(r)}
\end{equation}
which defines the T-duality-inspired regular spacetime considered throughout this work. Following Ref.~\cite{Nicolini:2019irw}, the parameter $\ell_0$ can also be related to the Regge slope parameter $\alpha'$ through $\ell_0=2\pi\sqrt{\alpha'}$, establishing a connection between the regularization of the geometry at short distances and the fundamental scale of string theory. Therefore, $\ell_0$ is expected to be associated with a microscopic scale, typically of the order of the Planck length. Furthermore, since the metric function depends on $\ell_0$ only through $\ell_0^2$, we shall consider $\ell_0\geq0$ without loss of generality. In the remainder of this section, we discuss its main geometric properties, including the recovery of the Schwarzschild limit, the regularity of the origin, and the horizon structure determined by the parameter $\ell_0$.

\subsubsection{Regularity and asymptotic behavior}
\label{sec:regularidade}

A fundamental feature of this geometry is its regularity in the central region. Expanding the metric function \eqref{eq-f(r)} as $r\rightarrow0$, we obtain
\begin{equation}
    f(r)
    =
    1-\frac{2M}{\ell_0^3}r^2
    +\mathcal{O}(r^4).
    \label{eq:f_origem}
\end{equation}
This behavior can be compared with the static form of the metric function of de Sitter spacetime,
\begin{equation}
    f_{\mathrm{dS}}(r)
    =
    1-\frac{\Lambda_{\mathrm{eff}}}{3}r^2,
\end{equation}
with the effective cosmological constant
\begin{equation}
    \Lambda_{\mathrm{eff}}
    =
    \frac{6M}{\ell_0^3}.
    \label{eq:lambda_eff}
\end{equation}
Therefore, near the central region, the geometry exhibits a de Sitter-like core, in contrast to the singular behavior of the Schwarzschild solution~\cite{Nicolini:2019irw, Konoplya:2023ahd, Lutfuoglu:2026rqe}. Regularity in this region can also be verified through curvature invariants. In particular, the Ricci scalar and the Kretschmann scalar remain finite at the origin, taking the values
\begin{equation}
    R(0)=\frac{24M}{\ell_0^3},
    \qquad
    K(0)=
    \frac{96M^2}{\ell_0^6},
    \label{eq:invariants_origin}
\end{equation}
where $K=R_{\mu\nu\rho\sigma}R^{\mu\nu\rho\sigma}$. Thus, for $\ell_0\neq0$, the curvature singularity present in the Schwarzschild solution is replaced by a regular core with finite curvature.

In the asymptotic regime, $r\gg\ell_0$, the metric function can be expanded as
\begin{equation}
    f(r)
    =
    1-\frac{2M}{r}
    +\frac{3M\ell_0^2}{r^3}
    +\mathcal{O}\left(\frac{\ell_0^4}{r^5}\right),
    \label{eq:f_assintotico}
\end{equation}
showing that the corrections associated with T-duality are suppressed at large distances. Furthermore, in the limit $\ell_0\rightarrow0$, one directly recovers $f(r)=1-2M/r$. Hence, $\ell_0$ controls the short-distance modifications of the geometry while preserving the Schwarzschild behavior in the classical regime.

\subsubsection{Horizon structure}
\label{sec:horizontes}

\begin{figure}[h!]
\centering
\includegraphics[width=0.48\textwidth]{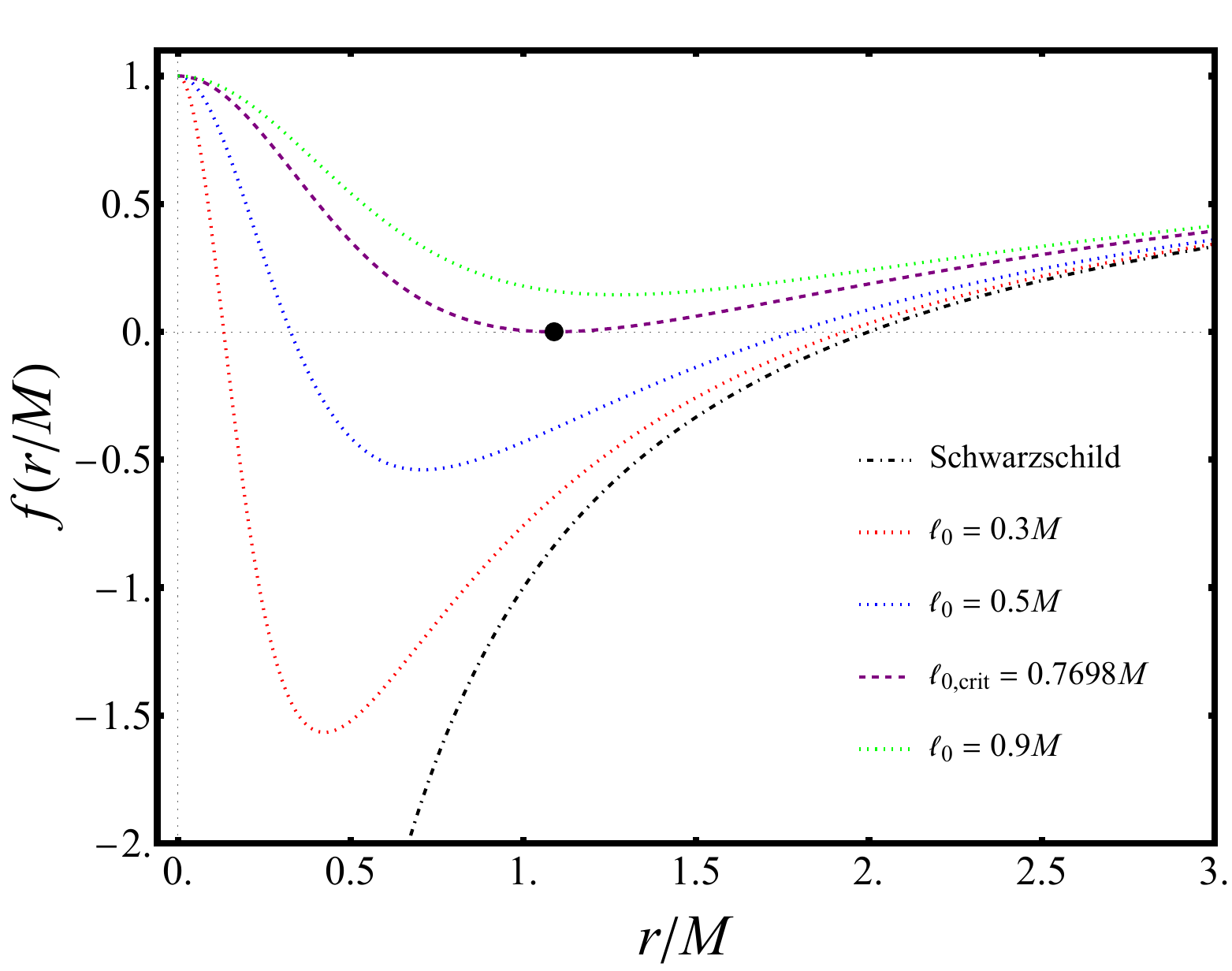}
\caption{\justifying Behavior of the lapse function as a function of $r/M$ for different values of the parameter $\ell_0$.  The  Schwarzschild solution, corresponding to $\ell_0=0$, is included for comparison. The black point indicates the degenerate horizon of the corresponding extremal configuration.}
\label{fig:f_r_l0}
\end{figure}

The horizon structure is determined by the positive roots of $f(r)$, which satisfy
\begin{equation}
    \left(r_{\pm}^{2}+\ell_0^{2}\right)^{3/2}
    =
    2Mr_{\pm}^{2},
    \label{eq:horizon_equation}
\end{equation}
where $r_-$ and $r_+$ correspond to the inner and outer horizons, respectively. The number of roots depends on the dimensionless ratio $\ell_0/M$. In the extremal case, the two horizons coincide and the metric function has a double root, such that
\begin{equation}
    f(r_{\rm ext})=0,
    \qquad
    f'(r_{\rm ext})=0.
    \label{eq:extremal_conditions}
\end{equation}
Since
\begin{equation}
    f'(r)
    =
    \frac{2Mr(r^2-2\ell_0^2)}
    {(r^2+\ell_0^2)^{5/2}},
\end{equation}
these conditions yield
\begin{equation}
    \ell_{0,\mathrm{crit}}
    =
    \frac{4M}{3\sqrt{3}}
    \simeq0.7698M,
    \quad
    r_{\rm ext}
    =
    \sqrt{2}\,\ell_{0,\mathrm{crit}}
    \simeq1.0887M.
    \label{eq:critical_values}
\end{equation}
Consequently, three distinct regimes can be identified: for $0<\ell_0<\ell_{0,\mathrm{crit}}$, there are two horizons, $r_-<r_+$; for $\ell_0=\ell_{0,\mathrm{crit}}$, they become degenerate, $r_-=r_+=r_{\rm ext}$; and, for $\ell_0>\ell_{0,\mathrm{crit}}$, the metric function has no positive roots, so that the regular geometry is horizonless~\cite{Nicolini:2019irw}. These transitions are illustrated in Fig.~\ref{fig:f_r_l0}, for different values of $\ell_0$. Furthermore, we observe that in the limit $\ell_0\rightarrow0$, the outer horizon approaches $r_+\rightarrow2M$, recovering the Schwarzschild horizon, while $r_-\rightarrow0$.

These properties show that the scale $\ell_0$ modifies both the central region and the horizon structure of the neutral
geometry while preserving the Schwarzschild limit at large distances. We now turn to the electrically charged extension
of the T-duality-inspired solution and investigate how the additional parameter $Q$ modifies these geometric properties.

\subsection{Charged T-duality black hole}
\label{subsec:charged}

The T-duality-inspired framework can also be extended to include an electric charge by incorporating the corresponding
regularized electromagnetic sector. The resulting static and spherically symmetric charged solution was obtained in
Ref.~\cite{Gaete:2022ukm} and can be written in the same Schwarzschild-like form introduced in Eq.~\eqref{eq-et}, with
the metric function
\begin{equation}
f(r)=1-\frac{2Mr^2}{(r^2+\ell_0^2)^{3/2}}+\frac{Q^2r^2}{(r^2+\ell_0^2)^2}F(r),
\label{fcharged}
\end{equation}
where
\begin{align}
F(r)=&\frac{5}{8}+\frac{3\ell_0^2}{8r^2}+\frac{3\pi}{16\ell_0}\sqrt{r^2+\ell_0^2}\nonumber\\
&\times\left\{1-\frac{(r^2+\ell_0^2)^{3/2}}{r^3}\left[\frac{2}{\pi}\arctan\left(\frac{r}{\ell_0}\right)\right]\right\}.
\label{eq:Fcharged}
\end{align}
The neutral geometry discussed in the previous subsection is recovered by setting $Q=0$. On the other hand, in the limit
$\ell_0\rightarrow0$, the T-duality corrections disappear and the metric reduces to the Reissner--Nordstr\"om solution,
\begin{equation}
f(r)\longrightarrow1-\frac{2M}{r}+\frac{Q^2}{r^2}.
\end{equation}
Therefore, the charged geometry provides a regularized extension of the Reissner--Nordstr\"om spacetime while continuously recovering the neutral T-duality solution in the limit $Q\rightarrow0$.

\subsubsection{Regularity and asymptotic behavior}
As in the uncharged case, it is important to examine the asymptotic behavior of the charged solution in order to assess how the electric charge modifies the geometric properties of the spacetime and, in particular, whether it affects the regularity of the central region. Expanding the metric function $f(r)$ around $r=0$, we obtain
\begin{equation}
    f(r) =1-r^2 \left(\frac{2 M}{\ell_0^3}-\frac{3 \pi  Q^2}{16 \ell_0^4}\right)+\mathcal{O}\left(r^4\right).
    \label{eq:fQ_origem}
\end{equation}
The leading behavior can again be written in the form
\begin{equation}
    f(r)=1-\frac{\Lambda_{\mathrm{eff}}^{(Q)}}{3}r^2+\mathcal{O}(r^4),
\end{equation}
where the effective cosmological constant is now given by
\begin{equation}
    \Lambda_{\mathrm{eff}}^{(Q)}  = \frac{6M}{\ell_0^3} -\frac{9\pi Q^2}{16\ell_0^4} = \frac{3\left(32M\ell_0-3\pi Q^2\right)}{16\ell_0^4}.
    \label{eq:lambda_eff_Q}
\end{equation}
Therefore, unlike the neutral case, the nature of the regular core depends on the interplay between the mass, the T-duality scale, and the electric charge. For $32M\ell_0>3\pi Q^2$, the central region is de Sitter-like, whereas for $32M\ell_0<3\pi Q^2$ it becomes anti-de Sitter-like. At the critical value $32M\ell_0=3\pi Q^2$, the quadratic contribution vanishes and the curvature at the center becomes zero.

Since the metric remains regular at $r=0$, we further investigate the regularity of the charged geometry by analyzing the behavior of the curvature invariants in the central region. The Ricci scalar and the Kretschmann scalar
evaluated at $r=0$ are given by
\begin{equation}
    \begin{aligned}
    R(0)&=\frac{3\left(32\ell_0M-3\pi Q^2\right)}{4\ell_0^4},
    \\
    K(0)&=\frac{3\left(32\ell_0M-3\pi Q^2\right)^2}{32\ell_0^8}.
    \end{aligned}
    \label{eq:invariantsQ_origin}
\end{equation}
These expressions can also be written as
\begin{equation}
    R(0)=4\Lambda_{\mathrm{eff}}^{(Q)},\qquad   K(0)=\frac{8}{3}
    \left(\Lambda_{\mathrm{eff}}^{(Q)}\right)^2,
\end{equation}
consistently with the maximally symmetric character of the geometry at the center. Therefore, the electric charge modifies the curvature in the central region without introducing any divergence. In particular, no curvature singularity arises at $r=0$.

In the asymptotic region, $r\gg\ell_0$, the metric function takes the form
\begin{equation}
\begin{aligned}
f(r)={}&1-\frac{2M}{r}+\frac{Q^2}{r^2}\\
&+\frac{1}{r^3}\left(3M\ell_0^2-\frac{9\pi Q^2\ell_0}{32}\right)+\mathcal{O}\left(\frac{1}{r^4}\right).
\end{aligned}
\label{eq:fQ_asymptotic}
\end{equation} 
The leading terms reproduce the Reissner--Nordstr\"om geometry, while the corrections induced by T-duality appear only at higher orders in the inverse radial coordinate. Therefore, the effects associated with the minimum length scale are suppressed at large distances, and the standard charged black hole behavior is recovered asymptotically.

\subsubsection{Horizon structure}
\label{subsubsec:horizontes}

The presence of the electric charge introduces an additional parameter into the horizon structure of the T-duality-inspired geometry. The horizons are determined by the positive roots of the metric function,
\begin{equation}
1-\frac{2Mr_{\pm}^2}{\left(r_{\pm}^2+\ell_0^2\right)^{3/2}}+\frac{Q^2r_{\pm}^2}{\left(r_{\pm}^2+\ell_0^2\right)^2}F(r_{\pm})=0.
\label{eq:horizon_equation_charged}
\end{equation}
In contrast to the neutral case, the horizon structure is now controlled by both dimensionless parameters $\ell_0/M$ and $Q/M$. Owing to the nontrivial radial dependence of $F(r)$, Eq.~\eqref{eq:horizon_equation_charged} cannot, in general, be solved analytically, and the horizon radii must therefore be determined numerically.

The extremal configuration occurs when the inner and outer horizons merge into a degenerate root. As in the neutral case, this configuration is determined by the simultaneous conditions
\begin{equation}
    f(r_{\rm ext})=0,
    \qquad
    f'(r_{\rm ext})=0.
    \label{eq:extremal_conditions_charged}
\end{equation}
For the neutral geometry, these conditions could be solved analytically, yielding explicit expressions for the critical value of $\ell_0$ and the corresponding extremal radius. In the charged case, however, the presence of the function $F(r)$ makes the extremality conditions transcendental, and the critical values must also be obtained numerically.

\begin{figure*}[!htb]
    \centering
    \includegraphics[width=0.49\linewidth]{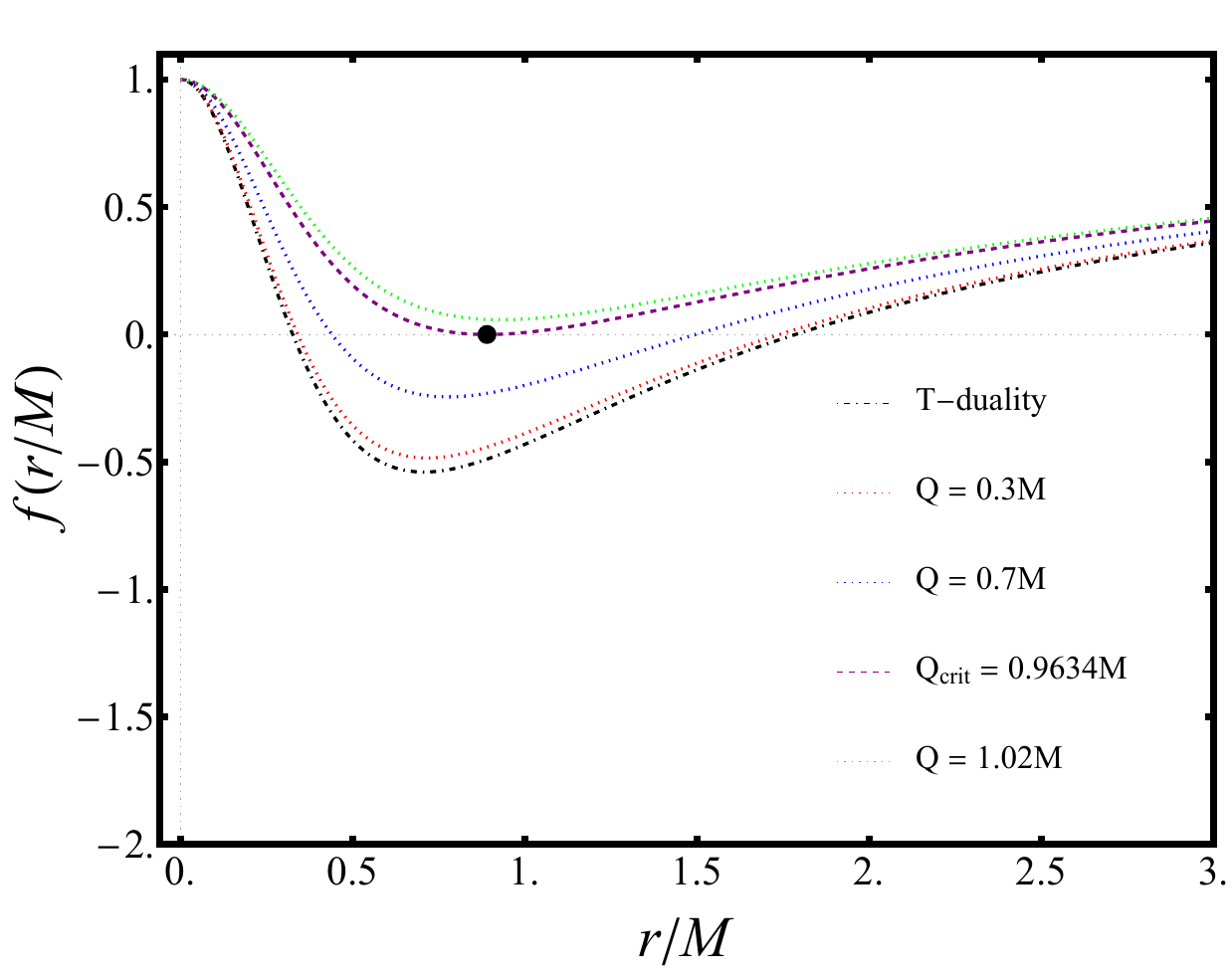}
    \hfill
    \includegraphics[width=0.49\linewidth]{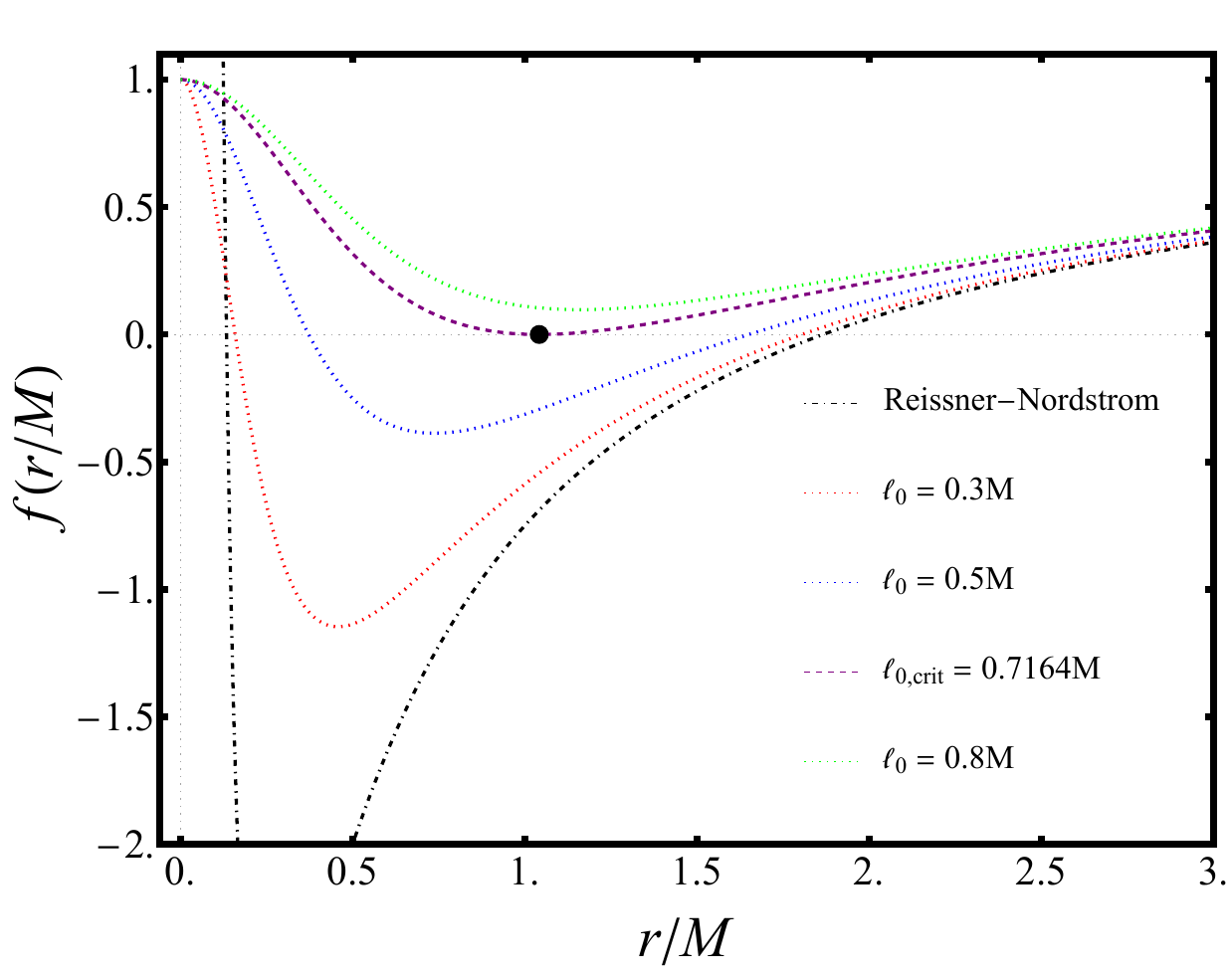}
    \caption{\justifying Behavior of the metric function $f(r)$ as a function of $r/M$ for the charged T-duality-inspired geometry. In the left panel, the T-duality parameter is fixed at $\ell_0=0.5M$ while the electric charge $Q$ is varied; the neutral T-duality solution, corresponding to $Q=0$, is shown for comparison. In the right panel, the electric charge is fixed at $Q=0.5M$ while $\ell_0$ is varied; the Reissner--Nordstr\"om solution, corresponding to $\ell_0=0$, is included for comparison. The black points indicate the degenerate horizons of the corresponding extremal configurations.}
    \label{fig:fQ_fixed_l0}
\end{figure*}

In Fig.~\ref{fig:fQ_fixed_l0}, we show the behavior of the metric function for a fixed value $\ell_0=0.5M$ and different values of the electric charge. The neutral T-duality solution, corresponding to $Q=0$, is included for comparison. As $Q$ increases, the two zeros of the metric function approach each other: the inner horizon moves outward, while the outer horizon shifts inward. At the critical value $Q_{\rm crit}\simeq0.9634M$, the two horizons merge into a degenerate horizon located at $r_{\rm ext}\simeq0.8903M$. For larger values of $Q$, the metric function no longer possesses positive roots and the geometry becomes horizonless. A complementary behavior is shown in Fig.~\ref{fig:fQ_fixed_l0}, where the electric charge is fixed at $Q=0.5M$ and the T-duality scale is varied. The Reissner--Nordstr\"om solution, corresponding to $\ell_0=0$, is shown for comparison. Increasing $\ell_0$ again reduces the separation between the two horizons. They become degenerate at $\ell_{0,\rm crit}\simeq0.7164M$, with the extremal radius $r_{\rm ext}\simeq1.0420M$. For larger values of $\ell_0$, the spacetime becomes horizonless.

These results show that both the electric charge and the T-duality scale play an important role in determining the causal structure of the geometry. For the representative parameter choices considered here, increasing either $Q$ or $\ell_0$ drives the spacetime toward extremality and eventually to a horizonless configuration. In particular, for $Q=0.5M$, the critical value
$\ell_{0,\rm crit}\simeq0.7164M$ is smaller than the neutral value $\ell_{0,\rm crit}\simeq0.7698M$, while for $\ell_0=0.5M$ the critical charge $Q_{\rm crit}\simeq0.9634M$ is smaller than the Reissner--Nordstr\"om extremal value $Q=M$.

\begin{figure}[h!]
\centering
\includegraphics[width=0.48\textwidth]{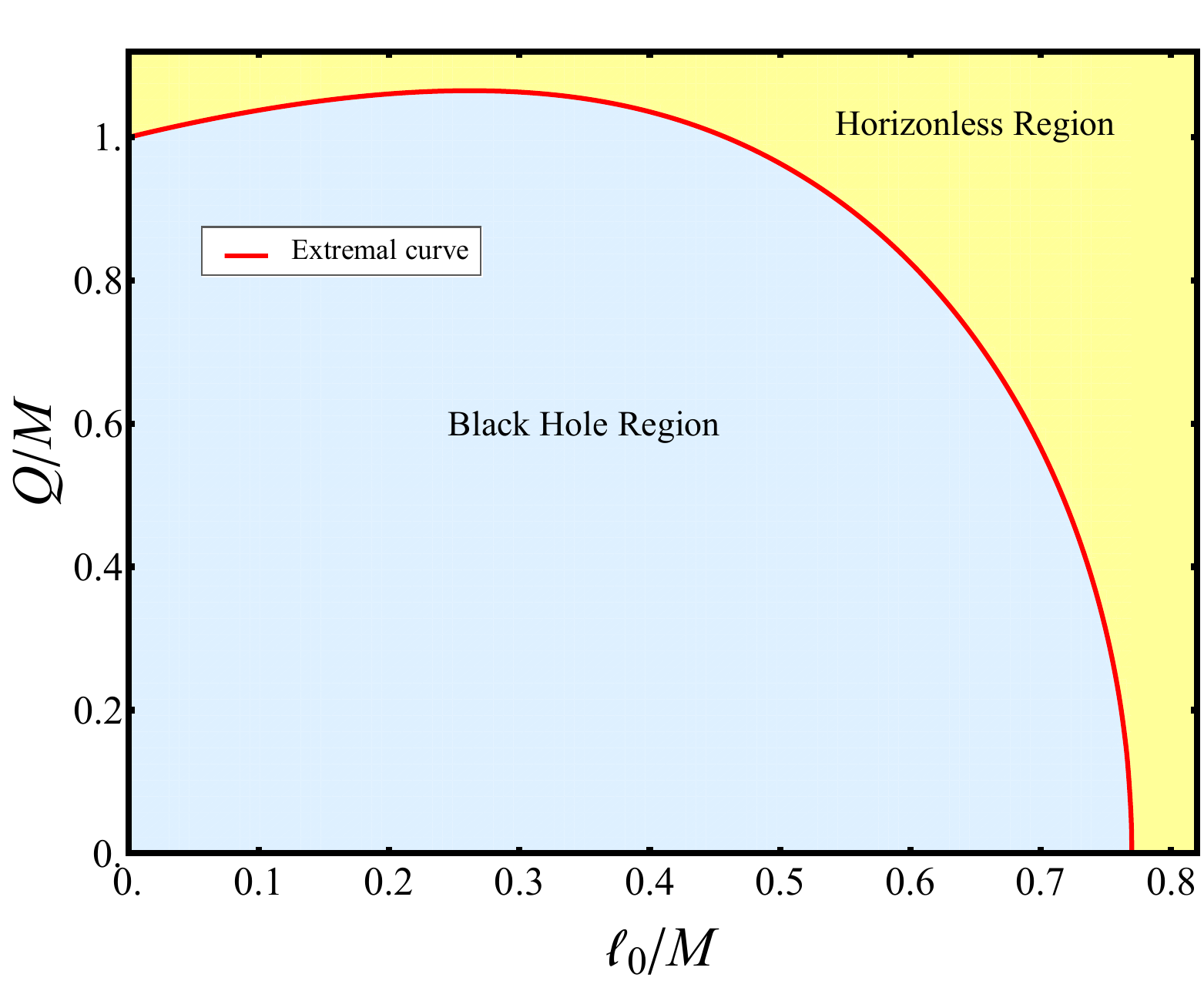}
\caption{\justifying Parameter space in the $(\ell_0/M,Q/M)$ plane. The red curve represents the extremal configurations, the blue region corresponds to black-hole configurations, whereas the yellow region represents horizonless configurations.}
\label{fig:l_0_Q_extremal}
\end{figure}

To complete our initial analyses, we present in Fig.~\ref{fig:l_0_Q_extremal} the parameter space $(\ell_0/M,Q/M)$, in which two distinct regions are identified. The solid red curve defines the extremal limit, where the inner and outer horizons coincide, forming a degenerate horizon. This boundary separates the black-hole region (light-blue shaded area), characterized by the presence of horizons, from the region of regular horizonless configurations (yellow shaded area). Interestingly, the extremal curve is not monotonic between these two limits. Starting from the Reissner--Nordstr\"om point, the critical charge initially increases with $\ell_0$ and reaches a maximum value of approximately $Q_{\rm crit}^{\rm max}\simeq 1.0651M$ at $\ell_0\simeq 0.2619M$, with the corresponding extremal radius given by $r_{\rm ext}\simeq 0.8806M$, before decreasing toward zero as the neutral extremal limit is approached.

In the next section, we investigate how these modifications affect the dynamics of massive and massless particles, with particular attention to circular orbits and the characteristic radii of the geometry.
\section{Geodesic motion}
\label{sec_III}

In this section, we analyze the motion of test particles in the neutral and electrically charged T-duality-inspired regular geometries. To this end, we start from the Lagrangian formulation of geodesic motion~\cite{Chandrasekhar:1985kt, Carroll:2004st}, from which we derive the equations of motion and introduce the effective potential.

\subsection{Lagrangian and equations of motion}

Geodesic motion can be obtained from the Lagrangian
\begin{equation}
\mathcal{L} = \frac{1}{2} g_{\mu\nu} \dot{x}^{\mu} \dot{x}^{\nu},
\end{equation}
where the dot denotes differentiation with respect to the affine parameter $\tau$, which corresponds to proper time in the case of timelike geodesics. For the spacetime considered in Eq.~\eqref{eq-et}, the Lagrangian takes the form
\begin{equation}
\mathcal{L} = \frac{1}{2} \left[ -f(r)\dot{t}^2 + \frac{\dot{r}^2}{f(r)} + r^2 \dot{\theta}^2 + r^2 \sin^2\theta\, \dot{\phi}^2 \right].
\end{equation}
Since the metric is independent of the coordinates $t$ and $\phi$, there are two conserved quantities associated with time-translation and rotational symmetries. Applying the Euler--Lagrange equations, we obtain
\begin{equation} \label{eq-lag}
\frac{\partial\mathcal{L}}{\partial\dot t}
=
-f(r)\dot t
=
-E \quad \text{and} \quad \frac{\partial\mathcal{L}}{\partial\dot\phi}
=
r^2\sin^2\theta\,\dot\phi
=
L,
\end{equation}
and therefore the equations of motion for $t$ and $\phi$ are given by
\begin{equation} \label{eqs_motion}
\dot{t} = \frac{E}{f(r)} \quad \text{and} \quad
\dot\phi
=
\frac{L}{r^2\sin^2\theta},
\end{equation}
where $E$ and $L$ correspond to the conserved energy and angular momentum, respectively.

\subsection{Radial dynamics}

The radial dynamics can be obtained from the normalization condition of the four-velocity,
\begin{equation}
    g_{\mu\nu}\dot{x}^{\mu}\dot{x}^{\nu}
    =
    -\kappa,
    \label{eq:normalization}
\end{equation}
where $\kappa=1$ for massive particles (timelike geodesics) and $\kappa=0$ for massless particles (null geodesics). Using the conserved quantities defined in Eq.~\eqref{eqs_motion}, the normalization condition can be written as
\begin{equation}
    -\frac{E^2}{f(r)}
    +\frac{\dot r^2}{f(r)}
    +r^2\dot{\theta}^{\,2}
    +\frac{L^2}{r^2\sin^2\theta}
    =
    -\kappa.
    \label{eq-end}
\end{equation}
Isolating the radial term, we obtain
\begin{equation}
    \dot r^2
    =
    E^2
    -
    f(r)
    \left[
        \kappa
        +r^2\dot{\theta}^{\,2}
        +\frac{L^2}{r^2\sin^2\theta}
    \right].
    \label{eq-full}
\end{equation}
Owing to the spherical symmetry of the spacetime, the geodesic motion can be restricted, without loss of generality, to the equatorial plane, $\theta=\pi/2$, so that $\dot{\theta}=0$. Thus, the radial equation reduces to
\begin{equation}
    \dot r^2
    =
    E^2
    -
    f(r)
    \left(
        \kappa+\frac{L^2}{r^2}
    \right).
    \label{eq-radial_general}
\end{equation}
This expression can be written in the form of an effective energy conservation equation,
\begin{equation}
    \dot r^2+V_{\mathrm{eff}}(r)=E^2,
    \label{eq-energy_effective}
\end{equation}
where the effective potential is defined as
\begin{equation}\label{eq-pot-geral}
    V_{\mathrm{eff}}(r)
    =
    f(r)
    \left(
        \kappa+\frac{L^2}{r^2}
    \right).
\end{equation}
Purely radial motion is obtained as the particular case $L=0$, for which
\begin{equation}
    \dot r^2+\kappa f(r)=E^2.
    \label{eq-radial}
\end{equation}
In what follows, we analyze timelike and null geodesics separately, together with the conditions for the existence and stability of circular orbits.

\subsection{Circular photon orbits}

Here, we analyze the motion of massless particles in the T-duality-induced regular spacetime. For null geodesics, for which $\kappa=0$, the effective potential in Eq.~\eqref{eq-pot-geral} is written as
\begin{equation}
V_{\mathrm{ph}}(r)=f(r)\frac{L^2}{r^2}.
\end{equation}
This potential determines the radial dynamics of photons with nonvanishing angular momentum. Null circular orbits are characterized by a constant radial coordinate and therefore satisfy $\dot r=0$ and $\ddot r=0$. Equivalently, from the radial equation, these conditions can be written as
\begin{equation}\label{eq-ci-ph}
E^2=V_{\mathrm{ph}}\big|_{r=r_{\mathrm{ph}}}
\quad
\text{and}
\quad
\frac{dV_{\mathrm{ph}}}{dr}\Bigg|_{r=r_{\mathrm{ph}}}=0.
\end{equation}
Null circular orbits are associated with the extrema of the effective potential $V_{\mathrm{ph}}$. The stability of circular orbits is determined by the sign of the second derivative of the effective potential. In particular, $V_{\mathrm{ph}}''(r_c)<0$ corresponds to a local maximum of the potential and, therefore, to an unstable null circular orbit, whereas $V_{\mathrm{ph}}''(r_c)>0$ characterizes a local minimum and a stable null circular orbit, where $r_{\mathrm{c}}$ is the radius of the circular orbit and a prime denotes differentiation with respect to the radial coordinate $r$. In the present work, we are particularly interested in the unstable outer branch, whose radius we denote by $r_{\mathrm{ph}}$. This branch corresponds to the photon sphere relevant to the dynamics observed in the exterior region of the black hole.

\begin{figure}[ht!]
\centering
\includegraphics[width=0.48\textwidth]{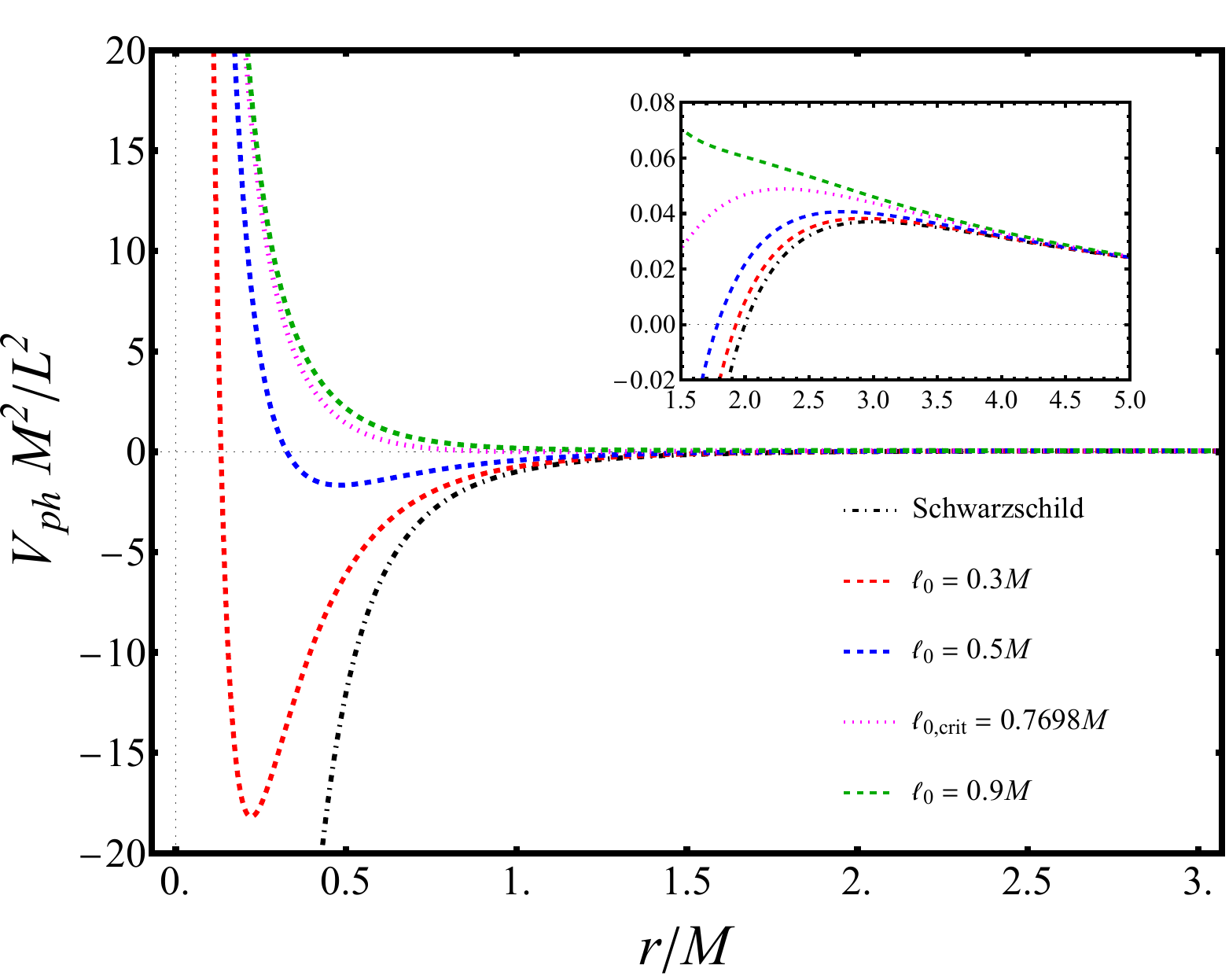}
\caption{\justifying Behavior of the effective potential for photons as a function of $r/M$ for different values of the parameter $\ell_0$, with the inset highlighting the region near the photon sphere.
}
\label{fig:V_ph_M_L}
\end{figure}

In Fig.~\ref{fig:V_ph_M_L}, we show the behavior of the effective potential for photons for different values of the parameter $\ell_0$. We observe that, in the limit $\ell_0\to0$, the potential exhibits a maximum located near $r_{\mathrm{ph}}=3M$, associated with the photon sphere of Schwarzschild spacetime. As $\ell_0$ increases, both the position and the height of this maximum vary, indicating that the T-duality corrections modify the structure of null orbits and the photon-capture dynamics in the vicinity of the regular black hole. For the extremal configuration $\ell_0=\ell_{0,\mathrm{crit}}$, the effective potential only approaches zero tangentially rather than crossing the horizontal axis, reflecting the critical structure of the extremal geometry and characterizing the transition between solutions with horizons and horizonless geometries. In particular, we find that increasing $\ell_0$ smooths the effective potential barrier, increases its maximum height, and shifts the position of the photon sphere. Furthermore, we observe that the regularization can give rise to local minima and, within the black hole sector considered here, these inner extrema lie inside the event horizon. 

\begin{figure}[ht!]
\centering
\includegraphics[width=0.48\textwidth]{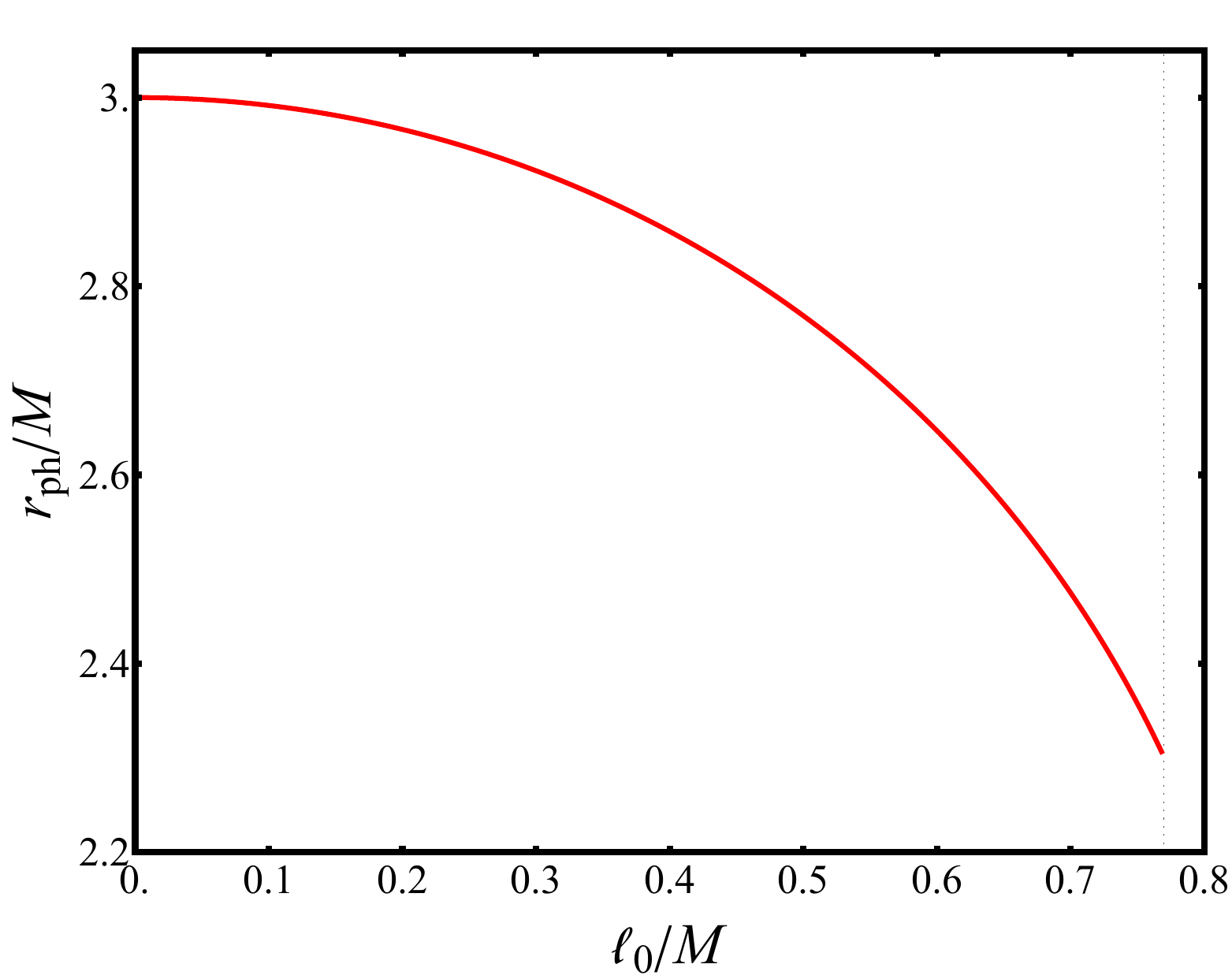}
\caption{\justifying Dependence of the photon-sphere radius on the parameter $\ell_0$. The Schwarzschild limit, for which $r_{\mathrm{ph}}=3M$, is recovered.}
\label{fig:g_r_U_ph_l_0}
\end{figure}

From the conditions given in Eq.~\eqref{eq-ci-ph}, we obtain
\begin{equation}
    r_{\mathrm{ph}}f'(r_{\mathrm{ph}})-2f(r_{\mathrm{ph}})=0.
   \label{eq:photon_tdual}
\end{equation}
The above equation can be solved numerically, allowing us to investigate the dependence of the radius of null circular orbits on the parameter $\ell_0$. In Fig.~\ref{fig:g_r_U_ph_l_0}, we show the dependence of the radius of the unstable outer null circular orbit on $\ell_0$. As $\ell_0$ increases, the photon-sphere radius decreases monotonically, indicating that the T-duality corrections shift the outer null circular orbit toward smaller values of the radial coordinate.

We now extend the analysis of null circular orbits to the charged T-duality-inspired geometry. Since the general condition in Eq.~\eqref{eq:photon_tdual} remains valid, the effect of the electric charge enters entirely through the metric function $f(r)$. In contrast to the neutral case, the radius of the circular photon orbits depends on both $\ell_0/M$ and $Q/M$, and the corresponding equation must in general be solved numerically. To isolate the effect of the electric charge, we fix $\ell_0=0.5M$ and vary $Q$.

\begin{figure}[ht!]
    \centering
    \includegraphics[width=1\linewidth]{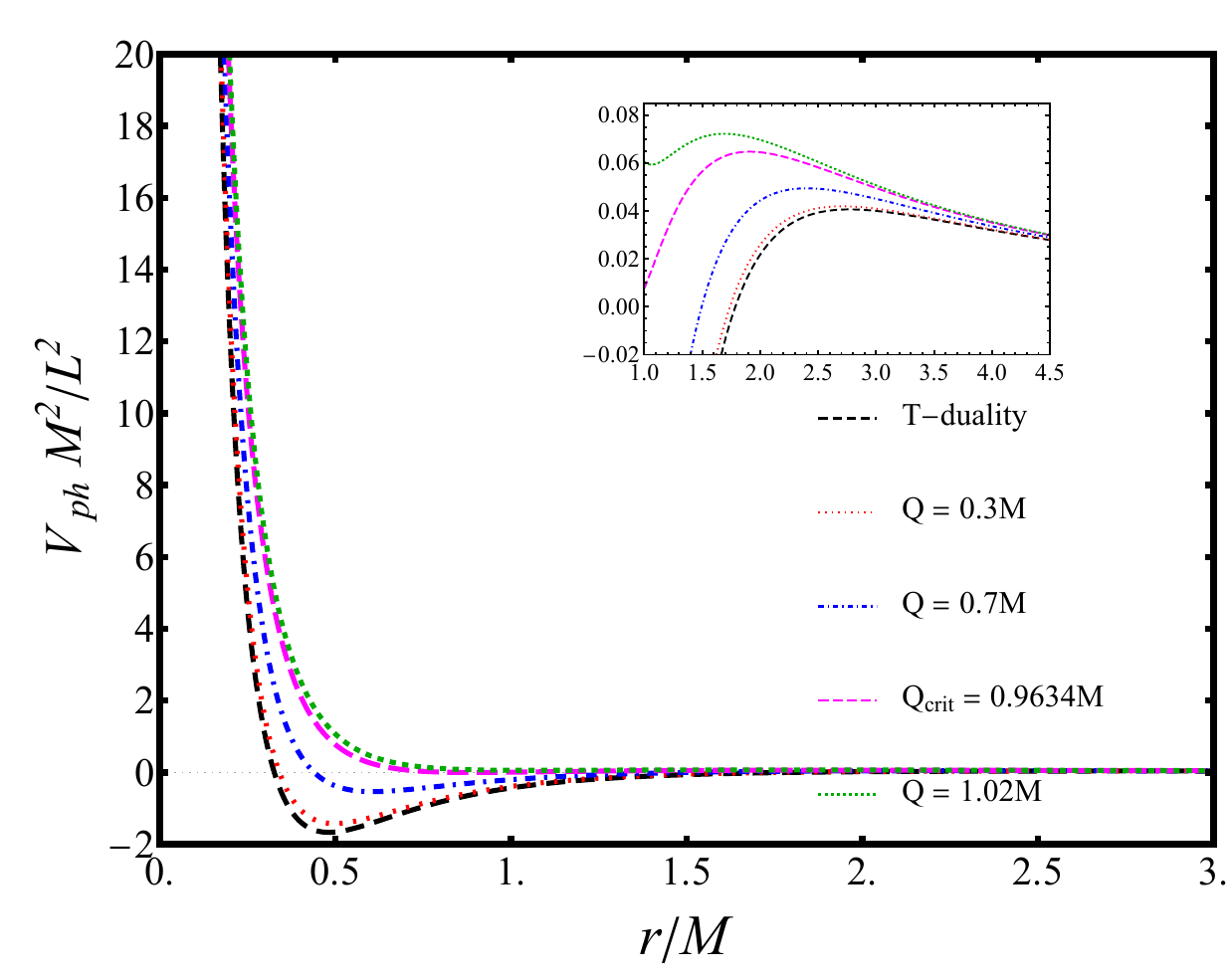}
    \caption{\justifying Behavior of the photon effective potential as a function of $r/M$ for fixed $\ell_0=0.5M$ and different values of the electric charge $Q$. The neutral T-duality solution corresponding to $Q=0$, is included for comparison. The inset highlights the region containing the circular photon orbits.} 
    \label{fig:Vph_charged}
\end{figure}

Figure~\ref{fig:Vph_charged} shows the effective potential for null geodesics for different values of the electric charge,
while keeping $\ell_0=0.5M$ fixed. The neutral T-duality solution, corresponding to $Q=0$, is included for comparison. As the charge increases, the outer maximum of the effective potential moves toward smaller radii and its height increases. Therefore, the electric charge modifies both the location of the unstable circular photon orbit and the height of the effective potential barrier. This behavior is more clearly seen in the enlarged panel, which highlights the region containing the circular photon orbits. For $Q=Q_{\rm crit}\simeq0.9634M$, the spacetime corresponds to the extremal black hole configuration discussed in Sec.~\ref{subsubsec:horizontes}. Nevertheless, increasing the charge beyond this value does not immediately eliminate the circular photon orbits. For instance, at $Q=1.02M$, although the geometry is already horizonless, the effective potential still exhibits an inner local minimum and an outer local maximum, corresponding respectively to stable and unstable circular null orbits. Thus, the disappearance of the event horizon and the disappearance of circular photon orbits occur at different values of the charge.

\begin{figure}[ht!]
    \centering
    \includegraphics[width=1\linewidth]{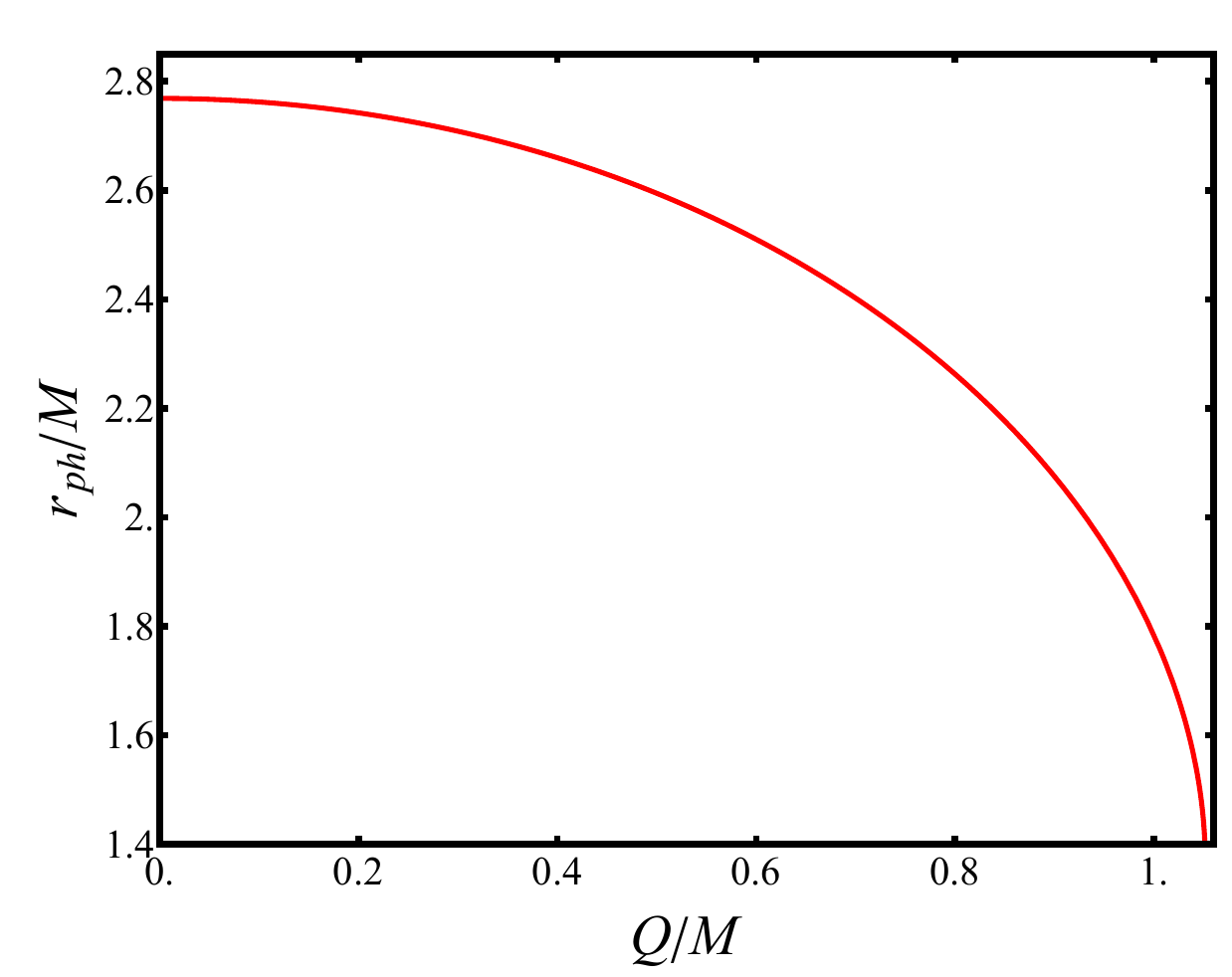}
    \caption{\justifying Dependence of the radius of the outer unstable circular photon orbit on the electric charge $Q/M$ for fixed $\ell_0=0.5M$. The radius decreases as the charge increases, and the branch terminates when the stable and unstable circular null orbits merge.}
\label{fig:rph_charged}
\end{figure}

The dependence of the radius of the outer unstable circular photon orbit on the electric charge is shown in
Fig.~\ref{fig:rph_charged}. Starting from the neutral T-duality value $r_{\rm ph}\simeq2.7687M$ at $Q=0$, the photon orbit radius decreases monotonically as the charge is increased. At the extremal black hole value $Q_{\rm crit}\simeq0.9634M$, we obtain $r_{\rm ph}\simeq1.9110M$. The unstable photon orbit persists beyond the black hole sector; for example, $r_{\rm ph}\simeq1.6922M$ at $Q=1.02M$, where the geometry is already horizonless. As the charge is increased further, the outer unstable branch approaches the inner stable branch. The two circular null orbits eventually merge at $Q_{\rm ph,crit}\simeq1.0517M$ and $r_{\rm ph,crit}\simeq1.3651M$, where the maximum and minimum of the effective potential coalesce. For $Q>Q_{\rm ph,crit}$, the condition for circular null motion admits no real-radius solution. Consequently, there exists a finite interval $Q_{\rm crit}<Q<Q_{\rm ph,crit}$, in which the spacetime is horizonless but still supports circular photon orbits.

\subsection{Circular orbits for massive particles}

Let us now consider the case of massive particles moving in the equatorial plane, for which $\kappa=1$. In this scenario, the effective potential in Eq.~\eqref{eq-pot-geral} takes the form
\begin{equation}
V_{m}(r)
=
f(r)
\left(
1+\frac{L^2}{r^2}
\right),
\label{eq:Vm}
\end{equation}
and, as in the photon case, the effective potential for massive particles also vanishes at the horizons, while in the limit $r \to \infty$ we have $V_{m}=1$.

As discussed above, the existence of circular orbits is determined by the conditions $\dot r=0$ and $\ddot r=0$, which, in terms of the effective potential, are equivalent to
\begin{equation}
E^2=V_{m}(r_c)
\quad \text{and} \quad
\frac{dV_{m}}{dr}\Bigg|_{r=r_c}=0.
\label{cond-orb-circ-mass}
\end{equation}
The stability of these orbits depends on the sign of the second derivative of the effective potential, namely, $V_m''(r_c)>0$ and $V_m''(r_c)<0$, which indicate, respectively, stable orbits, associated with local minima of the massive-particle potential where the particle remains confined around the compact object, and unstable orbits, corresponding to local maxima. This differs from the null case, in which the circular orbit located outside the event horizon corresponds to the outer photon sphere and is unstable.

From the conditions in Eq.~\eqref{cond-orb-circ-mass}, we can determine the energy $E$ and angular momentum $L$ associated with a circular orbit, given respectively by
\begin{equation}
E^2 = \frac{2 f(r_\mathrm{c})^2}{2 f(r_\mathrm{c}) - r_\mathrm{c} f'(r_\mathrm{c})}, \quad
L^2 = \frac{r_\mathrm{c}^3 f'(r_\mathrm{c})}{2 f(r_\mathrm{c}) - r_\mathrm{c} f'(r_\mathrm{c})}.
\label{eq:EL_circular}
\end{equation}
These equations explicitly show how the structure of the metric determines the dynamical properties of circular orbits.

\begin{figure}[ht!]
\begin{subfigure}{0.48\textwidth}
   
    \includegraphics[width=\textwidth]{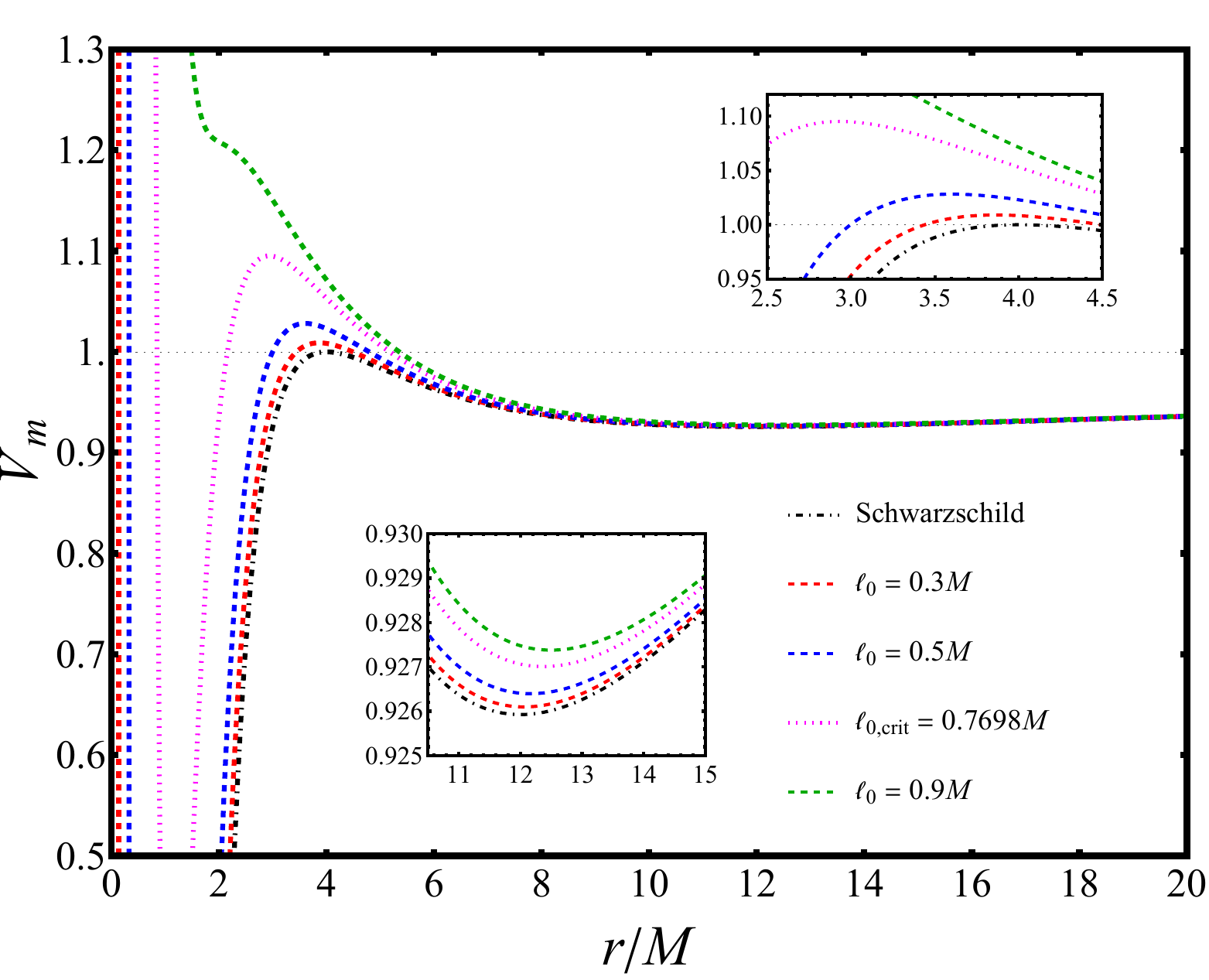}
    \label{fig:g_V_m_L=0.4_l_0}
\end{subfigure}
\hfill
\begin{subfigure}{0.48\textwidth}
    \centering
    \includegraphics[width=\textwidth]{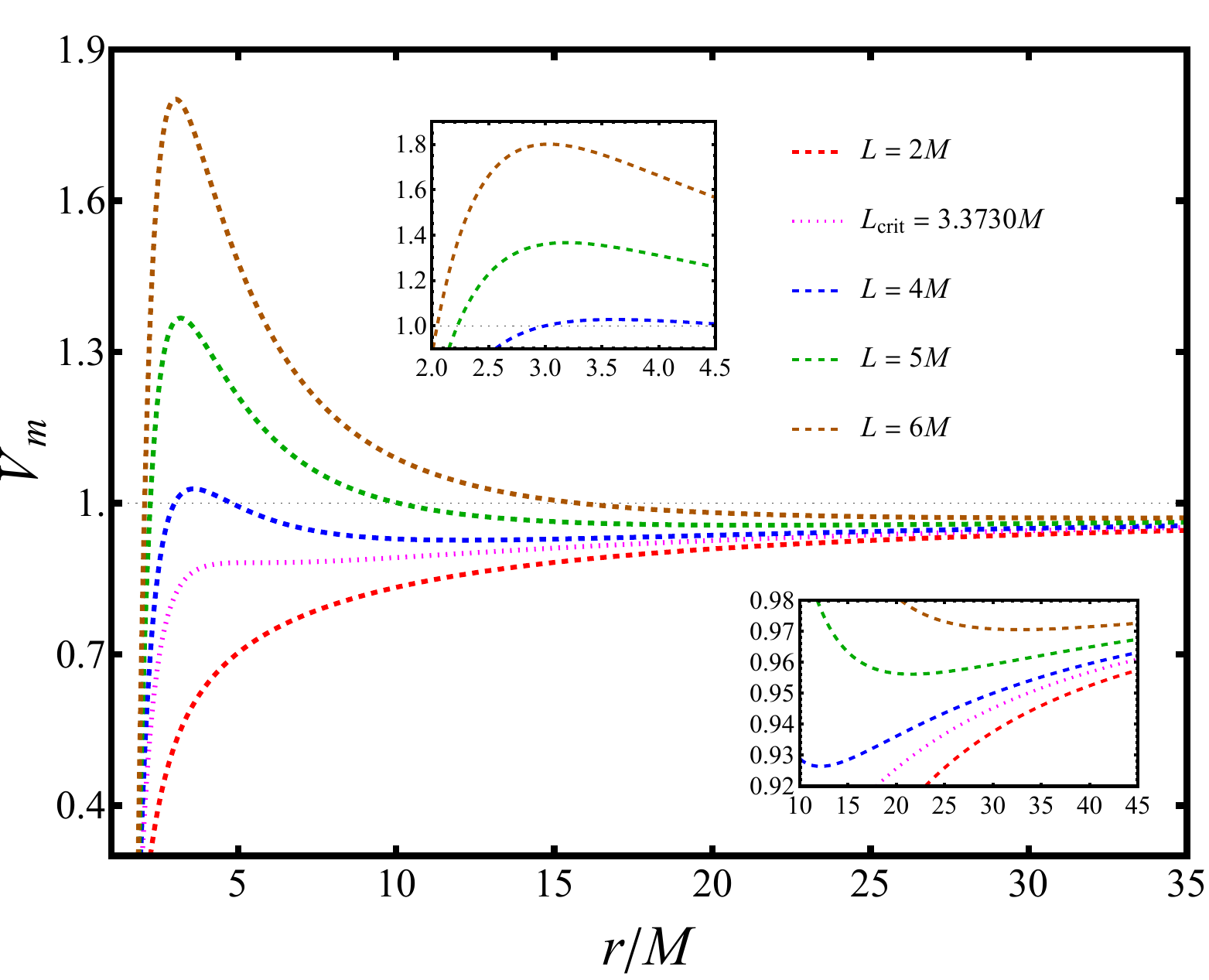}
    \label{fig:g_V_m_l_0=0.5_L}
\end{subfigure}
\caption{\justifying Behavior of the effective potential for massive particles as a function of $r/M$. In the upper panel, we show the curves for several values of the parameter $\ell_0$ with $L=4M$ fixed, while in the lower panel we show the curves for several values of $L$ with $\ell_0=0.5M$ fixed.}
\label{fig:V_m}
\end{figure}

In Fig.~\ref{fig:V_m}, we show the behavior of the  effective potential for massive particles for different values of the parameter $\ell_0$. In the Schwarzschild limit, we recover the well-known behavior of the relativistic effective potential, characterized by a negative divergence in the inner regions of the geometry and by a centrifugal barrier associated with the angular momentum of the particle. As $\ell_0$ increases, we observe modifications in the structure of the effective potential near the central region. In particular, the T-duality corrections smooth out the divergence present in the Schwarzschild case, reducing the depth of the potential and rendering the geometry regular at short distances. For values close to the extremal case, $\ell_0=\ell_{0,\mathrm{crit}}$, the potential develops a critical structure characterized by the presence of a local minimum in the inner regions of the geometry. This behavior indicates the possibility of stable bound orbits near the regular core, a feature absent in the Schwarzschild case. For the regime $\ell_0>\ell_{0,\mathrm{crit}}$, which corresponds to horizonless geometries, the effective potential is defined throughout the entire radial domain and no longer exhibits the negative divergence characteristic of the Schwarzschild case. For nonvanishing angular momentum, however, the centrifugal contribution $L^2/r^2$ produces a positive divergence as $r\rightarrow0$. Therefore, the regularization of the spacetime qualitatively modifies the inner structure of the effective potential, replacing the Schwarzschild-like behavior by an effective centrifugal barrier.

We now extend the analysis of the massive particle effective potential to the charged T-duality-inspired geometry. Since
Eq.~\eqref{eq:Vm} is written entirely in terms of the metric function $f(r)$, the electric charge enters through the charged solution given in Eq.~\eqref{fcharged}. To isolate its effect, we fix $\ell_0=0.5M$ and $L=4M$ and vary the electric charge $Q$, following the same representative values considered in the previous analyses.

\begin{figure}[ht!]
    \centering
    \includegraphics[width=1\linewidth]{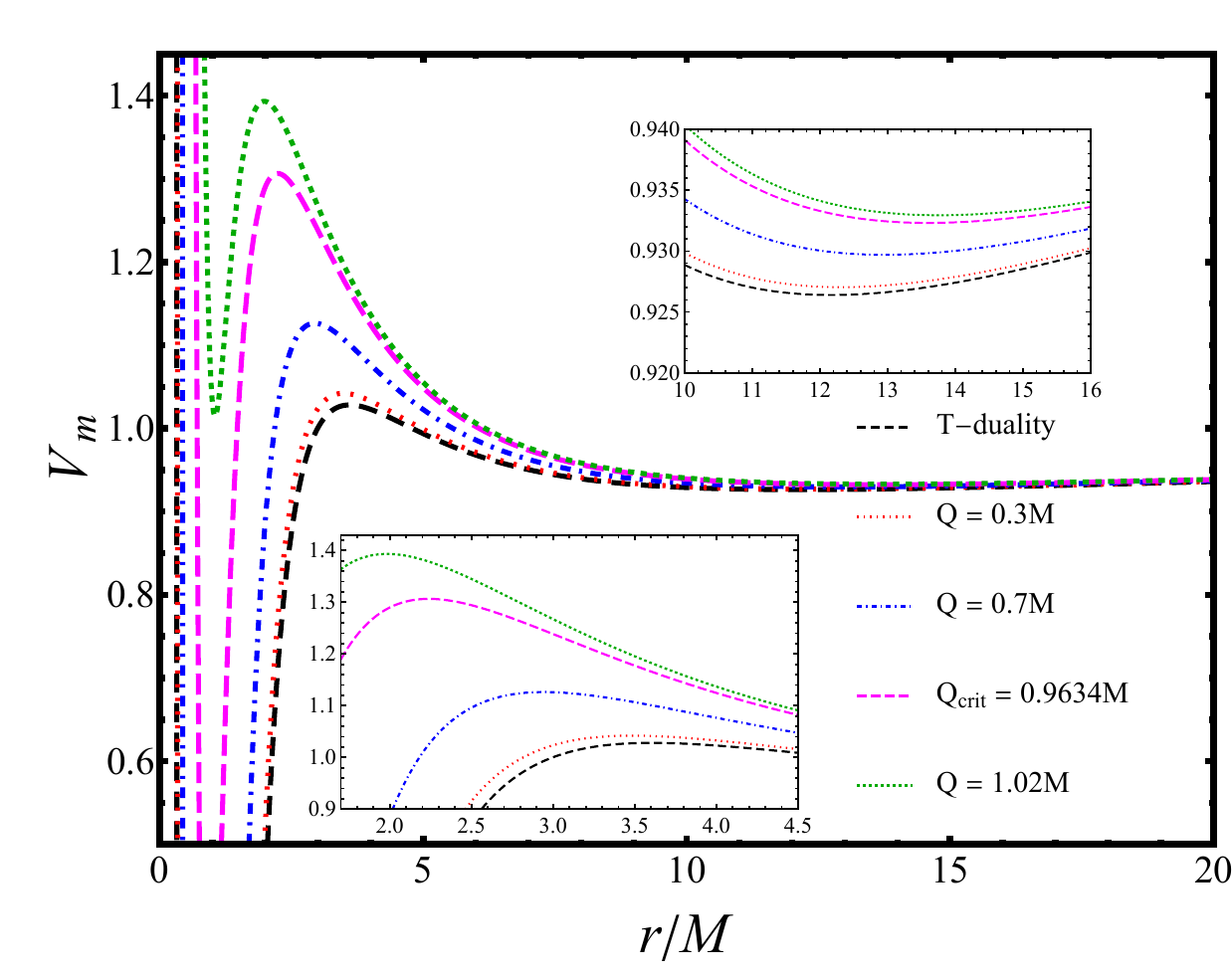}
    \caption{\justifying Behavior of the effective potential for massive particles as a function of $r/M$ for fixed
    $\ell_0=0.5M$ and $L=4M$, and different values of the electric charge $Q$. The neutral T-duality solution, corresponding to $Q=0$, is included for comparison. The enlarged panels highlight the regions containing the outer
    unstable maximum and the outer stable minimum of the effective potential.}
    \label{fig:Vm_charged}
\end{figure}

As shown in Fig.~\ref{fig:Vm_charged}, increasing the electric charge produces significant modifications in both extrema of the effective potential. The outer local maximum, associated with an unstable circular orbit, moves toward smaller radii and its height increases as $Q$ grows. By contrast, the outer local minimum, associated with a stable circular orbit, is shifted toward larger radii, while the corresponding potential well becomes progressively shallower. Therefore, the electric charge affects the stable and unstable branches in qualitatively different ways.

For $Q=Q_{\rm crit}\simeq0.9634M$, the geometry corresponds to the extremal black hole configuration, whereas the curve with $Q=1.02M$ belongs to the horizonless regime. Nevertheless, the effective potential still exhibits extrema in the latter case, showing that the disappearance of the event horizon does not immediately eliminate the circular timelike-orbit structure.

\subsubsection{Stable and unstable circular orbits}

The structure of the effective potential allows us to distinguish different regimes of circular orbits for massive particles. The solutions of the condition $V_m'(r_c)=0$ may correspond to local maxima or minima of the potential, characterizing unstable and stable circular orbits, respectively. The existence and location of these orbits depend on both the angular momentum $L$ and the parameter $\ell_0$.

\begin{figure}[h!]
\begin{subfigure}{0.48\textwidth}
   
    \includegraphics[width=\textwidth]{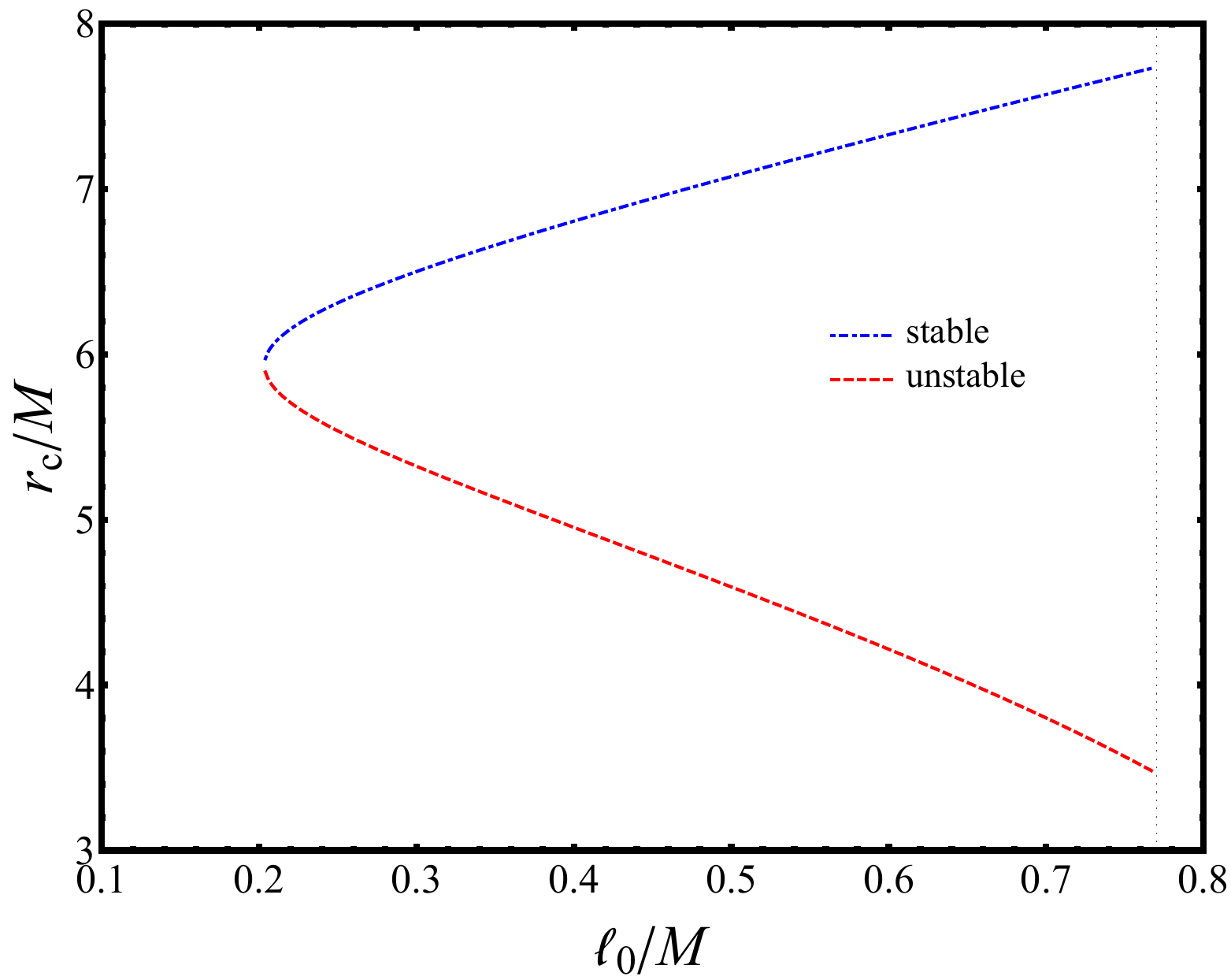}
    \label{fig:g_r_S_U_L=3.45_l_0}
\end{subfigure}
\hfill
\begin{subfigure}{0.48\textwidth}
    \centering
    \includegraphics[width=\textwidth]{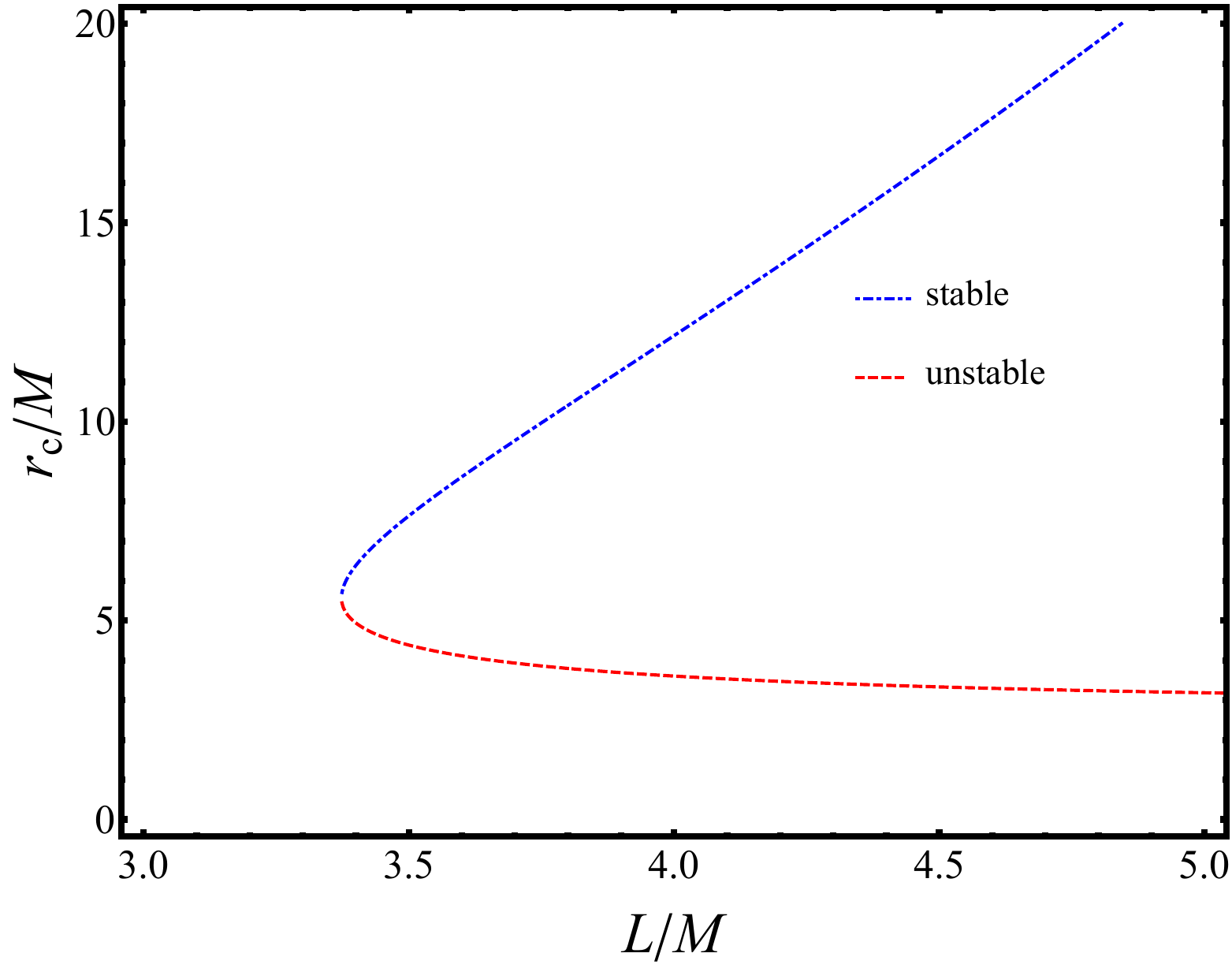}
    \label{fig:g_r_S_U_l_0=0.5_L}
\end{subfigure}
\caption{\justifying Radii of the stable and unstable circular orbits for massive particles. In the upper panel, we show the dependence of the circular-orbit radii on the parameter $\ell_0$, keeping the angular momentum fixed at $L=3.45M$. In the lower panel, we show their dependence on the angular momentum $L$, for $\ell_0=0.5M$.}
\label{fig:r_S_U}
\end{figure}

In Fig.~\ref{fig:r_S_U}, we show the radii of the stable and unstable circular orbits. In the upper panel, we fix the angular momentum at $L=3.45M$ and vary $\ell_0$. We observe that the two branches approach each other as $\ell_0$ decreases, merging at a critical value at which the distinction between a maximum and a minimum of the potential disappears. This point corresponds to a marginally stable circular orbit for the chosen value of $L$. In the lower panel, we fix $\ell_0=0.5M$ and analyze the dependence on the angular momentum. As $L$ decreases, the stable and unstable branches approach each other and merge at a critical value $L=L_{\rm crit}$. At this point, the conditions
\begin{equation}
    V_m'(r_c)=0,
    \qquad
    V_m''(r_c)=0
\end{equation}
are simultaneously satisfied, characterizing the marginally stable circular orbit. For $L<L_{\rm crit}$, this pair of circular orbits ceases to exist.

To analyze more directly how the electric charge modifies the circular orbit structure, we now fix $\ell_0=0.5M$ and
$L=3.45M$ and determine the radii of the exterior stable and unstable circular orbits as functions of $Q/M$. We restrict this analysis to the black hole sector, $0\leq Q\leq Q_{\rm crit}$.

\begin{figure}[ht!]
    \centering
    \includegraphics[width=1\linewidth]{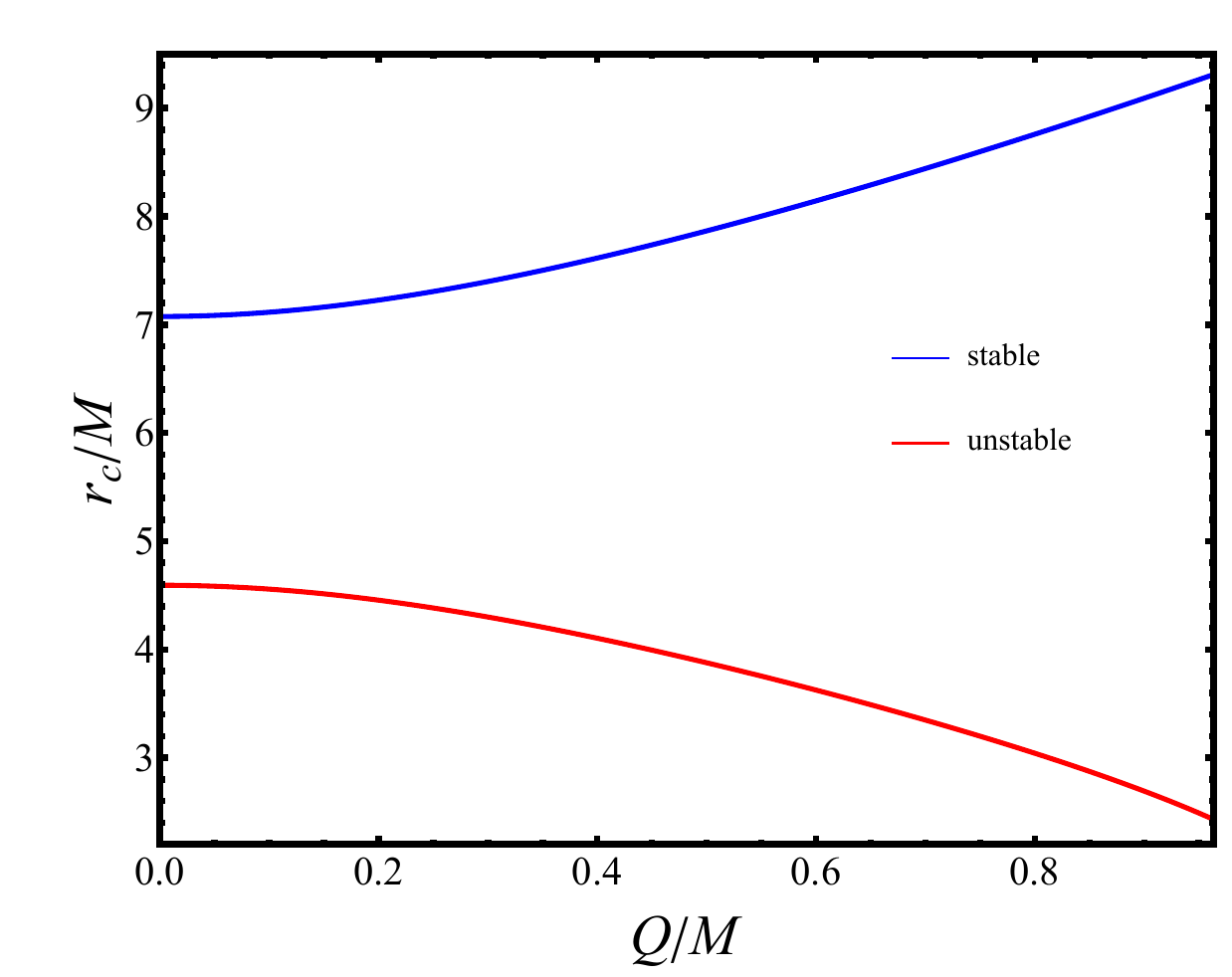}
    \caption{\justifying Radii of the exterior stable and unstable circular orbits for massive particles as functions
    of the electric charge $Q/M$, for fixed $\ell_0=0.5M$ and $L=3.45M$. The blue and red curves correspond, respectively, to the stable and unstable circular orbit branches.}
    \label{fig:rc_charged}
\end{figure}

Figure~\ref{fig:rc_charged} shows that the electric charge drives the two exterior circular orbit branches in opposite
directions. As $Q$ increases, the unstable circular orbit moves progressively toward smaller radii, whereas the stable circular orbit is shifted outward. Consequently, the radial separation between the two branches increases throughout the black hole sector. This behavior is consistent with the modifications of the maximum and minimum of the effective potential displayed in Fig.~\ref{fig:Vm_charged}.

\subsubsection{Innermost stable circular orbit}

The innermost stable circular orbit (ISCO) corresponds to the marginally stable circular orbit that separates the stable and unstable circular-orbit branches. Its location is determined by the conditions
\begin{equation}
    V_m'(r_{\mathrm{I}})=0,
    \qquad
    V_m''(r_{\mathrm{I}})=0.
    \label{eq:isco_conditions}
\end{equation}
From these conditions, we obtain the following equation for the ISCO radius:
\begin{equation}
    r_{\mathrm{I}} f(r_{\mathrm{I}})
    f''(r_{\mathrm{I}})
    -2r_{\mathrm{I}}
    \left[f'(r_{\mathrm{I}})\right]^2
    +3f(r_{\mathrm{I}})
    f'(r_{\mathrm{I}})
    =0.
    \label{eq:isco}
\end{equation}
This expression determines the innermost stable circular orbit directly in terms of the metric function $f(r)$. In the Schwarzschild limit, the above equation reproduces the standard result $r_{\rm ISCO}=6M$. For the regular metric function considered here, condition \eqref{eq:isco} leads to a nonlinear equation for $r_{\mathrm{ISCO}}$, which we solve numerically as a function of $\ell_0$.

\begin{figure}[h!]
\begin{subfigure}{0.48\textwidth}
   
    \includegraphics[width=\textwidth]{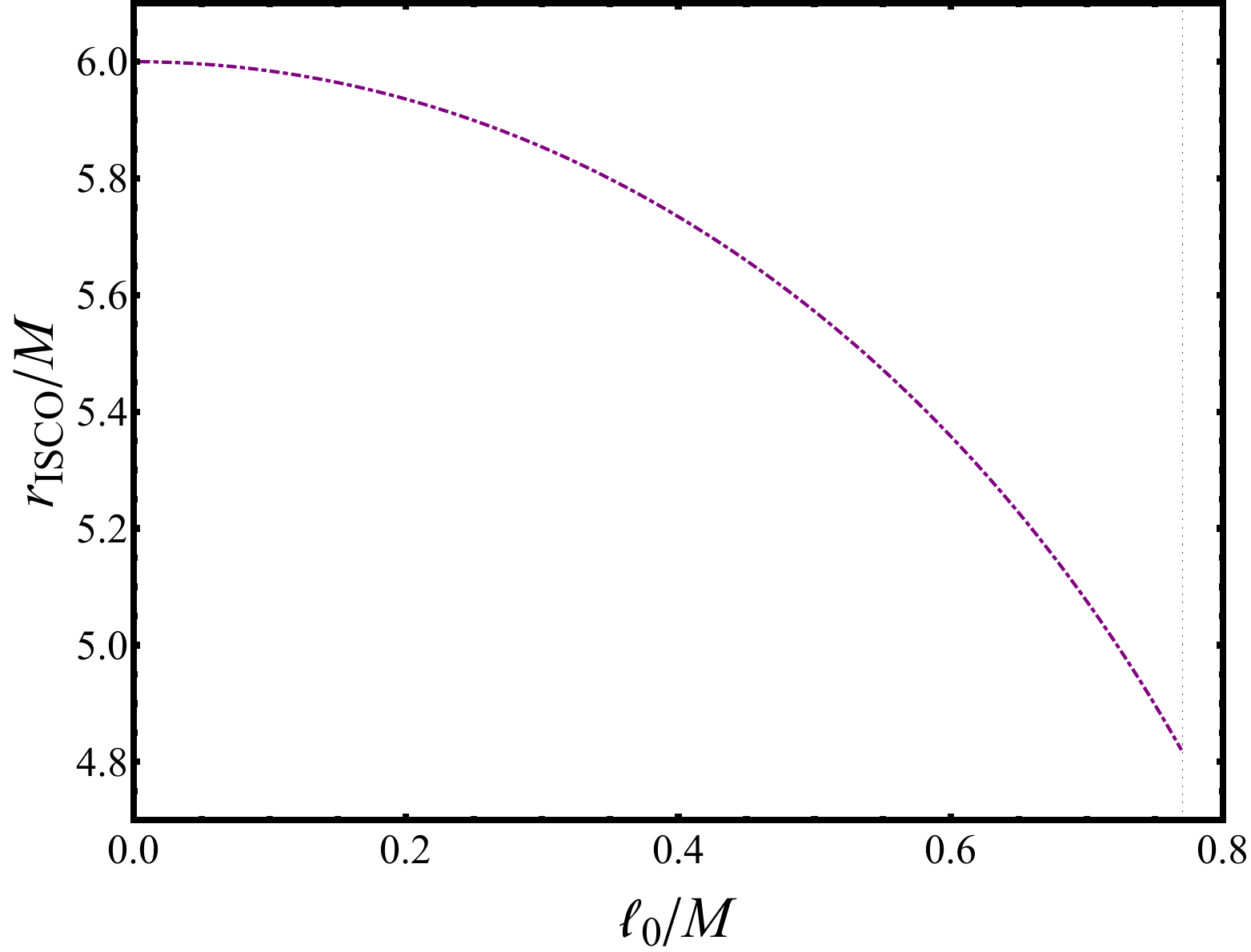}
    \label{fig:g_r_ISCO_l_0}
\end{subfigure}
\hfill
\begin{subfigure}{0.48\textwidth}
    \centering
    \includegraphics[width=\textwidth]{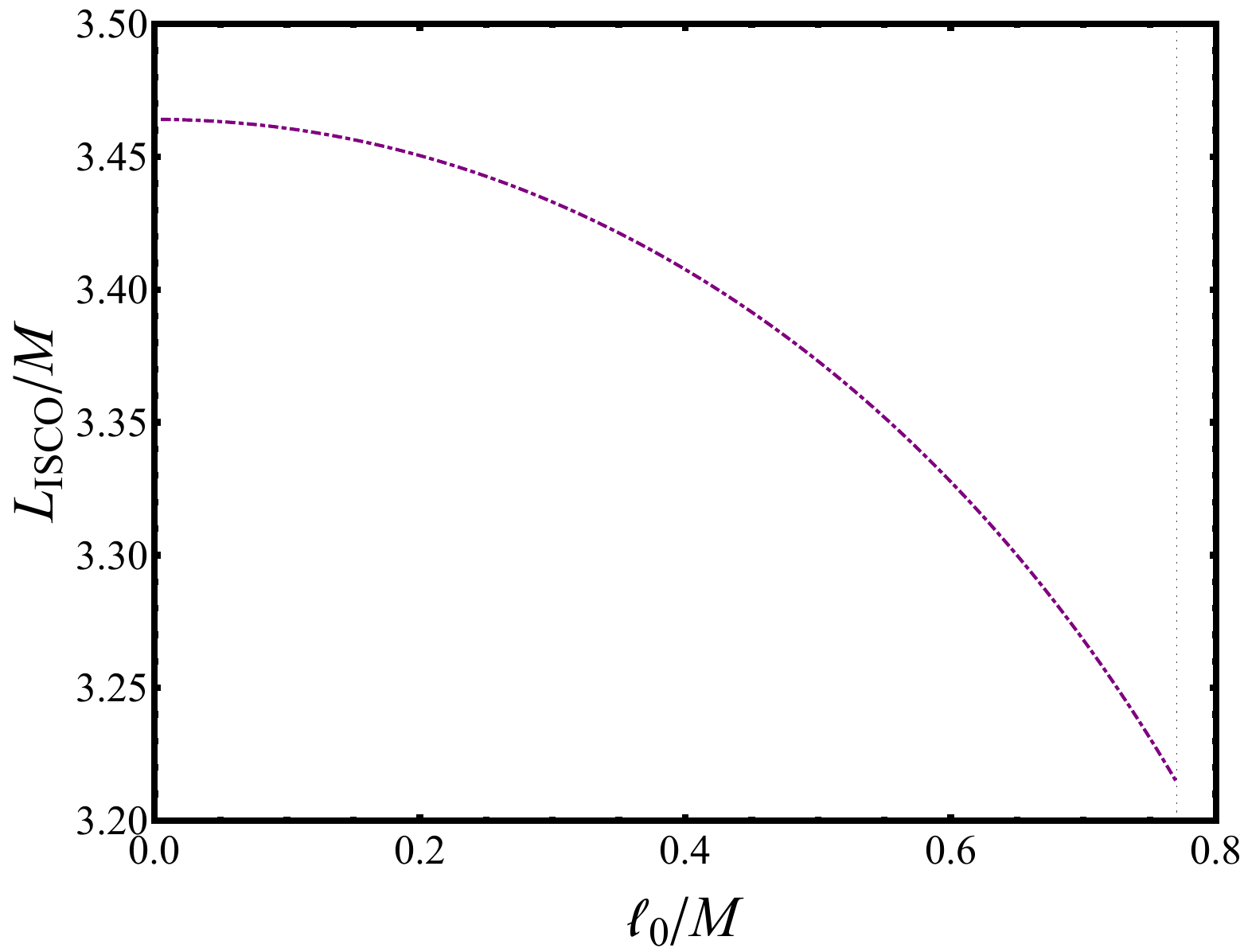}
    \label{fig:_L_ISCO_l_0}
\end{subfigure}
\hfill
\begin{subfigure}{0.48\textwidth}
    \centering
    \includegraphics[width=\textwidth]{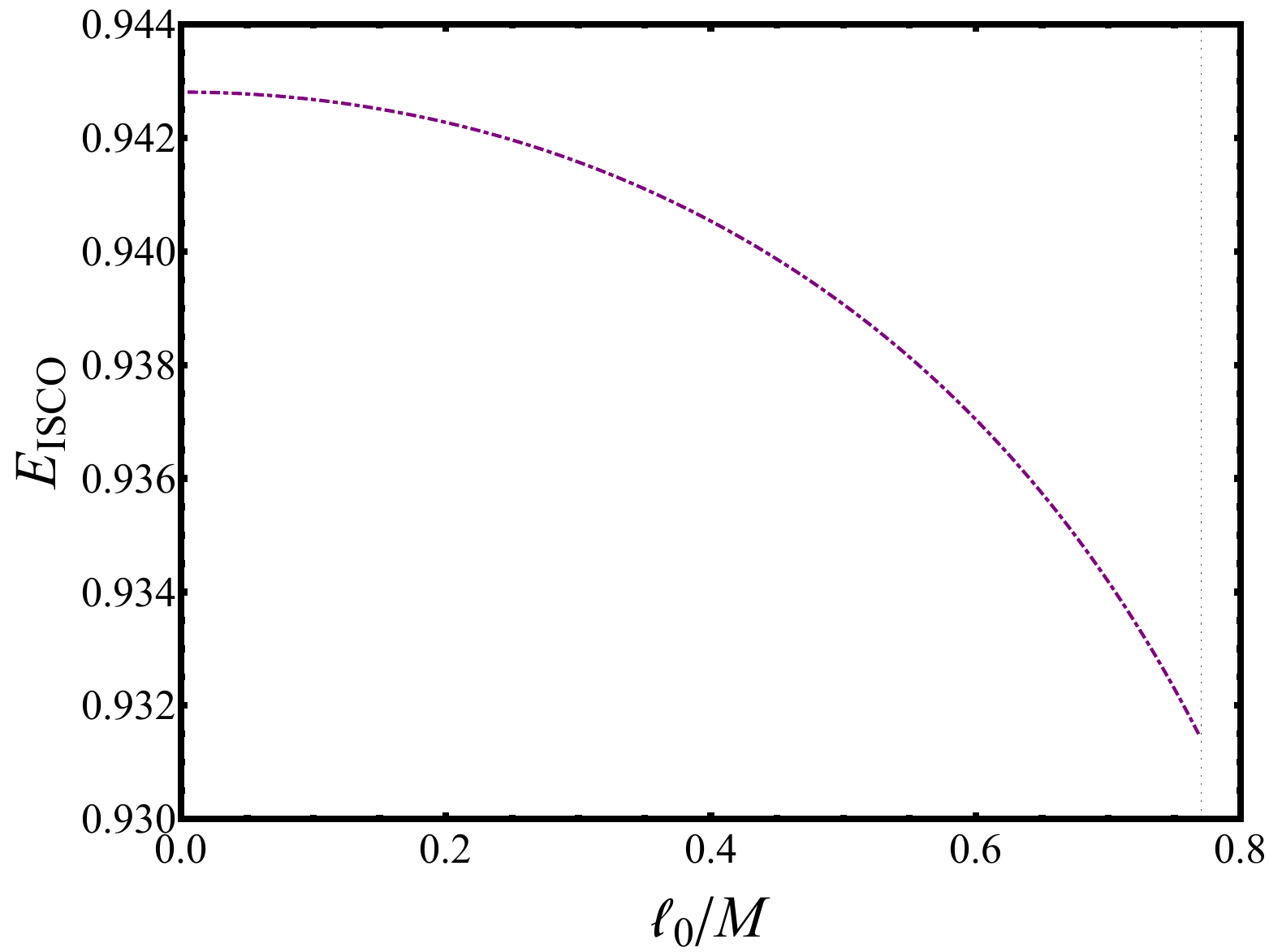}
    \label{fig:g_E_ISCO_l_0}
\end{subfigure}
\caption{\justifying Dependence of the radius $r_{\mathrm{ISCO}}$ (upper panel), angular momentum $L_{\mathrm{ISCO}}$ (middle panel), and energy $E_{\mathrm{ISCO}}$ (lower panel) on the parameter $\ell_0/M$. In the limit $\ell_0\rightarrow0$, the corresponding Schwarzschild values are recovered.}
\label{fig:r_L_E_ISCO}
\end{figure}

In Fig.~\ref{fig:r_L_E_ISCO}, we show the dependence of $r_{\mathrm{ISCO}}$, $L_{\mathrm{ISCO}}$, and $E_{\mathrm{ISCO}}$ on $\ell_0/M$. For small values of $\ell_0$, the corrections are mild and the results remain close to the Schwarzschild limit. However, as the parameter increases, we observe a monotonic decrease in all three quantities. In particular, the decrease in $r_{\mathrm{ISCO}}$ shows that the corrections associated with T-duality allow stable circular orbits to exist at smaller radii than in the Schwarzschild case. At the same time, the decrease in $L_{\mathrm{ISCO}}$ indicates that a smaller angular momentum is required to sustain the marginally stable orbit, while the decrease in $E_{\mathrm{ISCO}}$ corresponds to more strongly bound orbits.

We finally extend the ISCO analysis to the charged T-duality-inspired geometry. Since Eq.~\eqref{eq:isco} is written solely in terms of the metric function and its derivatives, the same condition applies directly to the charged solution. We therefore fix $\ell_0=0.5M$ and solve Eq.~\eqref{eq:isco} numerically as a function of $Q/M$ within the black hole sector. Once $r_{\rm ISCO}$ is determined, the corresponding values of $L_{\rm ISCO}$ and $E_{\rm ISCO}$ are obtained from the circular orbit relations given above.

\begin{figure}[ht!]
\begin{subfigure}{0.48\textwidth}
    \centering
    \includegraphics[width=\textwidth]{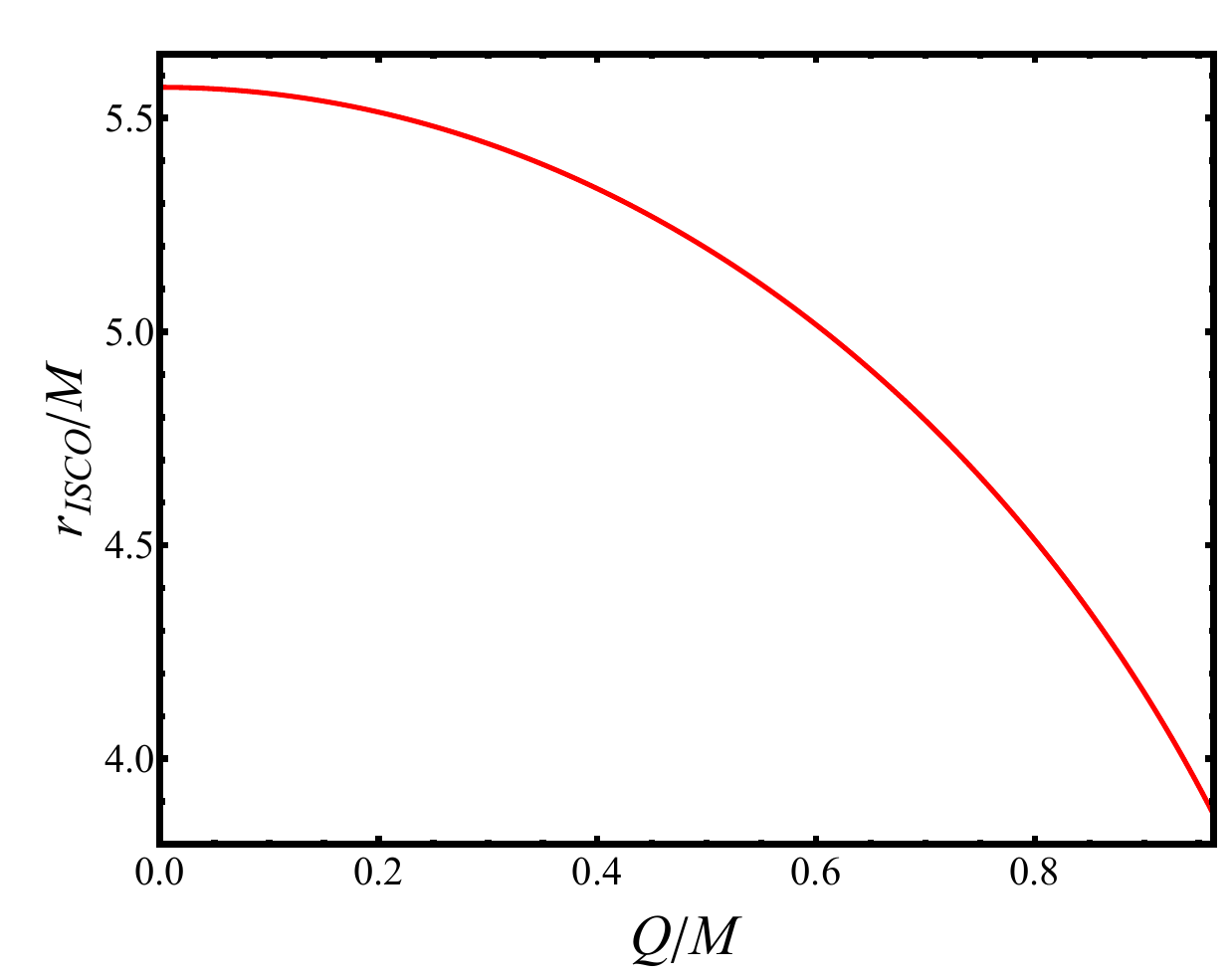}
\end{subfigure}
\hfill
\begin{subfigure}{0.48\textwidth}
    \centering
    \includegraphics[width=\textwidth]{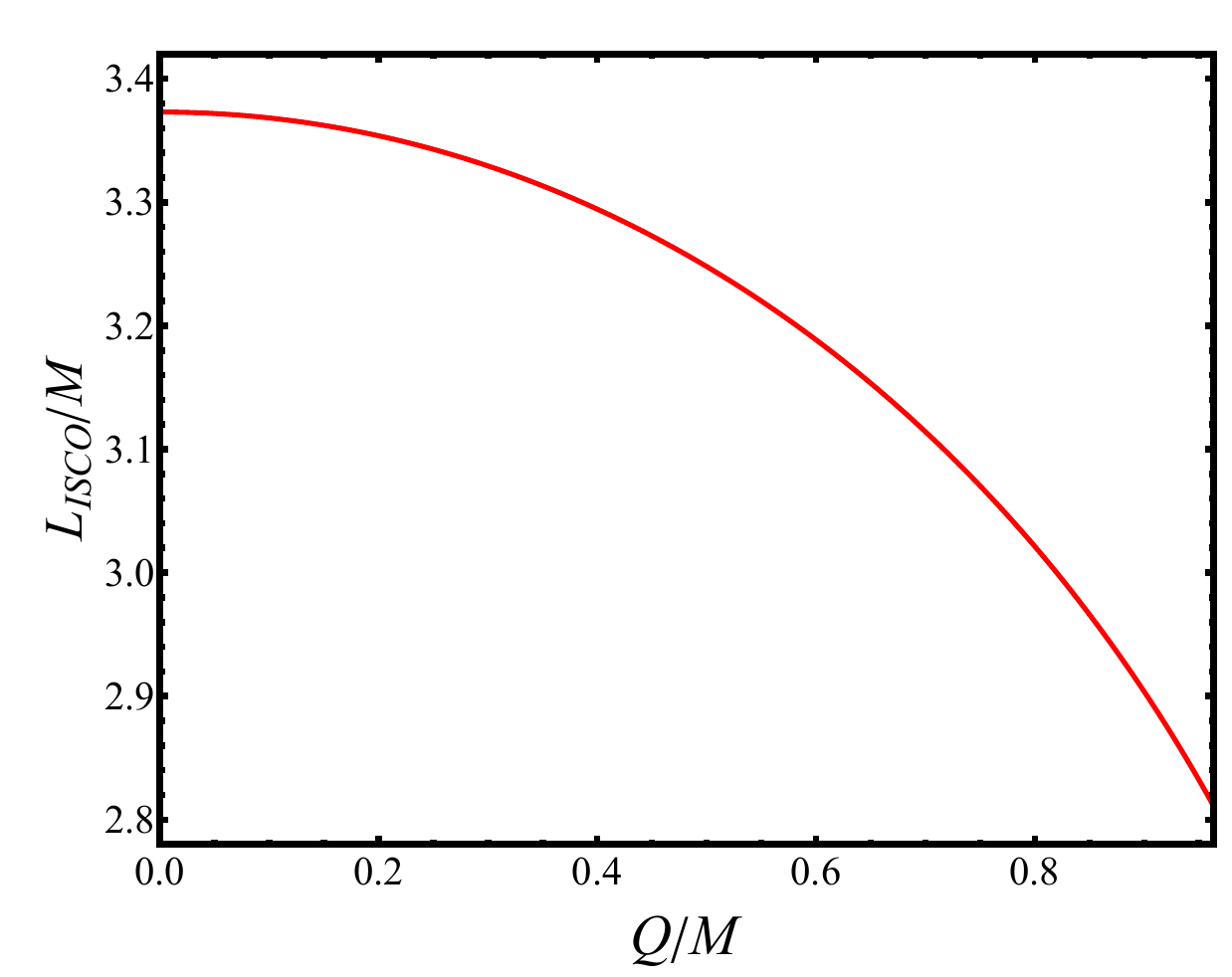}
\end{subfigure}
\vspace{0.15cm}
\begin{subfigure}{0.48\textwidth}
    \centering
    \includegraphics[width=\textwidth]{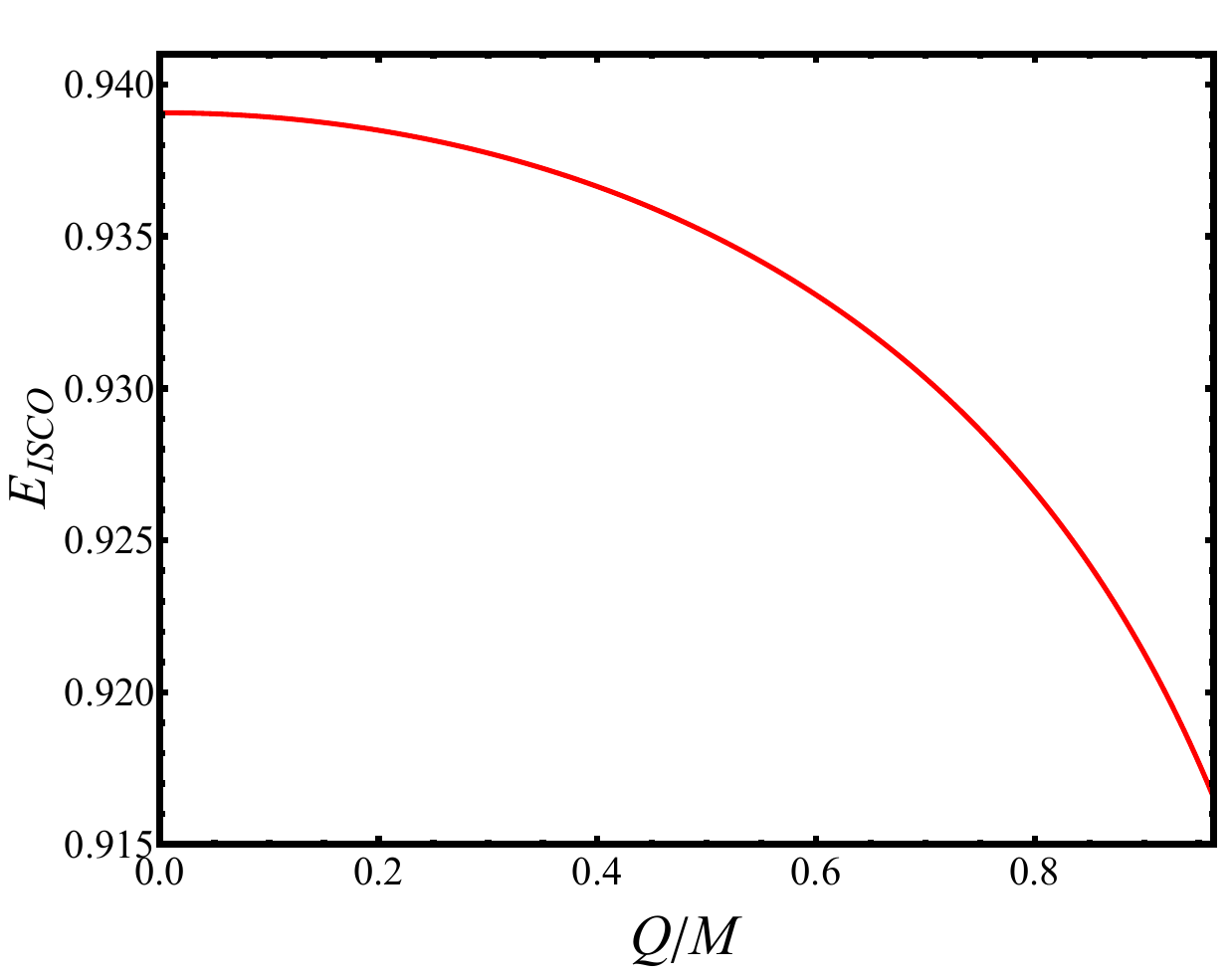}
\end{subfigure}
\caption{\justifying Dependence of the ISCO properties on the electric charge $Q/M$ for fixed $\ell_0=0.5M$. From top to bottom, the panels show the ISCO radius $r_{\mathrm{ISCO}}$, angular momentum $L_{\mathrm{ISCO}}$, and energy $E_{\mathrm{ISCO}}$, respectively. The curves extend throughout the black hole sector, up to the extremal configuration $Q=Q_{\rm crit}$.}
\label{fig:ISCO_charged}
\end{figure}

As shown in Fig.~\ref{fig:ISCO_charged}, all three ISCO quantities decrease monotonically as the electric charge is increased. The decrease in $r_{\rm ISCO}$ indicates that the marginally stable circular orbit moves progressively closer to the compact object. At the same time, $L_{\rm ISCO}$ decreases, showing that a smaller angular momentum is required to sustain the marginally stable orbit, while the decrease in $E_{\rm ISCO}$ indicates that the corresponding orbit becomes more strongly bound.

Therefore, the electric charge enhances the inward displacement of the ISCO already produced by the T-duality corrections. For fixed $\ell_0=0.5M$, increasing $Q$ from the neutral configuration toward the extremal black hole significantly changes the characteristic radius, angular momentum, and energy of the innermost stable circular orbit.

\subsection{Radial motion of massive particles}

Let us now consider the case of massive particles ($\kappa=1$) moving radially in the regularized geometry. Radial trajectories are characterized by vanishing angular momentum, $L=0$, so that the motion takes place along radial lines with no angular component. Equation~\eqref{eq-radial} therefore reduces to
\begin{equation}\label{eq-radial-mass}
\dot r^2 + f(r)=E^2 \implies \frac{dr}{d \tau}=\pm \sqrt{E^2-f(r)}.
\end{equation}
In this case, we see that the effective potential reduces simply to the metric function.

For a test particle undergoing radial free fall from rest at the position $b$~\cite{Martel:2001yf, Crispino:2016pnv}, we obtain the following relation for the energy:
\begin{equation}
    E=\sqrt{f(r=b)}.
\end{equation}
We can also obtain the Newtonian-like acceleration exerted by the T-duality regular black hole on the test particle, which is defined as~\cite{Crispino:2016pnv, Uniyal:2022ouc, Martel:2001yf, Liu:2022lrg}
$$
A^{(N)}\equiv \ddot{r}.
$$
From Eq.~\eqref{eq-radial-mass}, we infer that
\begin{equation}\label{eq:newtonian}
    A^{(N)}=-\frac{f'(r)}{2},
\end{equation}
This expression is valid for both the neutral and charged geometries.

For the charged metric function given in Eq.~\eqref{fcharged}, Eq.~\eqref{eq:newtonian}
takes the form
\begin{align}
A^{(N)}_Q={}&-\frac{Mr\left(r^2-2\ell_0^2\right)}{\left(r^2+\ell_0^2\right)^{5/2}}+\frac{Q^2r\left(r^2-\ell_0^2\right)}{\left(r^2+\ell_0^2\right)^3}F(r)\nonumber\\
&-\frac{Q^2r^2}{2\left(r^2+\ell_0^2\right)^2}F'(r).
\label{eq:radial_acc_charged}
\end{align}
The neutral T-duality result is recovered by setting $Q=0$ in
Eq.~\eqref{eq:radial_acc_charged}, yielding
\begin{equation}
    A^{(N)}
    =
    -\frac{Mr(r^2-2\ell_0^2)}
    {(r^2+\ell_0^2)^{5/2}}.
    \label{eq:radial_acc_tdual}
\end{equation}
In the limit $\ell_0\rightarrow0$, we recover $A^{N}\longrightarrow-M/r^2$, which formally coincides with the Newtonian radial gravitational acceleration.

\section{TIDAL FORCES}
\label{sec_IV}

In this section, we investigate the tidal forces experienced by test particles undergoing radial free fall in the T-duality-inspired regular spacetime. These effects can be characterized through the geodesic deviation equation, which describes the relative acceleration between neighboring geodesics and, consequently, the stretching or compression of an extended body in free fall. Considering the separation vector $\xi^{\mu}$ between two infinitesimally close geodesics, its spatial components satisfy the geodesic deviation equation~\cite{dinverno1992,hobson2006},
\begin{equation}
\frac{D^2 \xi^\mu}{D\tau^2}
=
K^\mu_{\nu}\, \xi^\nu,
\end{equation}
where $K^\mu{}_{\nu}$ is the tidal tensor, which can be expressed in terms of the Riemann tensor as
\begin{equation}
K^\mu_{\nu}=R^\mu_{\;\alpha\beta\nu}u^\alpha u^\beta,
\end{equation}
with $u^\mu$ denoting the four-velocity of the test particle, normalized according to $g_{\alpha\beta}u^\alpha u^\beta=-1$. Regarding the Riemann tensor, a set of independent nonvanishing components for the spherically symmetric spacetime considered here~\cite{Wald:1984rg} is given by
\begin{align}
R^r_{\;t r t} &= -\frac{f(r)\,f''(r)}{2}, \\
R^r_{\;\theta r \theta} &= \frac{r\,f'(r)}{2}, \\
R^r_{\;\phi r \phi} &= \frac{r\,f'(r)}{2}\,\sin^2\theta, \\
R^\theta_{\;t \theta t} &= -\frac{f(r)\,f'(r)}{2r}, \\
R^\phi_{\;t \phi t} &= -\frac{f(r)\,f'(r)}{2r}, \\
R^\theta_{\;\phi \theta \phi} &= -\left(1 - f(r)\right)\sin^2\theta,
\end{align}
expressed in the coordinate basis $(t,r,\theta,\phi)$.

\subsection{Tetrad formalism}

To interpret the effects of tidal forces measured by an observer in radial free fall toward a black hole, we project the Riemann tensor onto an orthonormal tetrad basis. The tetrad formalism~\cite{Crispino:2016pnv,Mitsou:2019nhj,Hong:2020bdb,Arora:2023ltv,Crispim:2025cql,Silva:2025hwl} defines a locally inertial basis at each point of spacetime. The relation between the coordinate basis $\vec{e}_{\mu}$ and the orthonormal tetrad basis $\vec{e}_{\hat{i}}$ is given by
\begin{equation}
\vec{e}_{\hat{i}} = \hat{e}_{\hat{i}}^{\,\,\mu}\,\vec{e}_{\mu},
\end{equation}
where $\hat{e}_{\hat{i}}^{\,\,\mu}$ are the tetrad components (hatted indices denote components in the orthonormal frame). The inverse metric can be expressed in terms of the tetrad vectors as~\cite{Mitsou:2019nhj}:
\begin{equation}
g^{\mu\nu} = \eta^{\hat{i}\hat{j}}\,\hat{e}_{\hat{i}}^{\,\,\mu}\,\hat{e}_{\hat{j}}^{\,\,\nu},
\end{equation}
where $\eta_{\hat{i}\hat{j}} = \mathrm{diag}(-1,1,1,1)$ is the Minkowski metric in the local orthonormal frame.

For an observer in radial free fall, we can choose the following set of tetrad vectors~\cite{Crispino:2016pnv,Lima:2020wcb,Crispim:2025cql}:
\begin{align}
\hat{e}_{\hat{t}}^{\;\mu} &= \left( \frac{E}{f(r)}, -\sqrt{E^2 - f(r)}, 0, 0 \right), \\
\hat{e}_{\hat{r}}^{\;\mu} &= \left( -\frac{\sqrt{E^2 - f(r)}}{f(r)}, E, 0, 0 \right), \\
\hat{e}_{\hat{\theta}}^{\;\mu} &= \left( 0, 0, \frac{1}{r}, 0 \right), \\
\hat{e}_{\hat{\phi}}^{\;\mu} &= \left( 0, 0, 0, \frac{1}{r\sin\theta} \right).
\end{align}
We can now express the components of the Riemann tensor in the tetrad frame, which also allows us to determine the components of the tidal tensor itself. In the orthonormal frame of an observer in radial free fall, the tidal tensor takes the diagonal form
\begin{equation}
K^{\hat{i}}_{\;\;\hat{j}} = \mathrm{diag}(k_1, k_2, k_2),
\end{equation}
where $k_1$ corresponds to the radial component and $k_2$ to the angular components. The components relevant to the tidal forces are those involving two local temporal directions, for which the above equations yield
\begin{equation}
R^{\hat{r}}_{\;\hat{t}\hat{r}\hat{t}}
=
-\frac{f''(r)}{2},
\quad
R^{\hat{\theta}}_{\;\hat{t}\hat{\theta}\hat{t}}
=
-\frac{f'(r)}{2r},
\quad
R^{\hat{\phi}}_{\;\hat{t}\hat{\phi}\hat{t}}
=
-\frac{f'(r)}{2r},
\end{equation}
and consequently,
\begin{equation}
k_1 = -\frac{f''(r)}{2},
\qquad
k_2 = -\frac{f'(r)}{2r}.
\end{equation}
In the following, we discuss in detail the radial and angular components of the tidal forces.

\subsection{Tidal-force effects along radial geodesics}

Returning to the geodesic deviation equation, now projected onto the observer's orthonormal frame, we have
\begin{equation}
\frac{d^2 \xi^{\hat{i}}}{d\tau^2} = K^{\hat{i}}_{\;\hat{j}} \, \xi^{\hat{j}}.
\end{equation}
Since the tidal tensor is diagonal, the equations decouple, and substituting the expressions for $k_1$ and $k_2$, we obtain
\begin{align}
\frac{d^2 \xi^{\hat{r}}}{d\tau^2}
&=
-\frac{f''(r)}{2} \, \xi^{\hat{r}}
\label{eq:geodesic_deviation_radial},
\\
\frac{d^2\xi^{\hat{\theta}}}{d\tau^2}
&=
-\frac{f'(r)}{2r} \, \xi^{\hat{\theta}}
\label{eq:geodesic_deviation_theta},
\\
\frac{d^2\xi^{\hat{\phi}}}{d\tau^2}
&=
-\frac{f'(r)}{2r} \, \xi^{\hat{\phi}}
\label{eq:geodesic_deviation_phi}.
\end{align}

We now specialize the general expressions for the tidal tensor to the T-duality-inspired geometries considered in this work. We first consider the charged solution introduced in Eq.~\eqref{fcharged}. Substituting the corresponding metric function into the general expressions for $k_1$ and $k_2$, we obtain
\begin{align}
k_1^{(Q)}={}&\frac{M\left(2\ell_0^4-11\ell_0^2r^2+2r^4\right)}{\left(r^2+\ell_0^2\right)^{7/2}}\nonumber\\
&+Q^2\Bigg[\frac{-\ell_0^4+8\ell_0^2r^2-3r^4}{\left(r^2+\ell_0^2\right)^4}F(r)\nonumber\\
&\qquad+\frac{2r\left(r^2-\ell_0^2\right)}{\left(r^2+\ell_0^2\right)^3}F'(r)-\frac{r^2}{2\left(r^2+\ell_0^2\right)^2}F''(r)\Bigg].
\label{eq:k1_charged}
\end{align}

\begin{align}
k_2^{(Q)}={}&\frac{M\left(2\ell_0^2-r^2\right)}{\left(r^2+\ell_0^2\right)^{5/2}}\nonumber\\
&+Q^2\left[\frac{r^2-\ell_0^2}{\left(r^2+\ell_0^2\right)^3}F(r)-\frac{r}{2\left(r^2+\ell_0^2\right)^2}F'(r)\right].
\label{eq:k2_charged}
\end{align}

The neutral T-duality geometry is recovered by setting $Q=0$. In this limit, Eqs.~\eqref{eq:k1_charged} and
\eqref{eq:k2_charged} reduce to
\begin{equation}\label{eq-k1}
    k_1=\frac{M(2\ell_0^4 - 11\ell_0^2 r^2 + 2r^4)}{(r^2 + \ell_0^2)^{7/2}},
\end{equation}
and
\begin{equation}\label{eq-k2}
    k_2= \frac{M(2\ell_0^2 - r^2)}{(r^2 + \ell_0^2)^{5/2}}.
\end{equation}

As a further consistency check, if the T-duality scale is removed while the electric charge is kept finite, $\ell_0\rightarrow0$, the charged geometry reduces to the Reissner--Nordstr\"om spacetime. Accordingly, the tidal components become
\begin{equation}
    k_1^{\rm RN}=\frac{2M}{r^3}-\frac{3Q^2}{r^4},\qquad k_2^{\rm RN}=-\frac{M}{r^3}+\frac{Q^2}{r^4}.
    \label{eq:k1-k2-RN}
\end{equation}

Finally, by further setting $Q=0$, or equivalently by taking $\ell_0\rightarrow0$ in the neutral solution, we recover the
Schwarzschild behavior
\begin{equation}\label{k1-k2-schw}
k_1 = \frac{2M}{r^3}, \quad k_2 = -\frac{M}{r^3}.
\end{equation}

To examine the behavior of the tidal forces at large distances, we expand the charged tidal components in the asymptotic regime $r\gg\ell_0$. This gives
\begin{align}
k_1^{(Q)}(r\gg\ell_0)\approx&\frac{2M}{r^3}-\frac{3Q^2}{r^4}\nonumber\\
&-\frac{18M\ell_0^2}{r^5}+\frac{27\pi Q^2\ell_0}{16r^5} +\mathcal{O}\left(\frac{1}{r^6}\right),\label{eq:k1_asymp_charged}\\
k_2^{(Q)}(r\gg\ell_0)\approx&-\frac{M}{r^3}+\frac{Q^2}{r^4}\nonumber\\
&+\frac{9M\ell_0^2}{2r^5}-\frac{27\pi Q^2\ell_0}{64r^5}+\mathcal{O}\left(\frac{1}{r^6}\right).\label{eq:k2_asymp_charged}
\end{align}
The leading terms reproduce the Reissner--Nordstr\"om tidal components given in Eq.~\eqref{eq:k1-k2-RN}, while the T-duality corrections appear only at higher orders and vanish asymptotically.

For $Q=0$, Eqs.~\eqref{eq:k1_asymp_charged} and
\eqref{eq:k2_asymp_charged} reduce to the asymptotic behavior
of the neutral T-duality geometry,
\begin{equation}
k_1(r \gg \ell_0)
\approx
\frac{2M}{r^3}
-
\frac{18M\ell_0^2}{r^5}
+
\mathcal{O}\left(\frac{\ell_0^4}{r^7}\right),
\end{equation}
and
\begin{equation}
k_2(r \gg \ell_0)
\approx
-\frac{M}{r^3}
+
\frac{9M\ell_0^2}{2r^5}
+
\mathcal{O}\left(\frac{\ell_0^4}{r^7}\right),
\end{equation}
where the leading terms exactly reproduce the Schwarzschild behavior in Eq.~\eqref{k1-k2-schw}, while the corrections are controlled by powers of $\ell_0^2/r^2$ and vanish asymptotically.

We next examine the behavior of the tidal components near the regular core. Expanding the charged expressions around
$r=0$, we obtain
\begin{align}
k_1^{(Q)}(r\to0)\approx{}&\frac{2M}{\ell_0^3}-\frac{3\pi Q^2}{16\ell_0^4}\nonumber\\
&-\left[\frac{18M}{\ell_0^5}-\frac{Q^2}{\ell_0^6}\left(\frac{6}{5}+\frac{27\pi}{16}\right)\right]r^2+\mathcal{O}(r^4),
\label{eq:k1_origin_charged}\\
k_2^{(Q)}(r\to0)\approx{}&\frac{2M}{\ell_0^3}-\frac{3\pi Q^2}{16\ell_0^4}\nonumber\\
&+\left[-\frac{6M}{\ell_0^5}+\frac{Q^2}{\ell_0^6}\left(\frac{2}{5}+\frac{9\pi}{16}\right)\right]r^2+\mathcal{O}(r^4).
\label{eq:k2_origin_charged}
\end{align}
Therefore, both tidal components remain finite at the regular center and approach the same value,
\begin{equation}
\lim_{r\to0}k_1^{(Q)} = \lim_{r\to0}k_2^{(Q)}=\frac{2M}{\ell_0^3}-\frac{3\pi Q^2}{16\ell_0^4}=\frac{\Lambda_{\rm eff}^{(Q)}}{3}.
\label{eq:tidal_origin_charged}
\end{equation} 
Hence, the electric charge preserves the isotropy of the tidal field at the center, while modifying both its magnitude and its sign. As discussed in Sec.~\ref{subsec:charged}, for $\Lambda_{\rm eff}^{(Q)}>0$ the regular core is de Sitter-like and the two tidal components are positive at $r=0$, whereas for $\Lambda_{\rm eff}^{(Q)}<0$ the core is anti-de Sitter-like and both components are negative. At $\Lambda_{\rm eff}^{(Q)}=0$, the tidal components vanish at the center.

For $Q=0$, Eqs.~\eqref{eq:k1_origin_charged} and \eqref{eq:k2_origin_charged} reduce to
\begin{equation} \label{eq-exp-r01}
k_1(r \to 0)
\approx
\frac{2M}{\ell_0^3}
-
\frac{18Mr^2}{\ell_0^5}
+
\mathcal{O}(r^4),
\end{equation}
and
\begin{equation}\label{eq-exp-r02}
k_2(r \to 0)
\approx
\frac{2M}{\ell_0^3}
-
\frac{6Mr^2}{\ell_0^5}
+
\mathcal{O}(r^4),
\end{equation}
recovering the isotropic tidal field associated with the de Sitter-like core of the neutral T-duality geometry.

The radial component $k_1$ exhibits an interesting structure, with two sign changes. Starting from Eq.~\eqref{eq-k1}, one readily obtains the following roots,
$$
r_{k_1}^{\,\,(1)}
=
\frac{\ell_0}{2}
\sqrt{11-\sqrt{105}},
\quad
r_{k_1}^{\,\,(2)}
=
\frac{\ell_0}{2}
\sqrt{11+\sqrt{105}},
$$
together with two local extrema,
$$
r_{k_1}^{\,\,(\mathrm{min})}
=
\frac{\ell_0}{2}
\sqrt{21-\sqrt{345}}, 
\quad
r_{k_1}^{\,\,(\mathrm{max})}
=
\frac{\ell_0}{2}
\sqrt{21+\sqrt{345}}.
$$
This means that a freely falling observer crosses successive regions of radial stretching and compression before reaching the regular core.

On the other hand, from Eq.~\eqref{eq-k2}, we find that the angular component $k_2$ exhibits a single sign change at $r_{k_2}=\sqrt{2}\,\ell_0$, separating an outer region of angular compression ($k_2<0$) from an inner region of angular stretching ($k_2>0$). In addition, $k_2$ has a global minimum at $r_{k_2}^{\,\,(\mathrm{min})}=2\ell_0$, which corresponds to the maximum angular compression along the radial trajectory. In particular, there exists an intermediate region in which both components of the tidal tensor satisfy $k_1<0$ and $k_2<0$, characterizing simultaneous compression in the radial and angular directions. This region is bounded by the sign change of the angular component, $r_{k_2}$, and the outer root of the radial component, $r_{k_1}^{\,\,(2)}$, such that $r_{k_2}<r<r_{k_1}^{\,\,(2)}$.

The analytical expressions obtained above rely on the comparatively simple form of the tidal components in the neutral geometry. In the charged case, the presence of $F(r)$ and its derivatives in Eqs.~\eqref{eq:k1_charged} and \eqref{eq:k2_charged} makes the equations determining the zeros and extrema of the tidal components considerably more involved, and simple closed-form expressions for the corresponding characteristic radii are no longer available. We therefore determine these radii numerically from
\begin{equation}
k_1^{(Q)}(r)=0, \qquad \frac{d k_1^{(Q)}}{dr}=0,
\label{eq:k1_char_charged}
\end{equation}
for the radial component, and
\begin{equation}
k_2^{(Q)}(r)=0, \qquad \frac{d k_2^{(Q)}}{dr}=0,
\label{eq:k2_char_charged}
\end{equation}
for the angular component. These conditions will be used below to identify the characteristic tidal radii of the charged geometry and to analyze their relation to the horizon structure.

\begin{figure}[!ht]
\begin{subfigure}{0.48\textwidth}  \includegraphics[width=\textwidth]{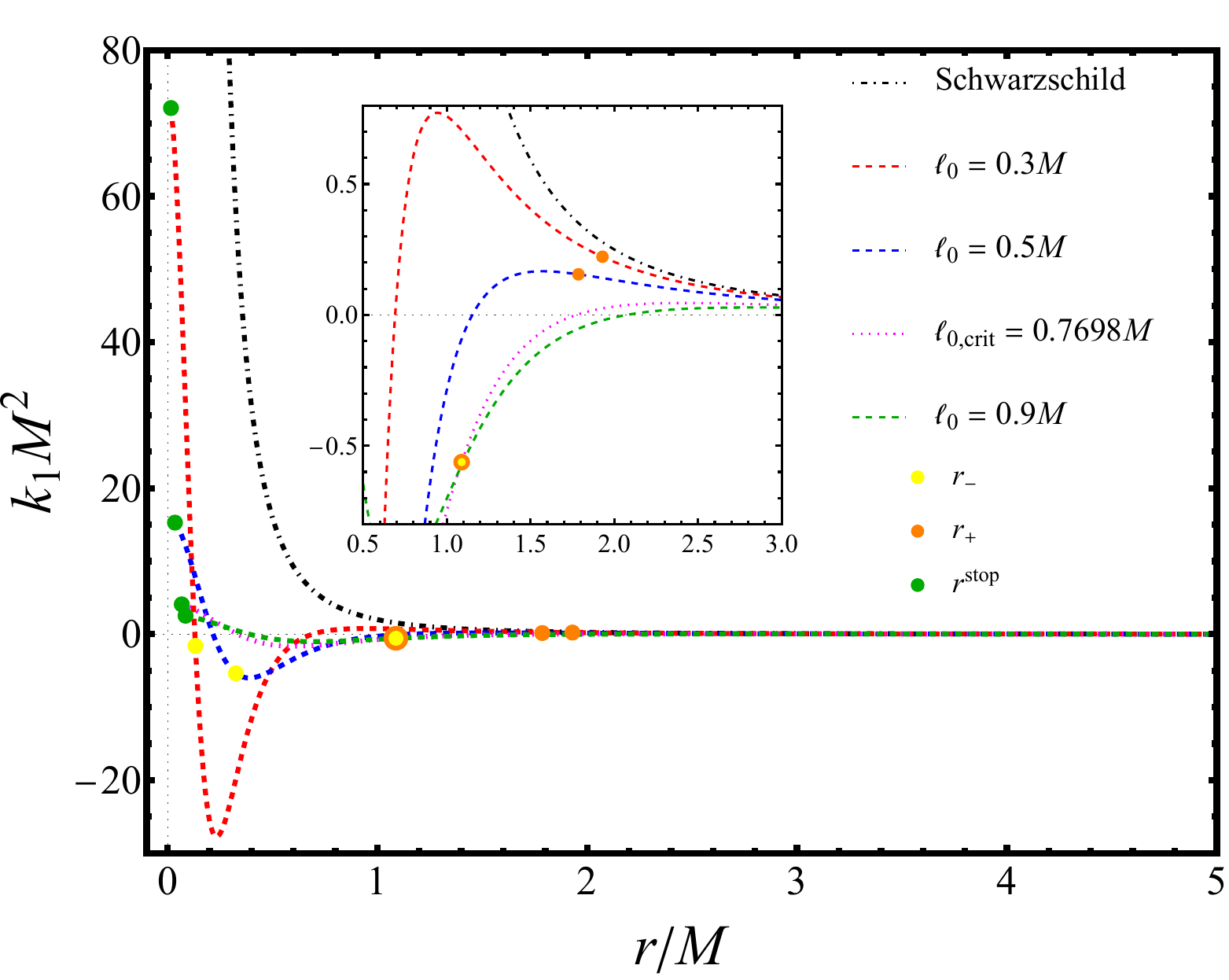}
    \label{fig:g__k1}
\end{subfigure}
\hfill
\begin{subfigure}{0.48\textwidth}
    \centering
    \includegraphics[width=\textwidth]{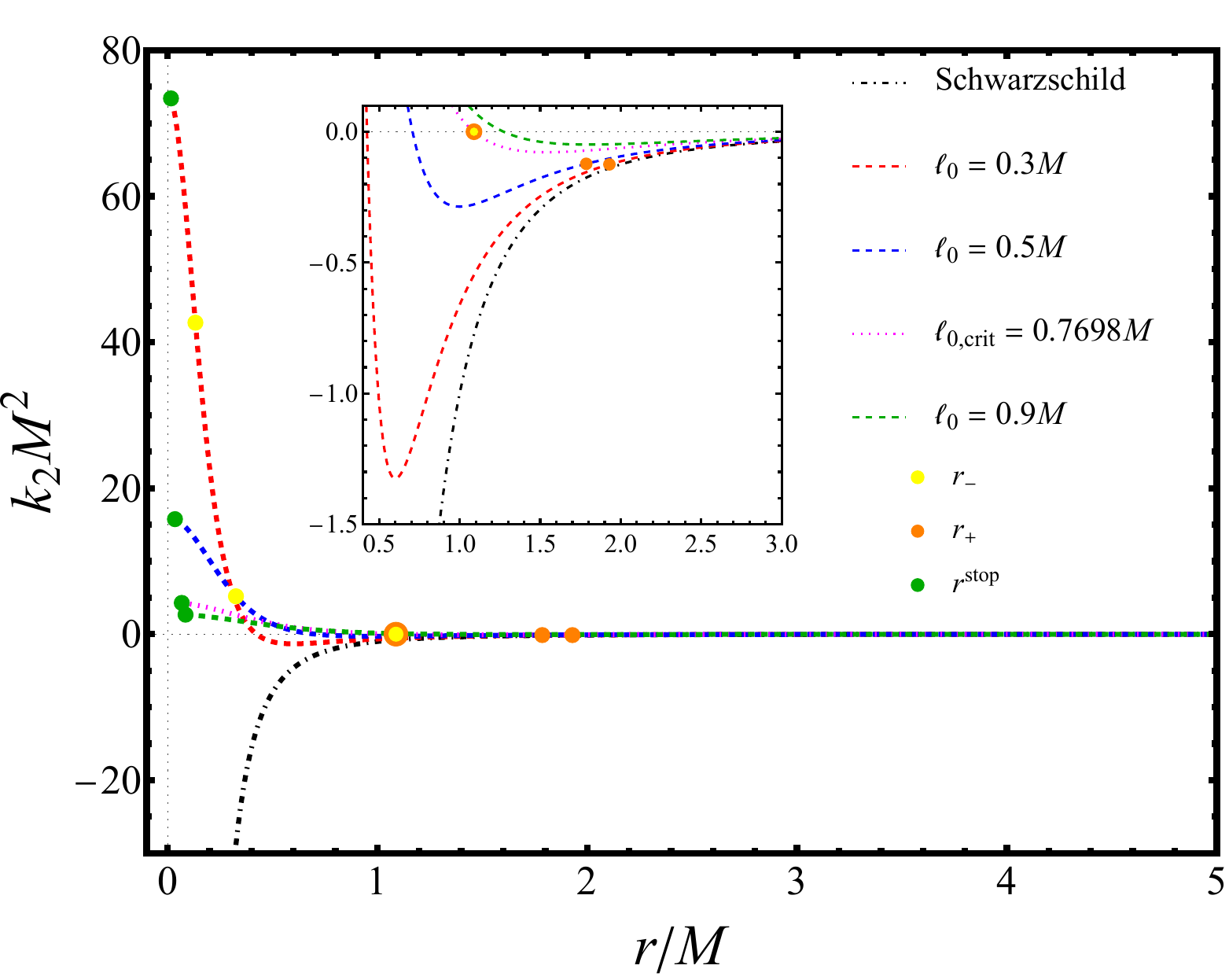}
    \label{fig:g__k2}
\end{subfigure}
\caption{\justifying Radial $k_1$ (upper panel) and angular $k_2$ (lower panel) components of the tidal tensor as functions of $r/M$, for different values of $\ell_0/M$. The Schwarzschild solution is shown for comparison. The markers indicate the positions of the inner $r_-$ and outer $r_+$ horizons, as well as the turning radius $r^{\mathrm{stop}}$. The insets enlarge the regions near the sign changes of the tidal components.
}
\label{fig:k1_k2}
\end{figure}

In Fig.~\ref{fig:k1_k2}, we show the behavior of the radial and angular components of the tidal forces for different values of the parameter $\ell_0$. Unlike the Schwarzschild solution, both components remain finite throughout the geometry. The plots show that the characteristic points associated with the sign changes and extrema of $k_1$ and $k_2$ shift toward larger radii as $\ell_0$ increases. For sufficiently small values of $\ell_0$, the main characteristic points remain hidden behind the outer horizon. However, as $\ell_0$ increases toward $\ell_{0,\mathrm{crit}}$, part of this structure emerges into the exterior region. As a result, some of these characteristic points become accessible to external observers already in the subcritical regime, even before the two horizons merge. 

\begin{figure}[!ht]
\begin{subfigure}{0.48\textwidth}
    \centering
    \includegraphics[width=\textwidth]{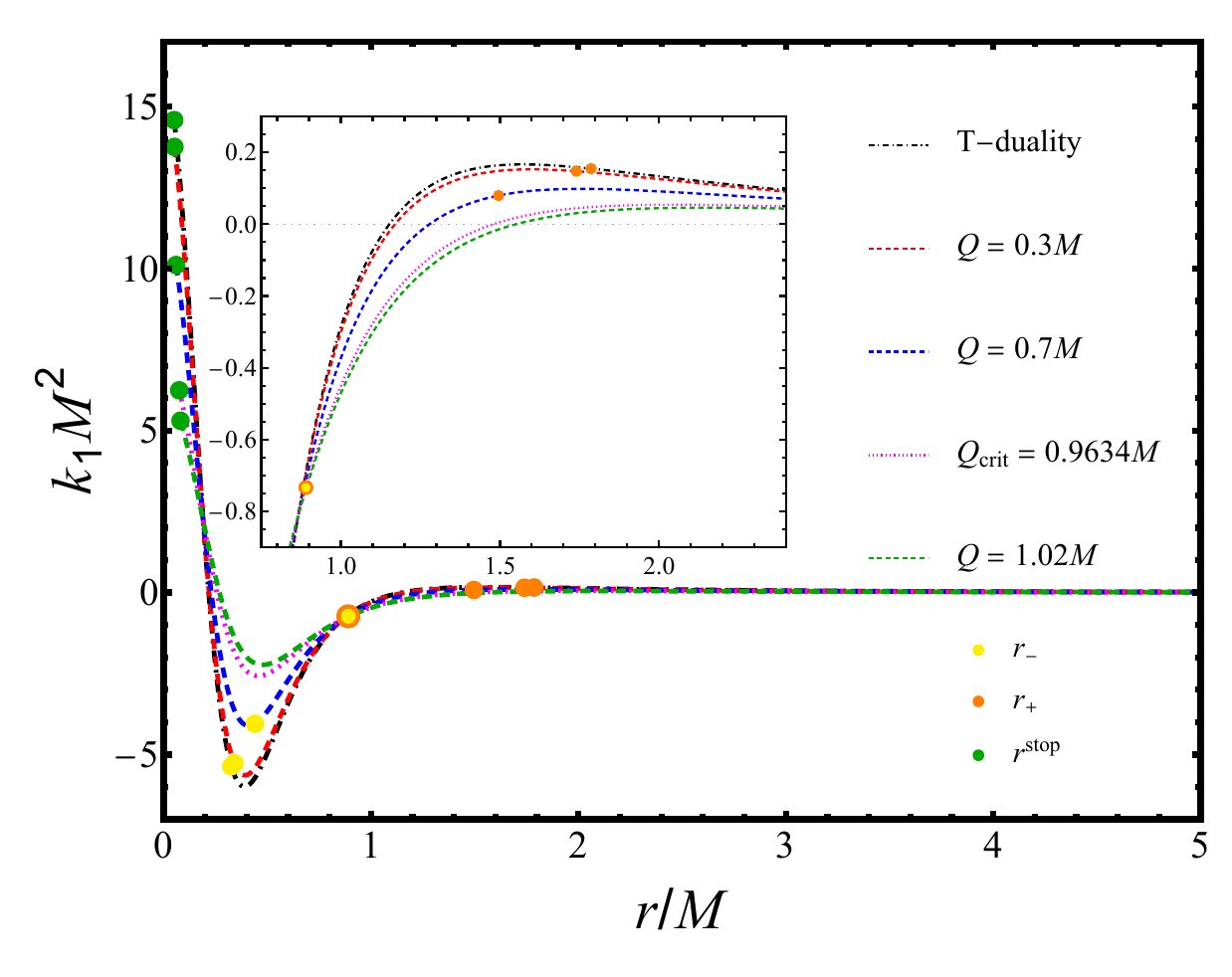}
    \label{fig:g_k1_charged}
\end{subfigure}
\hfill
\begin{subfigure}{0.48\textwidth}
    \centering
    \includegraphics[width=\textwidth]{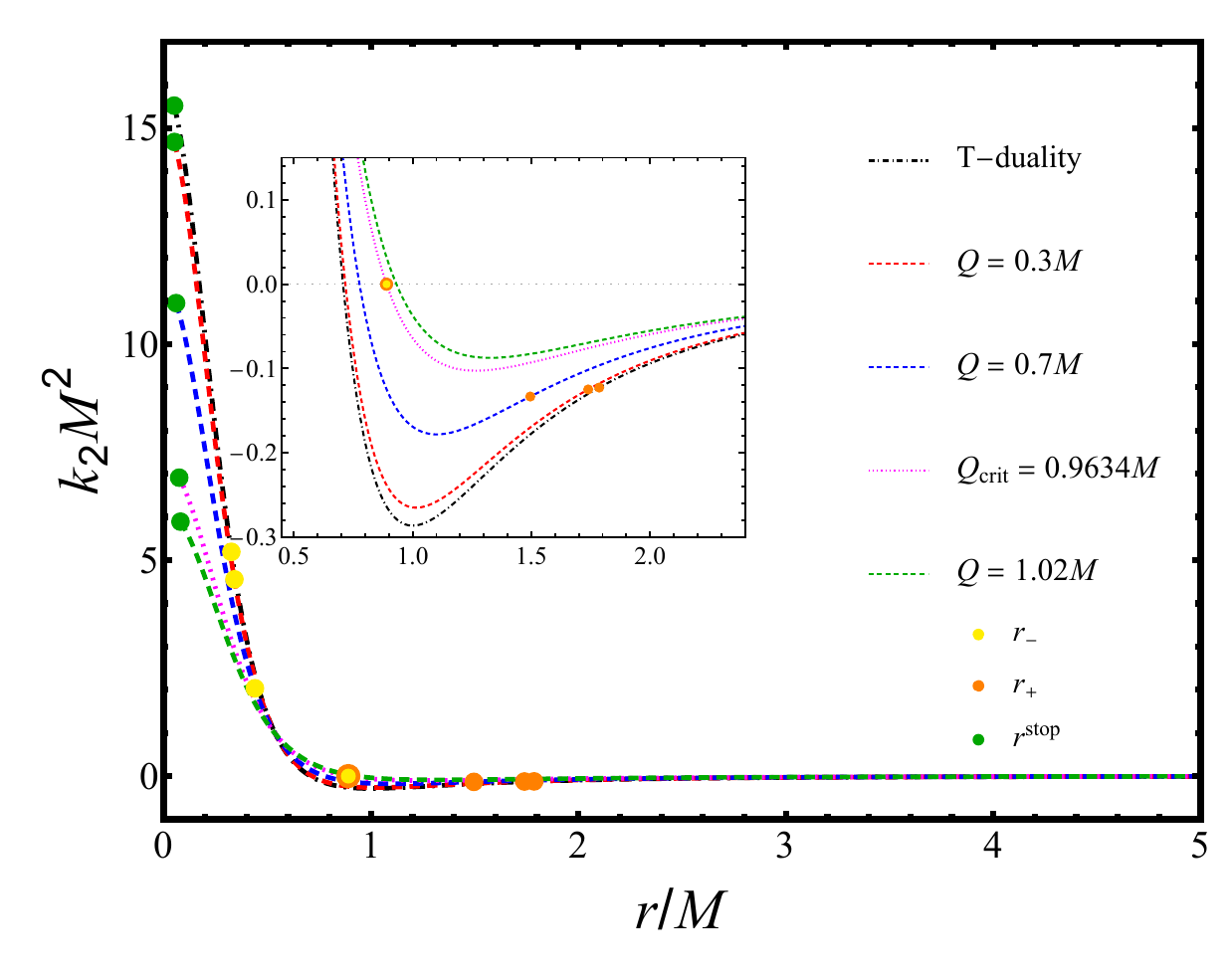}
    \label{fig:g_k2_charged}
\end{subfigure}
\caption{\justifying 
Radial $k_1$ (upper panel) and angular $k_2$ (lower panel) components of the tidal tensor as functions of $r/M$, for fixed $\ell_0=0.5M$ and different values of $Q/M$. The neutral T-duality solution, $Q=0$, is shown for comparison. The insets enlarge the regions around the sign changes and extrema of the tidal components. }
\label{fig:k1_k2_charged}
\end{figure}

We now examine how the electric charge modifies the tidal force profiles. In Fig.~\ref{fig:k1_k2_charged}, we fix $\ell_0=0.5M$ and show the radial and angular tidal components for different values of $Q/M$. The neutral T-duality solution is included for comparison. As in the neutral case, the curves are shown only along the portion of the radial trajectory accessible to a particle released from rest at $r=b$, terminating at the corresponding turning radius $r^{\rm stop}$. The electric charge produces a systematic modification of both tidal components. For the radial component, increasing $Q$ reduces the magnitude of the intermediate compressive region and also suppresses the subsequent positive maximum. A similar trend is observed for the angular component, whose negative minimum becomes progressively shallower as the charge is increased. Moreover, the characteristic zeros and extrema are displaced as $Q/M$ varies, as can be seen more clearly in the enlarged panels. For the subcritical configurations, the relevant tidal structure must be interpreted together with the positions of the Cauchy and event horizons. At the critical charge the two horizons coincide, whereas the configuration with $Q=1.02M$ is horizonless. These results motivate a direct comparison between the characteristic tidal radii and the causal structure of the charged geometry.

\begin{figure}[!ht]
\centering
\includegraphics[width=0.48\textwidth]{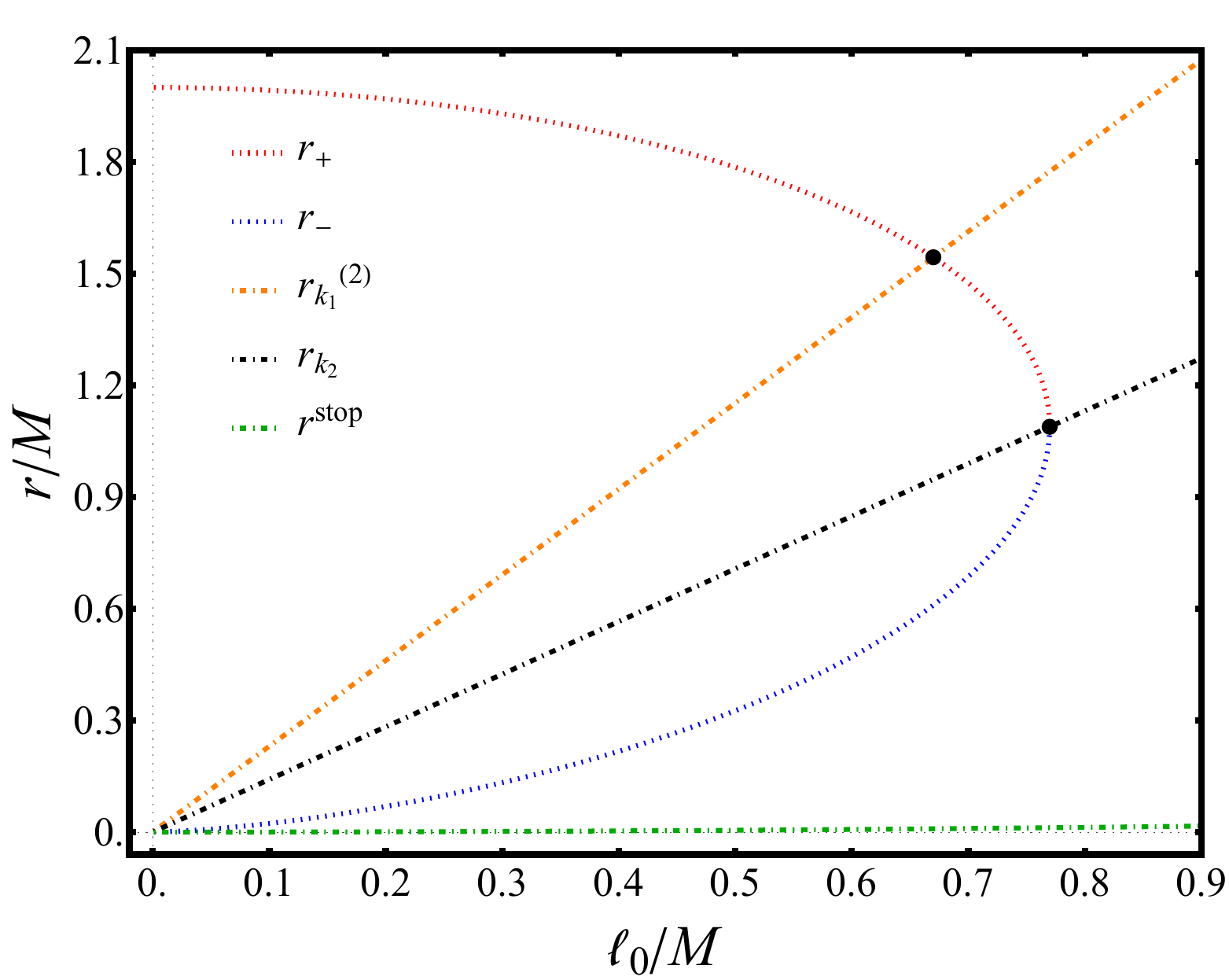}
\caption{\justifying Comparison between the horizons $r_-$ and $r_+$, the characteristic radii associated with the sign changes of the tidal components,  $r_{k_1}^{\,\,(2)}$ and $r_{k_2}$, and the turning point $r^{\rm stop}$. The intersections with $r_+$ indicate the values of $\ell_0$ for which these structures move into the region outside the event horizon. For these results, we adopt $b=50M$.
}
\label{fig:g__R}
\end{figure}

To clarify these transitions, in Fig.~\ref{fig:g__R} we compare the horizons $r_\pm$ with the characteristic radii associated with the sign changes of the tidal components, $r_{k_1}^{\,\,(2)}$ and $r_{k_2}$. We also indicate the turning radius $r^{\mathrm{stop}}$ associated with a massive particle released from rest at $r=b$. Since $E^2=f(b)$, the turning point is determined by the condition $f(r^{\mathrm{stop}})=f(b)$, corresponding to $\dot r=0$ during the radial motion. For $\ell_0=0.5M$, all these characteristic points lie below $r_+$; therefore, the corresponding tidal-force effects can only be experienced by a freely falling observer after crossing the event horizon. In the critical case, the sign change of $k_2$ occurs exactly at the degenerate horizon, while part of the structure associated with $k_1$ already lies in the exterior region. Finally, for $\ell_0>\ell_{0,\mathrm{crit}}$, the absence of horizons makes the entire tidal force structure directly accessible.

We now perform the analogous comparison for the charged geometry. Since the characteristic radii of the tidal components can no longer be expressed in a simple closed form, they are determined numerically from Eqs.~\eqref{eq:k1_char_charged} and \eqref{eq:k2_char_charged}. Fixing $\ell_0=0.5M$, we compare the Cauchy and event horizons, $r_-$ and $r_+$, with the outer zero of the radial tidal component, $r_{k_1}^{\,\,(2)}$, the zero of the angular component, $r_{k_2}$, and the turning radius $r^{\rm stop}$ as functions of $Q/M$. As the electric charge increases, the causal structure becomes progressively more compact: the event horizon $r_+$ moves inward, whereas the Cauchy horizon $r_-$ moves outward, until they merge at the extremal configuration $Q=Q_{\mathrm{crit}}\simeq0.9634M$, for $\ell_0=0.5M$, where $r_-=r_+\simeq0.8903M$. At the same time, the characteristic radii associated with the tidal force sign changes move outward. In particular, $r_{k_1}^{\,\,(2)}$ increases with $Q$ and crosses the event horizon at $Q/M\simeq0.8194$. Therefore, beyond this value, the outer transition of the radial tidal force already occurs in the exterior region. By contrast, $r_{k_2}$ remains inside the event horizon throughout the nonextremal black hole sector and reaches the degenerate horizon precisely at extremality, $r_{k_2}=r_-=r_+$. The turning radius $r^{\mathrm{stop}}$ also increases with the charge, although much more slowly, and remains well inside the horizon throughout the black hole branch. For $Q>Q_{\mathrm{crit}}$, the horizons disappear while $r_{k_1}^{\,\,(2)}$, $r_{k_2}$, and $r^{\mathrm{stop}}$ remain well defined, so that the corresponding tidal force structure is no longer hidden behind an event horizon.

\begin{figure}[!ht]
\centering
\includegraphics[width=0.48\textwidth]
{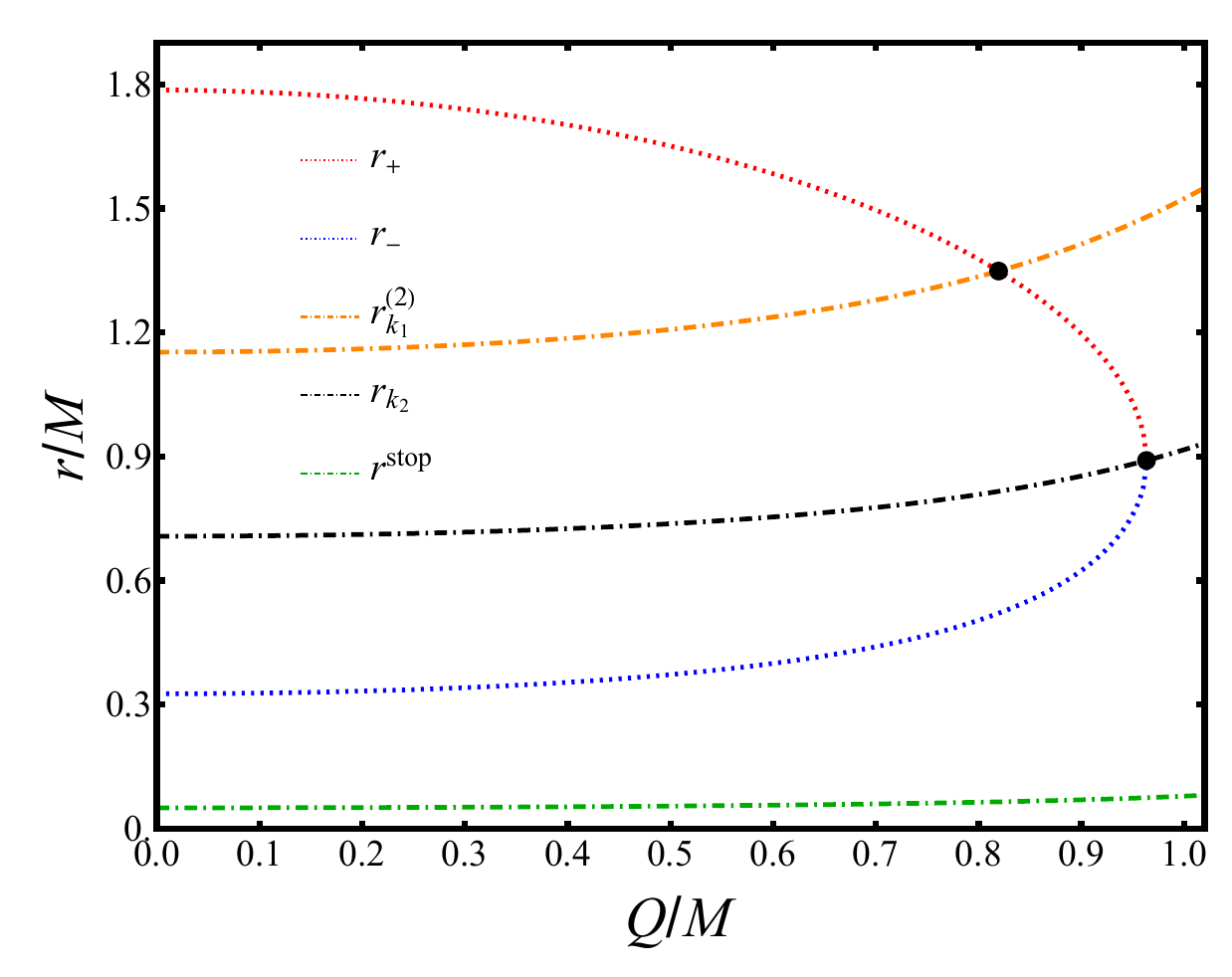}
\caption{\justifying Comparison between the horizons $r_-$ and $r_+$, the characteristic radii associated with the sign changes of the charged tidal components, $r_{k_1}^{\,\,(2)}$ and $r_{k_2}$, and the turning radius $r^{\rm stop}$ as functions of $Q/M$, for fixed $\ell_0=0.5M$. We adopt $b=50M$. The black points indicate the intersections of the characteristic tidal radii with the event horizon.}
\label{fig:g__R_charged}
\end{figure}

\section{GEODESIC DEVIATION}
\label{sec_V}

Although the components $k_1$ and $k_2$ allow us to locally identify the regions of stretching and compression produced by tidal forces, they do not directly describe the deformation accumulated by a body along its trajectory. Therefore, to investigate this evolution, it is necessary to analyze the geodesic deviation vector, which measures the separation between neighboring particles in free fall. In this section, we present this formalism and discuss in detail each component of the separation vector.

\subsection{Evolution of the separation vector and initial conditions (IC)}

As shown above, Eqs.~\eqref{eq:geodesic_deviation_radial}, \eqref{eq:geodesic_deviation_theta}, and \eqref{eq:geodesic_deviation_phi} represent the geodesic deviation equations in the observer's orthonormal frame. We now solve these equations to obtain the geodesic deviation vectors for radial free-fall geodesics as functions of $r$. To rewrite them as differential equations with respect to $r$, we use Eq.~\eqref{eq-radial-mass} together with the chain rule, obtaining the differential equations for the radial ($\xi^{\hat{r}}$) and angular ($\xi^{\hat{i}}$) components of the separation vector:
\begin{align}
(E^2 - f)\,\xi^{\hat{r}\prime\prime} - \frac{f'}{2}\,\xi^{\hat{r}\prime} + \frac{f''}{2}\,\xi^{\hat{r}} &= 0, \label{edo_rad} \\
(E^2 - f)\,\xi^{\hat{i}\prime\prime} - \frac{f'}{2}\,\xi^{\hat{i}\prime} + \frac{f'}{2r}\,\xi^{\hat{i}} &= 0,\label{edo_ang}
\end{align}
where $\hat{i}=\{\hat{\theta}, \,\hat{\phi}\}$. The general solutions of Eqs.~\eqref{edo_rad} and \eqref{edo_ang} can then be written, respectively, as~\cite{Crispino:2016pnv,Sharif:2018gzj,Lima:2020wcb}
\begin{align}
\xi^{\hat{r}}(r) &= \sqrt{E^2 - f}\left[ C_1 + C_2 \int \frac{dr}{(E^2 - f)^{3/2}} \right], \label{qr-evo-1} \\
\xi^{\hat{i}}(r) &= \left[ C_3 + C_4 \int \frac{dr}{r^2 (E^2 - f)^{1/2}} \right] r, \label{qr-evo-2}
\end{align}
where $C_1$, $C_2$, $C_3$, and $C_4$ are integration constants. To determine these constants, we numerically evaluate Eqs.~\eqref{qr-evo-1} and \eqref{qr-evo-2}. For this purpose, we impose initial conditions at an initial radius $r=b$, from which the radial infall begins. Following the literature on tidal forces in black-hole spacetimes~\cite{Crispino:2016pnv,Sharif:2018gzj,Lima:2020wcb, Silva:2025hwl}, we consider the following two distinct scenarios:
\begin{equation}
    \xi^{\hat{\alpha}}(b) > 0, \quad \dot{\xi}^{\hat{\alpha}}(b) = 0, \tag{IC I}
    \label{ici}
\end{equation}
\begin{equation}
    \xi^{\hat{\alpha}}(b) = 0, \quad \dot{\xi}^{\hat{\alpha}}(b) > 0, \tag{IC II}
    \label{icii}
\end{equation}
where $\xi^{\hat{\alpha}}(b)$ denotes the component of the separation vector evaluated at $r=b$, with $b>r_+$ and $\hat{\alpha}=\{\hat{r},\hat{i}\}$. Initial condition \eqref{ici} represents a body initially located at the radial position $r=b$, with no initial relative velocity between its constituent particles. On the other hand, condition \eqref{icii} describes a scenario in which the system has a nonvanishing initial separation rate between neighboring geodesics, corresponding to an initial isotropic expansion~\cite{Crispino:2016pnv,Sharif:2018gzj,Lima:2020wcb, Silva:2025hwl}. In the following subsections, we numerically analyze the evolution of the radial and angular components of the geodesic deviation vector for both sets of initial conditions. We first investigate their dependence on the T-duality parameter $\ell_0$ and on the initial radial position $b$ in the neutral geometry, and then extend the analysis to the charged solution in order to determine how the electric charge modifies the accumulated radial and angular deformations.

\subsection{Radial and angular components of the geodesic deviation vector in the IC I scenario}

\begin{figure}[h!]
\begin{subfigure}{0.50\textwidth}
   
    \includegraphics[width=\textwidth]{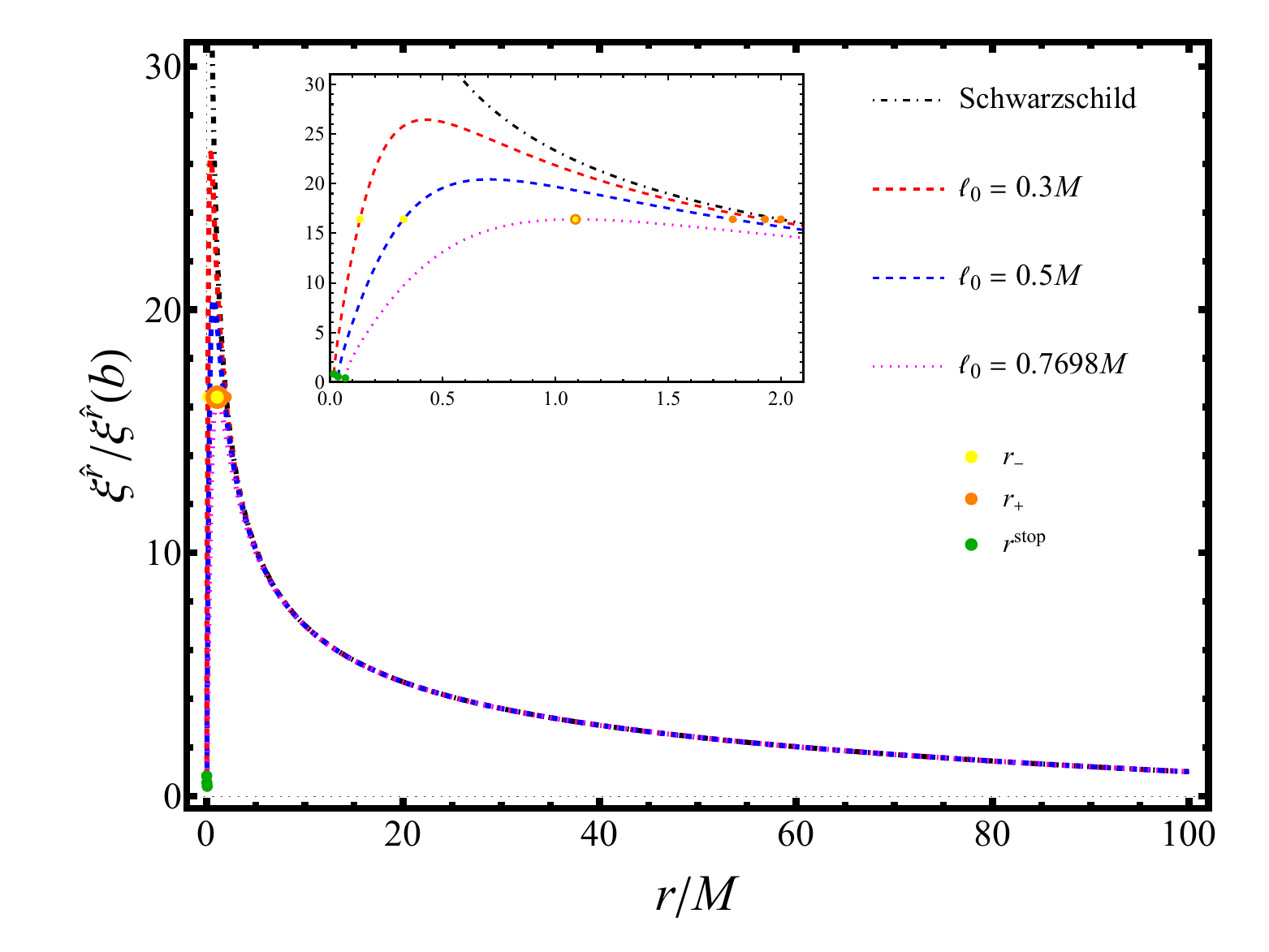}
    \label{fig:xi-radial-ic-i-b}
\end{subfigure}
\hfill
\begin{subfigure}{0.50\textwidth}
    \centering
    \includegraphics[width=\textwidth]{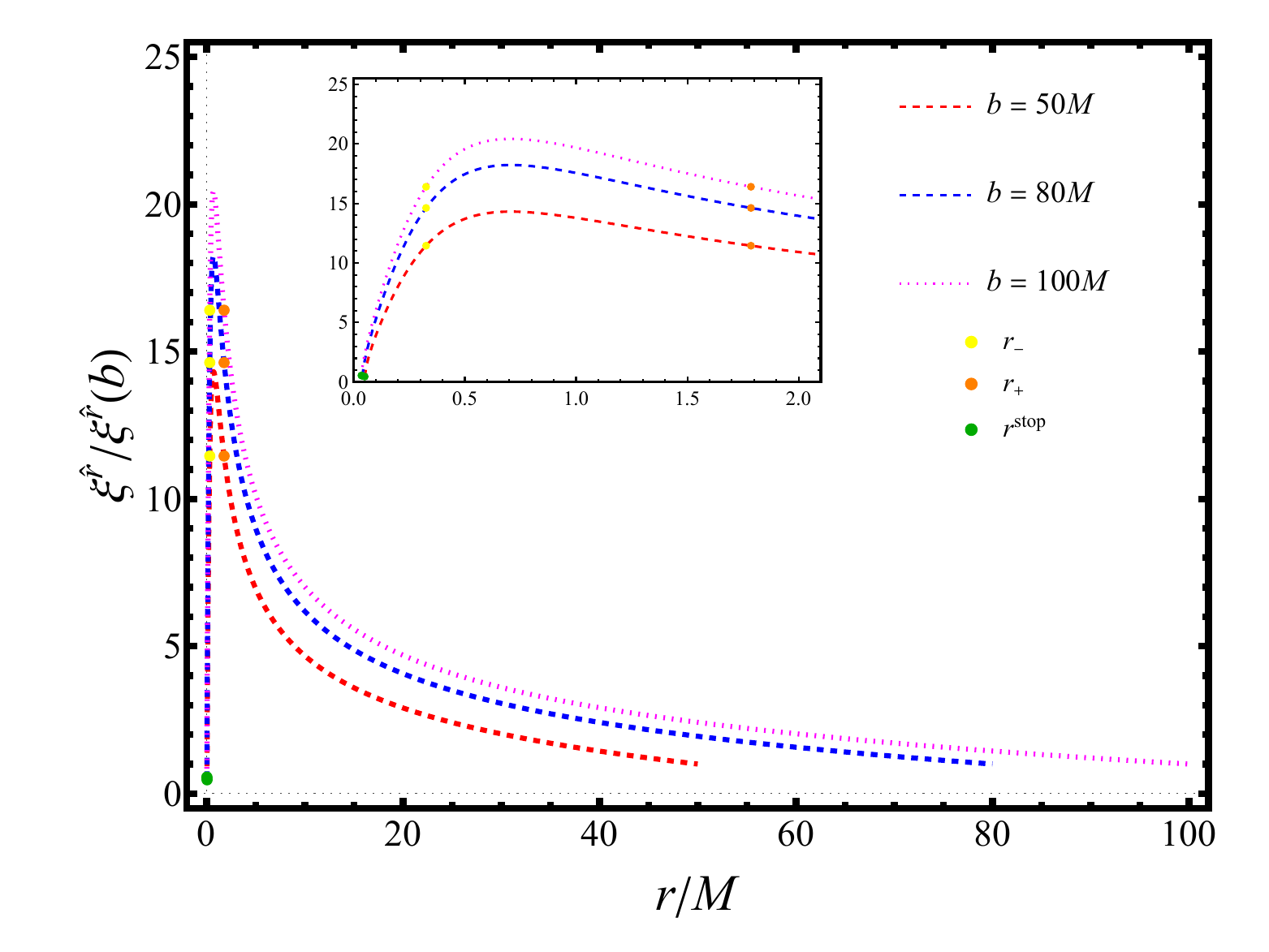}
    \label{fig:xi-radial-ic-i-l0}
\end{subfigure}
\caption{\justifying  Behavior of the normalized radial component of the geodesic deviation vector, $\xi^{\hat r}/\xi^{\hat r}(b)$, as a function of $r/M$ for a radially freely falling frame in the T-duality regular black hole for the scenario \eqref{ici}. In the upper panel, we consider the initial position $b=100M$ for different values of $\ell_0$. In the lower panel, we fix $\ell_0=0.5M$ and vary the value of $b$. The radii $r_-$, $r_+$, and $r^{\mathrm{stop}}$ are also shown in each panel.}
\label{fig:xi-radial-b}
\end{figure}

In Fig.~\ref{fig:xi-radial-b}, we present the evolution of the radial component of the geodesic deviation vector for the scenario \eqref{ici}. Starting from the initial radius $r=b$ and following the trajectory along the direction of infall, $\xi^{\hat r}$ initially increases, corresponding to the radial stretching of the congruence of geodesics. In the inner regions, this component reaches a maximum and subsequently decreases, reflecting the changes in the behavior of the radial component of the tidal force. For nonzero values of $\ell_0$, the maximum stretching amplitude is reduced relative to the Schwarzschild case, highlighting the regularizing effect associated with the scale $\ell_0$. Furthermore, for fixed $\ell_0$, increasing the initial radius $b$ leads to larger amplitudes of $\xi^{\hat r}$ without significantly altering the overall structure of the curves.

\begin{figure}[ht]
    \centering
    \includegraphics[width=\columnwidth]{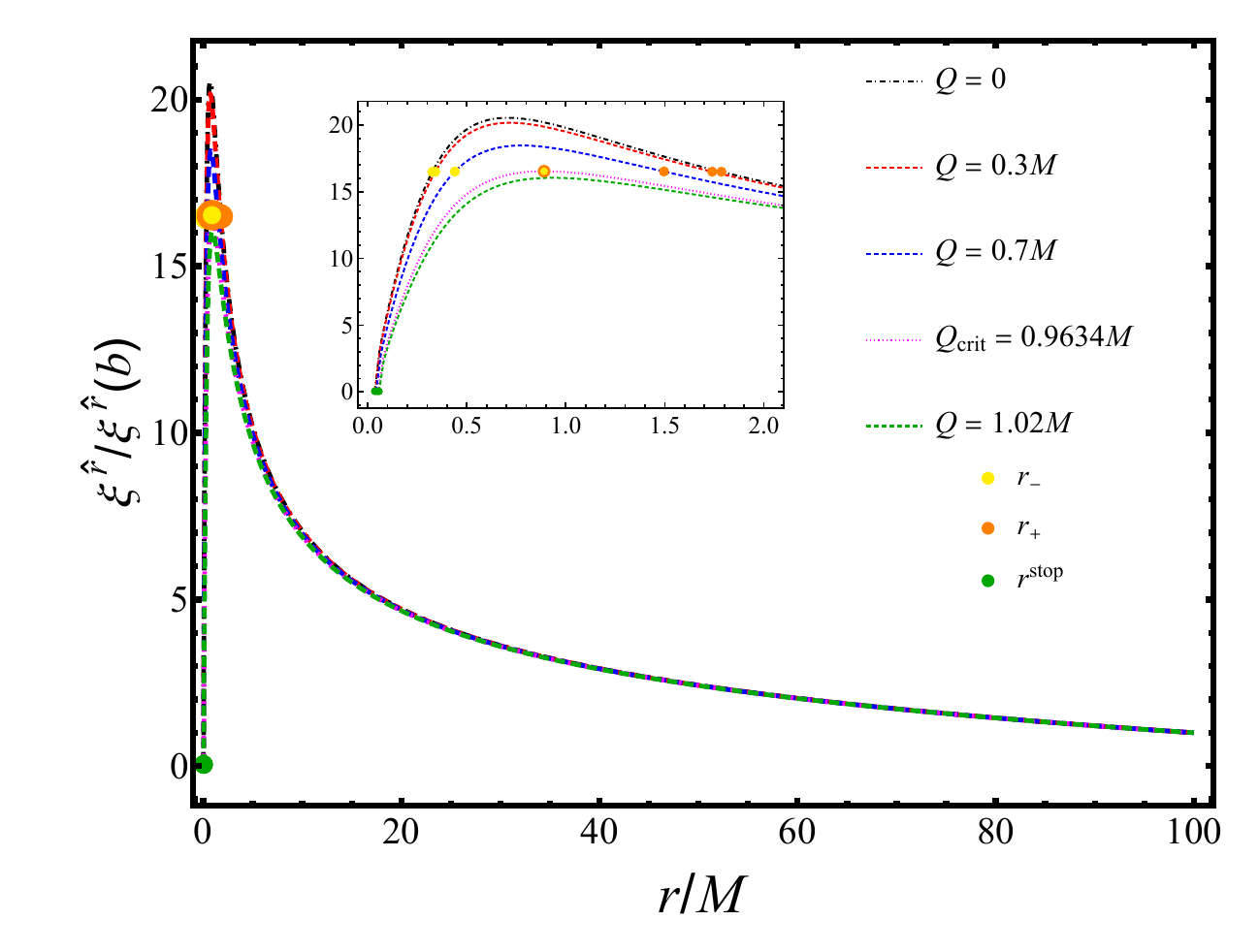}
    \caption{\justifying Behavior of the normalized radial component of the geodesic deviation vector, $\xi^{\hat r}/\xi^{\hat r}(b)$, as a function of $r/M$ for the charged T-duality-inspired geometry in the scenario \eqref{ici}. We fix $\ell_0=0.5M$ and $b=100M$ and consider different values of the electric charge $Q$. The radii $r_-$, $r_+$, and $r^{\mathrm{stop}}$ are also indicated whenever they exist.}
    \label{fig:xi-radial-Q-ic-i}
\end{figure}

To isolate the effect of the electric charge on the accumulated radial deformation, we now fix $\ell_0=0.5M$ and $b=100M$ and vary $Q$ from the neutral configuration to the horizonless regime. The corresponding results are shown in Fig.~\ref{fig:xi-radial-Q-ic-i}. The curves remain nearly indistinguishable at large radii, where the charge-dependent corrections are subdominant, and begin to separate only in the strong-field region. As $Q$ increases, the maximum value of $\xi^{\hat r}$ is progressively reduced. Therefore, the electric charge suppresses the accumulated radial stretching experienced by the congruence during the infall. This behavior persists smoothly from the two-horizon black hole configurations through the extremal case and into the horizonless regime.

\begin{figure}[h!]
\centering
\begin{subfigure}{0.50\textwidth}
    \centering
    \includegraphics[width=\textwidth]{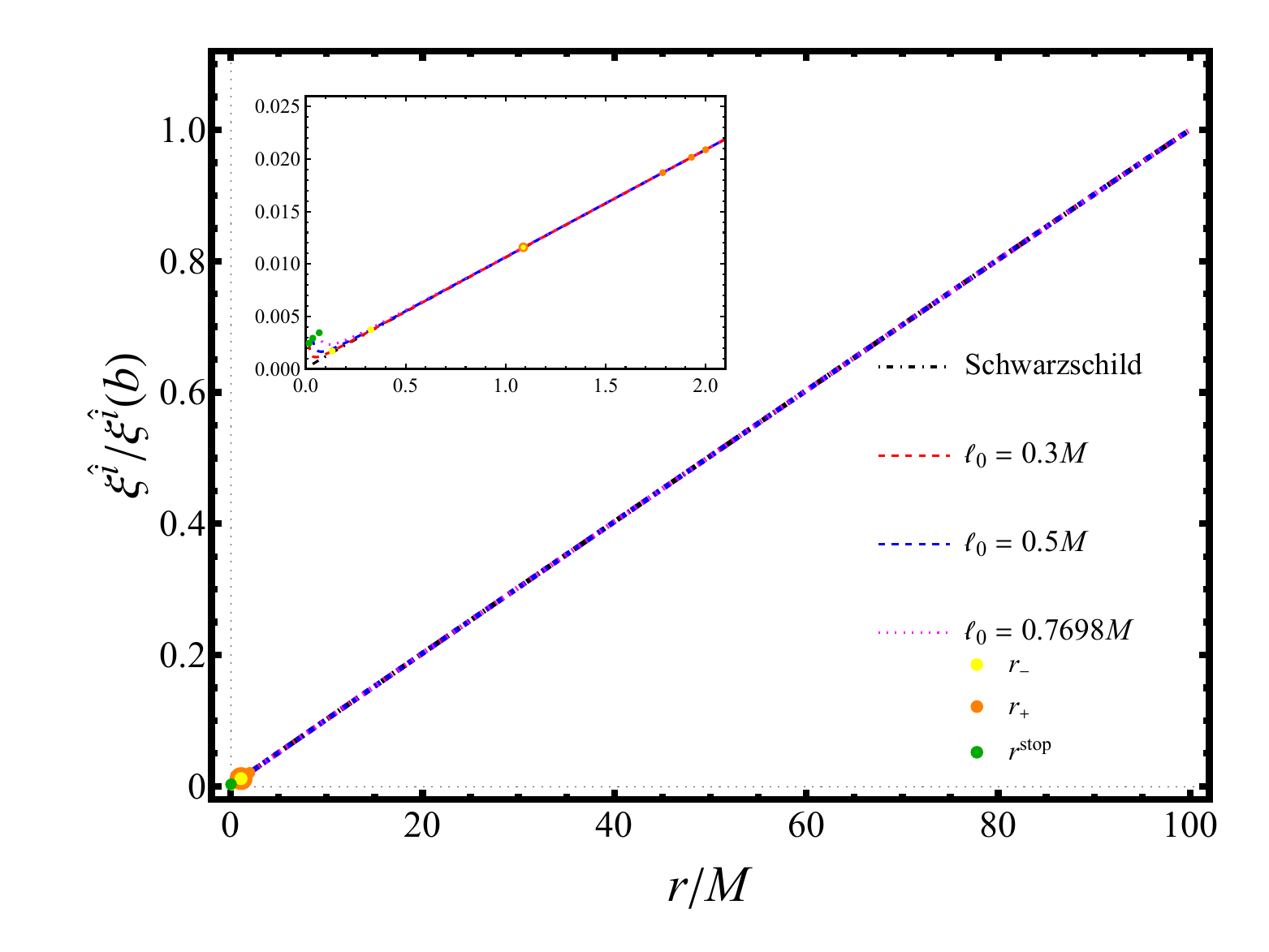}
    \label{fig:xi-angular-ic-i-l0}
\end{subfigure}
\hfill
\begin{subfigure}{0.50\textwidth}
    \centering
    \includegraphics[width=\textwidth]{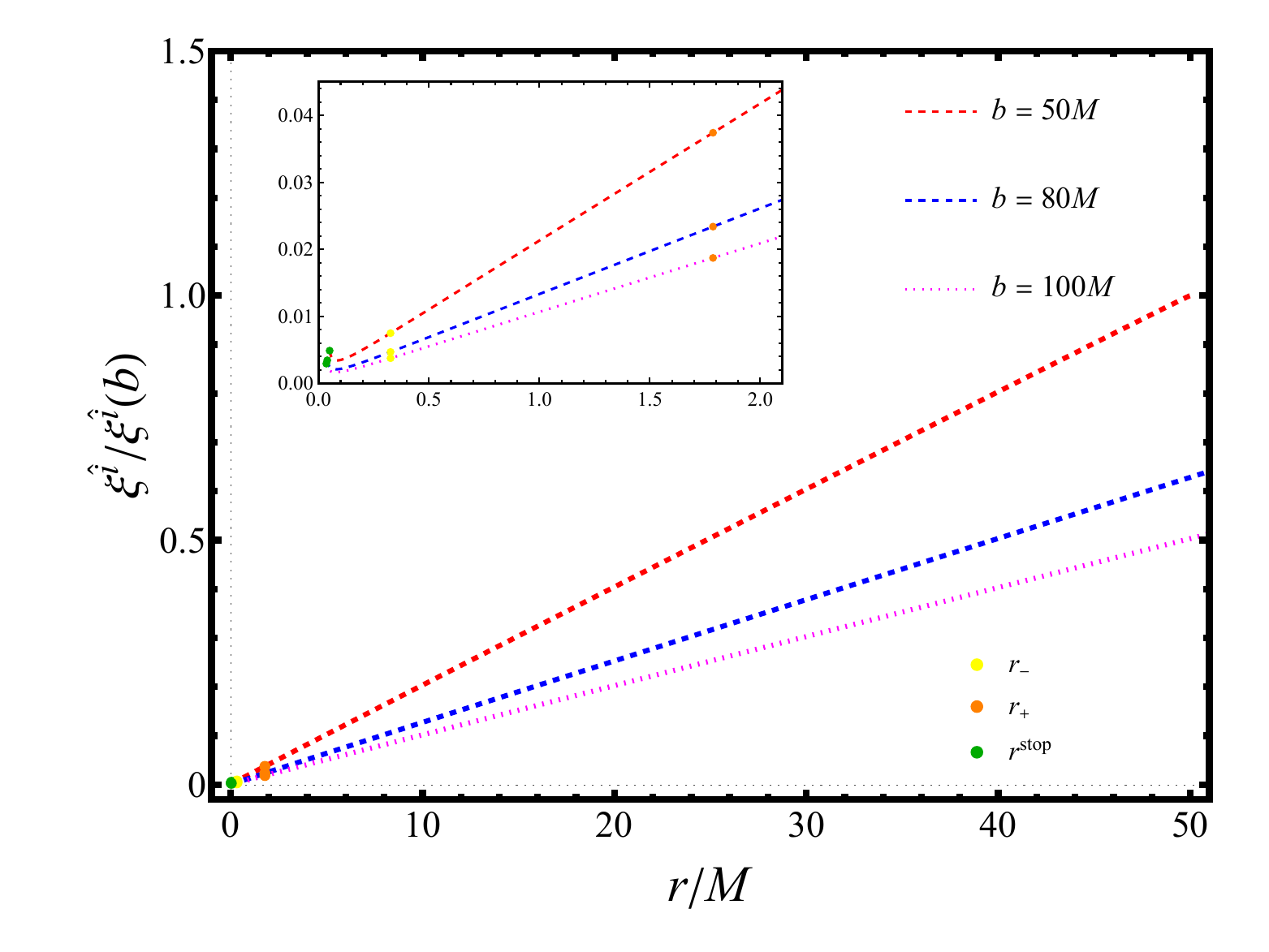}
    \label{fig:xi-angular-ic-i-b}
\end{subfigure}
\caption{\justifying Behavior of the normalized angular component of the geodesic deviation vector, $\xi^{\hat i}/\xi^{\hat i}(b)$, as a function of $r/M$ for a radially freely falling frame in the T-duality regular black hole for the scenario \eqref{ici}. In the upper panel, we consider the initial position $b=100M$ for different values of $\ell_0$. In the lower panel, we fix $\ell_0=0.5M$ and vary the value of $b$. The radii $r_-$, $r_+$, and $r^{\mathrm{stop}}$ are also shown in each panel.}
\label{fig:xi-angular-lo}
\end{figure}

The angular component of the geodesic deviation vector, shown in Fig.~\ref{fig:xi-angular-lo}, exhibits an approximately linear behavior as a function of $r/M$. However, along the direction of infall, this dependence corresponds to a progressive decrease in $\xi^{\hat i}$ as the particle approaches the central region. This behavior characterizes the transverse compression of the congruence of geodesics throughout most of the trajectory. For fixed $b$, the dependence on the parameter $\ell_0$ is relatively weak, as indicated by the near overlap of the curves in the upper panel. On the other hand, varying the initial radius $b$ modifies the rate of evolution of the angular component, as shown in the lower panel.

The positions of the horizons $r_-$ and $r_+$, as well as the turning radius $r^{\mathrm{stop}}$, are also indicated in Figs.~\ref{fig:xi-radial-b} and \ref{fig:xi-angular-lo}, allowing the evolution of the deviation vector to be located relative to the causal structure of the geometry. These results can be interpreted together with Fig.~\ref{fig:g__R}, where we compare the horizons with the radii associated with the sign changes of the tidal components $k_1$ and $k_2$. While these radii delimit the different regions of stretching and compression along the radial infall, Figs.~\ref{fig:xi-radial-b} and \ref{fig:xi-angular-lo} show the deformation accumulated by the congruence of geodesics as it crosses these regions. In particular, the corrections associated with T-duality have a stronger effect on the radial component, reducing the maximum stretching relative to the Schwarzschild case, whereas the angular component exhibits a comparatively weak dependence on $\ell_0$ in the scenario \eqref{ici}.

\begin{figure}[h!]
    \centering
    \includegraphics[width=\columnwidth]{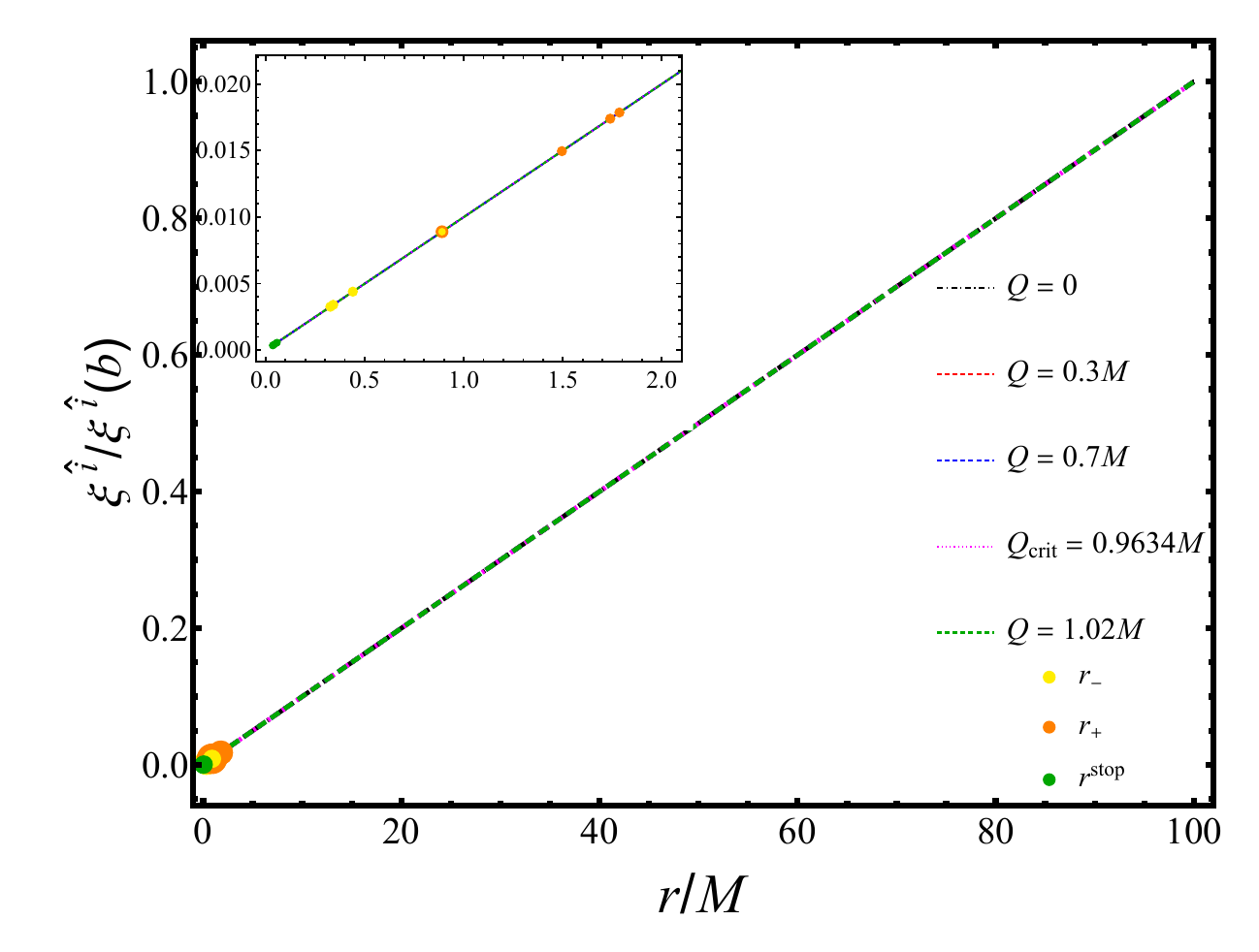}
    \caption{\justifying
    Behavior of the normalized angular component of the geodesic deviation vector, $\xi^{\hat i}/\xi^{\hat i}(b)$, as a function of $r/M$ for the charged T-duality-inspired geometry in the scenario \eqref{ici}. We fix $\ell_0=0.5M$ and $b=100M$ and consider different values of the electric charge $Q$. The radii $r_-$, $r_+$, and $r^{\mathrm{stop}}$ are also indicated whenever they exist.}
    \label{fig:xi-angular-Q-ic-i}
\end{figure}

The influence of the electric charge on the angular component for the same values $\ell_0=0.5M$ and $b=100M$ is shown in Fig.~\ref{fig:xi-angular-Q-ic-i}. In contrast with the radial component, the angular deviation in the scenario \eqref{ici} is essentially insensitive to the electric charge. The curves corresponding to all values of $Q$ overlap and display the same linear dependence on $r$. In fact, the numerical result follows $\xi^{\hat i}/\xi^{\hat i}(b)=r/b$, consistently with the linear branch of Eq.~\eqref{qr-evo-2} selected by the adopted initial conditions. Consequently, in this scenario the charge modifies the characteristic radii $r_-$, $r_+$, and $r^{\mathrm{stop}}$, but does not appreciably alter the angular deformation accumulated along the trajectory.

\subsection{Radial and angular components of the geodesic deviation vector in the IC~II scenario}

\begin{figure}[h!]
\begin{subfigure}{0.50\textwidth}
  
    \includegraphics[width=\textwidth]{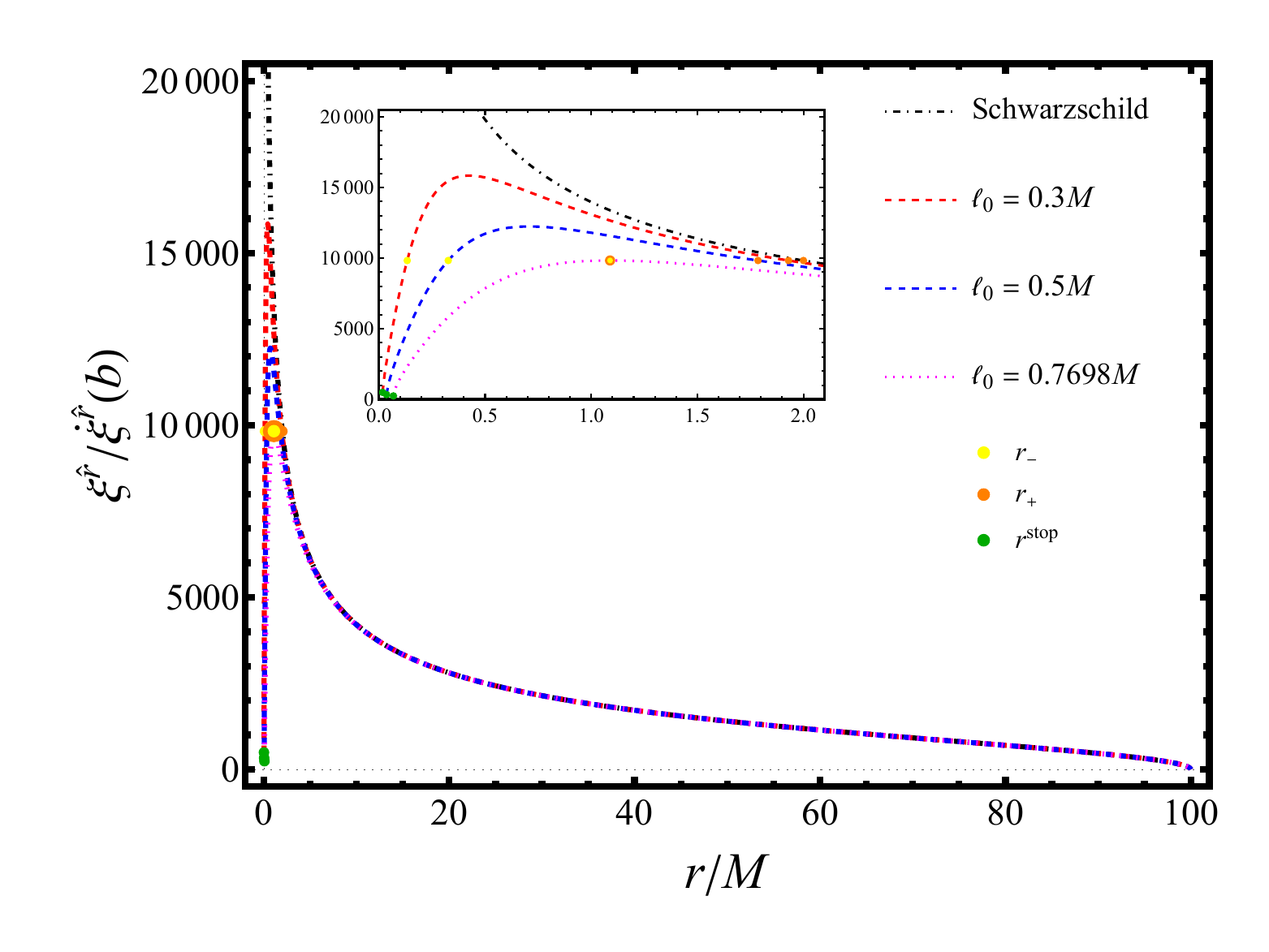}
    \label{fig:xi-radial-ic-ii-b}
\end{subfigure}
\hfill
\begin{subfigure}{0.50\textwidth}
    \centering
    \includegraphics[width=\textwidth]{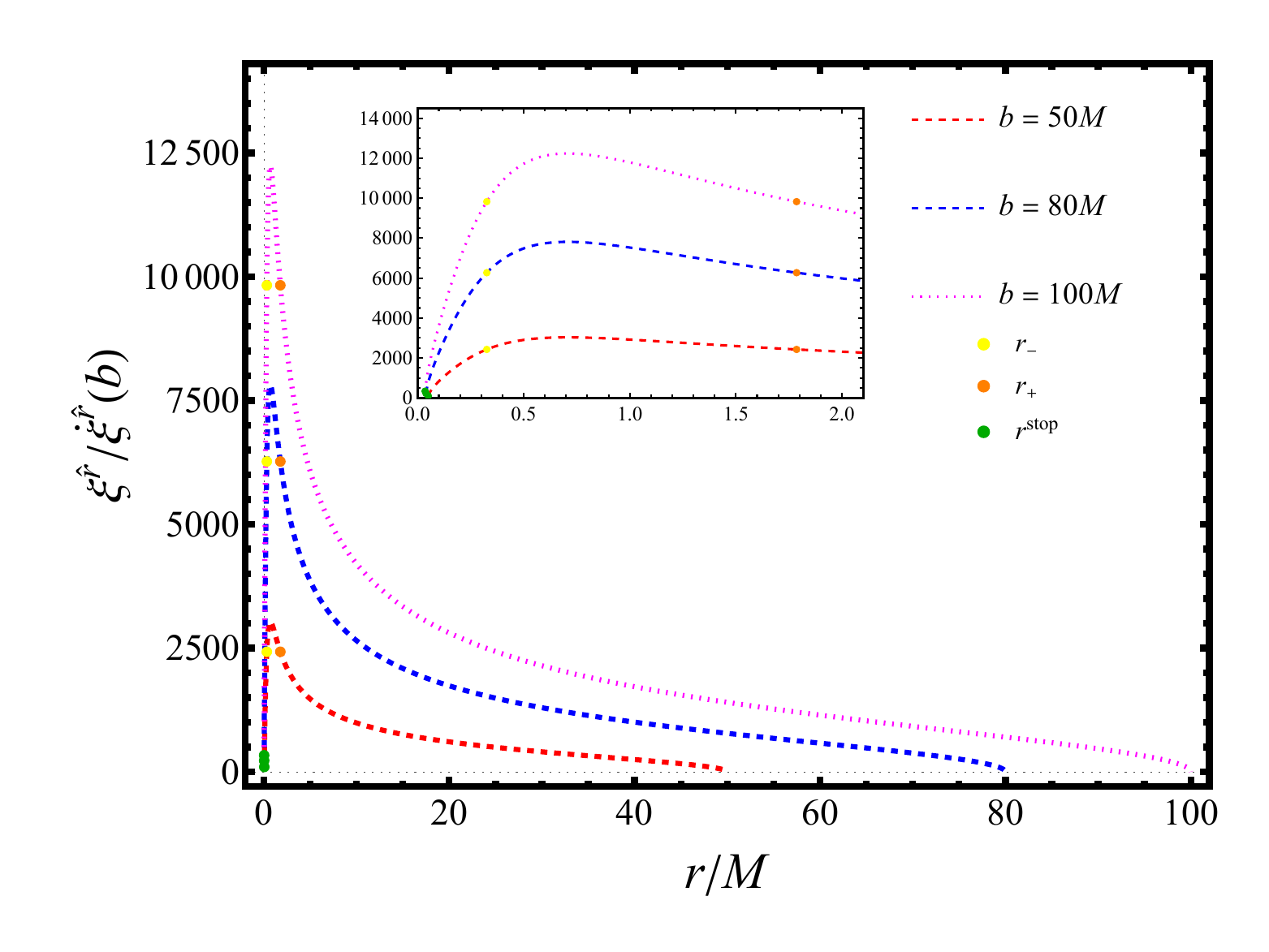}
    \label{fig:xi-radial-ic-ii-l0}
\end{subfigure}
\caption{\justifying Behavior of the normalized radial component of the geodesic deviation vector, $\xi^{\hat r}/\dot{\xi}^{\hat r}(b)$, as a function of $r/M$ for a radially freely falling frame in the T-duality regular black hole for the scenario \eqref{icii}. In the upper panel, we consider the initial position $b=100M$ for different values of $\ell_0$. In the lower panel, we fix $\ell_0=0.5M$ and vary the value of $b$. The radii $r_-$, $r_+$, and $r^{\mathrm{stop}}$ are also shown in each panel.}
\label{fig:xi-radial-b-ic-ii}
\end{figure}

In Fig.~\ref{fig:xi-radial-b-ic-ii}, we present the evolution of the radial component of the geodesic deviation vector for the scenario \eqref{icii}. Starting from $r=b$, the component $\xi^{\hat r}$ increases significantly, reaches a maximum in the inner region, and subsequently decreases as the particle approaches the regular core. This behavior is qualitatively similar to that observed in the scenario \eqref{ici}, although the amplitudes are considerably larger due to the initial condition $\dot{\xi}^{\hat r}>0$. For nonzero values of $\ell_0$, the maximum value of $\xi^{\hat r}$ is progressively reduced as the regularization parameter increases, showing that the corrections induced by T-duality soften the accumulated radial stretching. The lower panel further shows that increasing the initial radius $b$ leads to larger amplitudes without qualitatively changing the overall structure of the curves.

\begin{figure}[h!]
    \centering
    \includegraphics[width=\columnwidth]{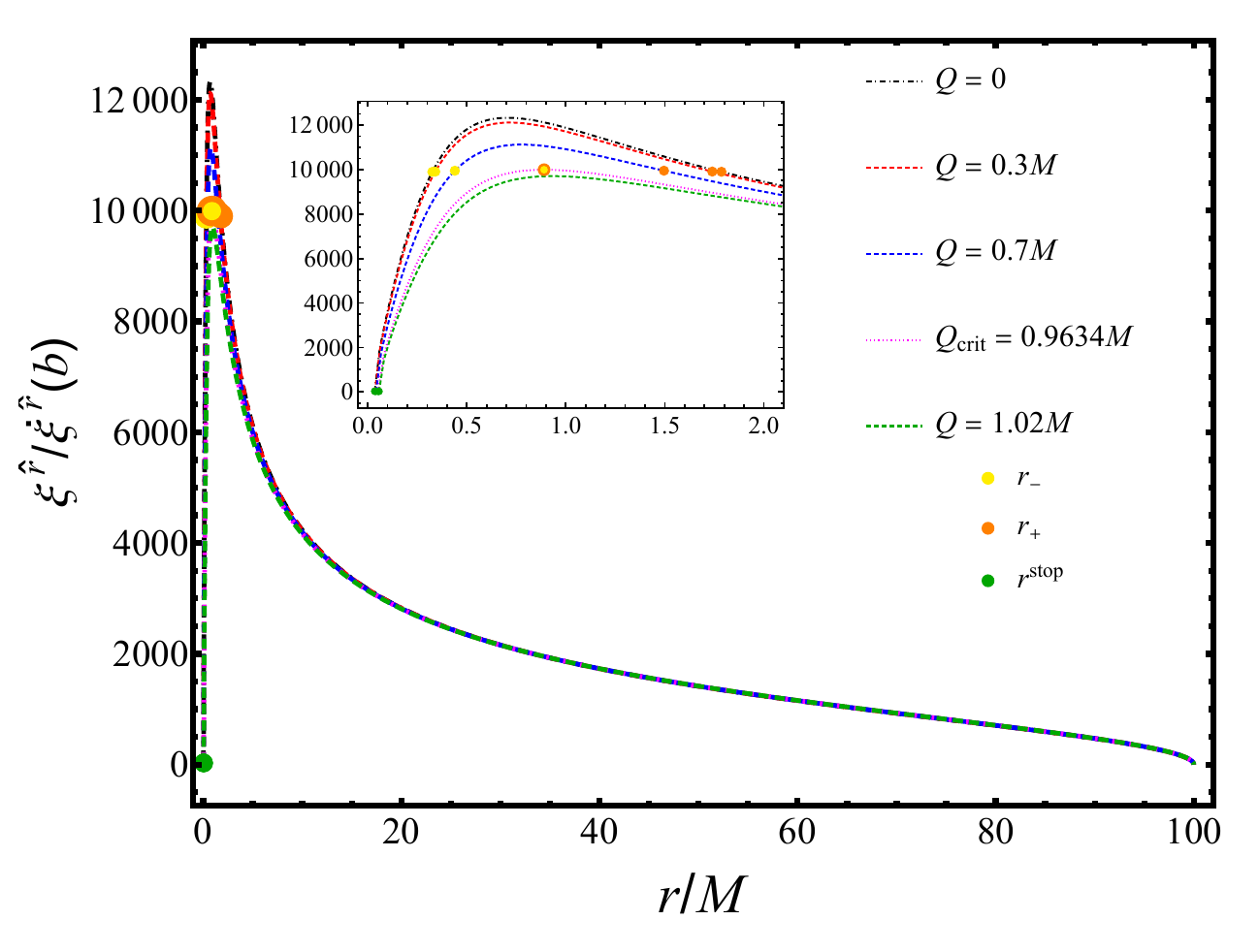}
    \caption{\justifying
    Behavior of the normalized radial component of the geodesic deviation vector, $\xi^{\hat r}/\dot{\xi}^{\hat r}(b)$, as a function of $r/M$ for the charged T-duality-inspired geometry in the scenario \eqref{icii}. We fix $\ell_0=0.5M$ and $b=100M$ and consider different values of the electric charge $Q$. The radii $r_-$, $r_+$, and $r^{\mathrm{stop}}$ are also indicated whenever they exist.}
    \label{fig:xi-radial-Q-ic-ii}
\end{figure}

We next investigate whether the same charge dependence persists for the initial conditions \eqref{icii}. Keeping $\ell_0=0.5M$ and $b=100M$ fixed, Fig.~\ref{fig:xi-radial-Q-ic-ii} shows the radial component for different values of $Q$. The qualitative behavior found in the neutral geometry is preserved: the radial separation grows during the infall, reaches a maximum in the strong-field region, and subsequently decreases toward $r^{\mathrm{stop}}$. The effect of the charge is again concentrated in the inner region, while the curves practically coincide at large radii. Increasing $Q$ systematically reduces the maximum value of $\xi^{\hat r}$, showing that the suppression of the accumulated radial stretching by the electric charge is not specific to the initial conditions \eqref{ici}, but is also present in the scenario \eqref{icii}.

\begin{figure}[h!]
\centering

\begin{subfigure}{0.50\textwidth}
    \centering
    \includegraphics[width=\textwidth]{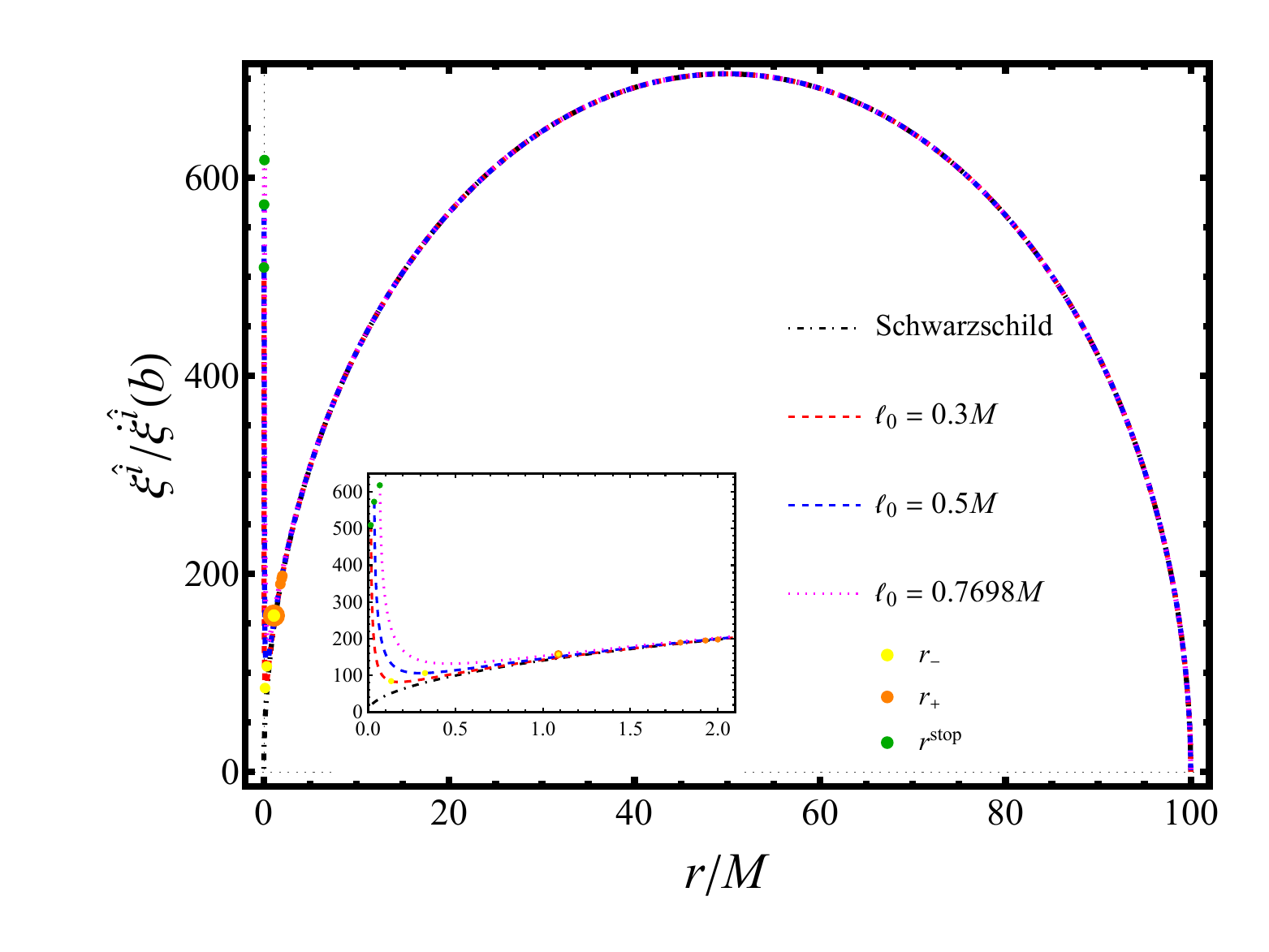}
    \label{fig:xi-angular-ic-ii-l0}
\end{subfigure}
\hfill
\begin{subfigure}{0.50\textwidth}
    \centering
    \includegraphics[width=\textwidth]{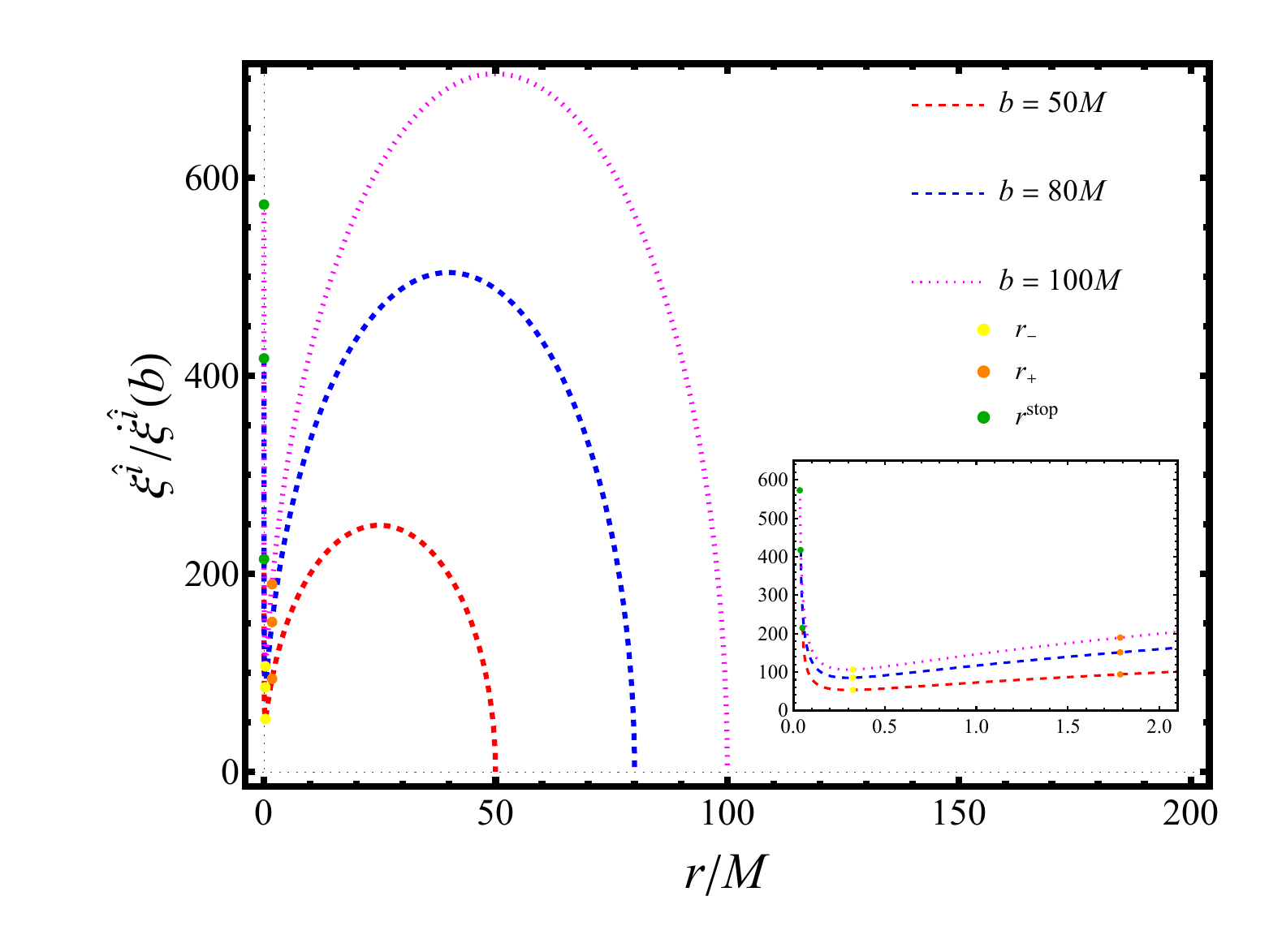}
    \label{fig:xi-angular-ic-ii-b}
\end{subfigure}

\caption{\justifying Behavior of the normalized angular component of the geodesic deviation vector, $\xi^{\hat i}/\dot{\xi}^{\hat i}(b)$, as a function of $r/M$ for a radially freely falling frame in the T-duality regular black hole for the scenario \eqref{icii}. In the upper panel, we consider the initial position $b=100M$ for different values of $\ell_0$. In the lower panel, we fix $\ell_0=0.5M$ and vary the value of $b$. The radii $r_-$, $r_+$, and $r^{\mathrm{stop}}$ are also shown in each panel.}
\label{fig:xi-angular-l0-ic-ii}
\end{figure}

The angular component in the scenario \eqref{icii}, shown in Fig.~\ref{fig:xi-angular-l0-ic-ii}, exhibits a richer structure than that found in \eqref{ici}. The angular separation increases during the infall, reaches a maximum, and then begins to decrease around $r=b/2$. Subsequently, $\xi^{\hat i}$ decreases significantly as the particle moves toward the inner regions, exhibiting another change in behavior near the regular core. This nonmonotonic evolution reflects the accumulated effect of the different regions of angular compression and stretching determined by the sign of $k_2$. For fixed $b$, the curves corresponding to different values of $\ell_0$ remain very close throughout most of the trajectory, becoming more clearly separated in the inner region. On the other hand, for fixed $\ell_0$, increasing $b$ produces larger amplitudes of the angular component while preserving its qualitative behavior.

The positions of the horizons $r_-$ and $r_+$, as well as the turning radius $r^{\mathrm{stop}}$, are also indicated in Figs.~\ref{fig:xi-radial-b-ic-ii} and \ref{fig:xi-angular-l0-ic-ii}. Together with Fig.~\ref{fig:g__R}, these markers allow us to relate the accumulated evolution of the deviation vector to the tidal-force structure along the trajectory. In particular, while the sign changes of $k_1$ and $k_2$ locally delimit the regions of radial and angular stretching and compression, respectively, the components $\xi^{\hat r}$ and $\xi^{\hat i}$ encode the accumulated effect of the passage of the geodesic congruence through these different regions. Compared with the scenario \eqref{ici}, the scenario \eqref{icii} exhibits considerably larger amplitudes and a more pronounced angular structure as a consequence of the nonvanishing initial separation rate between the geodesics.

\begin{figure}[h!]
    \centering
    \includegraphics[width=\columnwidth]{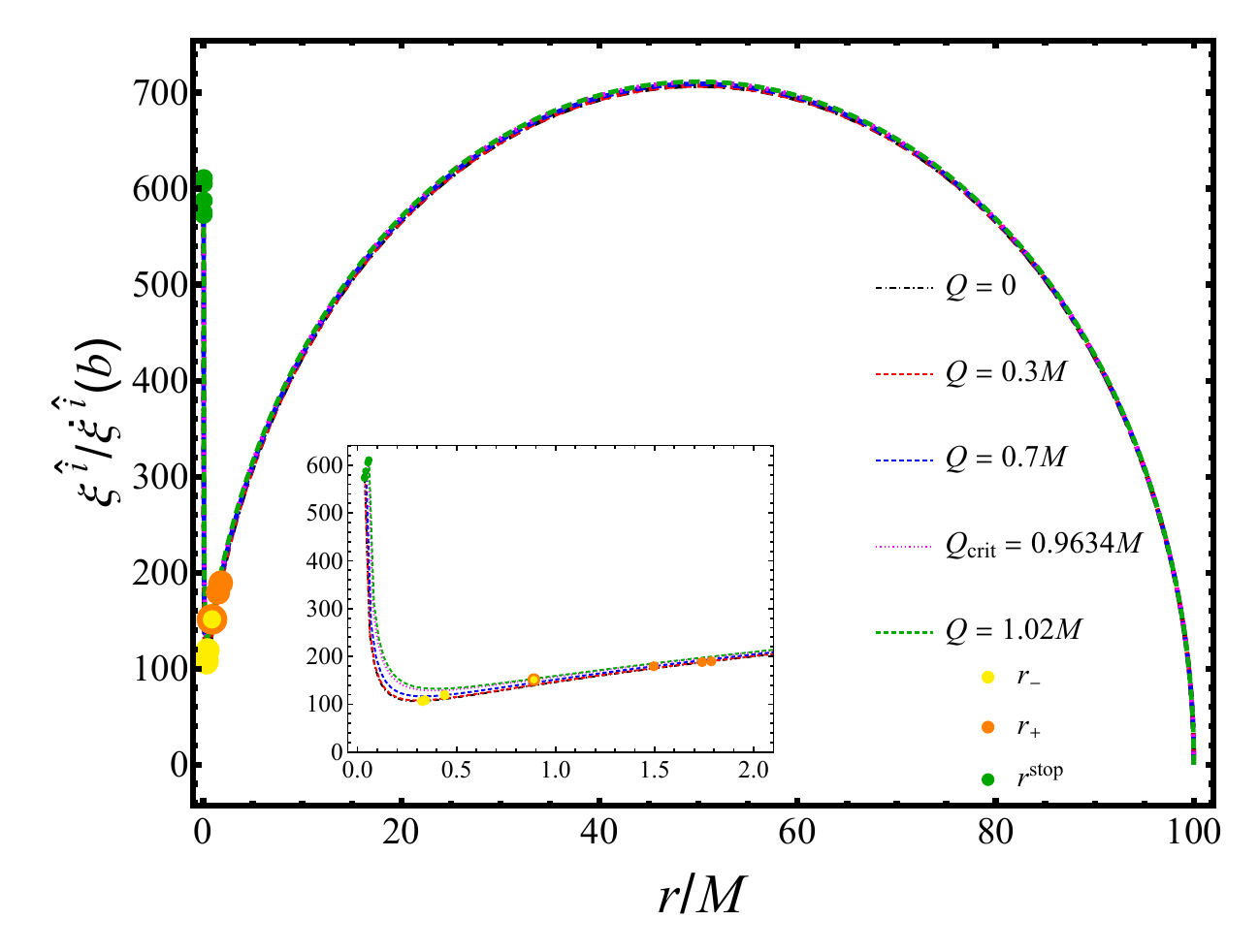}
    \caption{\justifying Behavior of the normalized angular component of the geodesic deviation vector, $\xi^{\hat i}/\dot{\xi}^{\hat i}(b)$, as a function of $r/M$ for the charged T-duality-inspired geometry in the scenario \eqref{icii}. We fix $\ell_0=0.5M$ and $b=100M$ and consider different values of the electric charge $Q$. The radii $r_-$, $r_+$, and $r^{\mathrm{stop}}$ are also indicated whenever they exist.}
    \label{fig:xi-angular-Q-ic-ii}
\end{figure}

The corresponding charge dependence of the angular component is displayed in Fig.~\ref{fig:xi-angular-Q-ic-ii}, again for $\ell_0=0.5M$ and $b=100M$. The nonmonotonic structure of the angular deviation is remarkably robust against the introduction of the electric charge. The curves remain almost indistinguishable throughout most of the trajectory, including the broad maximum at intermediate radii, and noticeable differences appear only in the innermost region. Thus, unlike the radial component, the accumulated angular deformation is only weakly affected by $Q$. The charge mainly changes the evolution close to the regular core and shifts the positions of the horizons and of $r^{\mathrm{stop}}$. Moreover, the angular deviation remains smooth as the system passes from the nonextremal black-hole sector to the extremal configuration and subsequently to the horizonless regime.

\section{CONCLUSIONS}
\label{sec_VI}

In this work, we investigated the geodesic dynamics and tidal effects in neutral and electrically charged regular spacetimes inspired by string T-duality. Starting from the regularized geometries characterized by the fundamental length scale $\ell_0$, we analyzed how the interplay between the T-duality scale and the electric charge modifies the spacetime structure and the motion of massive and massless test particles, with particular emphasis on circular orbits, radial free fall, and the deviations from the corresponding classical geometries. We also examined the tidal forces experienced by radially freely falling observers and the corresponding evolution of the geodesic deviation vector, allowing us to characterize both the local tidal response and the accumulated deformation of neighboring geodesics throughout the infall.

At the geometric level, the T-duality scale $\ell_0$ regularizes the central region while preserving the appropriate classical limits. In the neutral case, the Schwarzschild singularity is replaced by a regular de Sitter-like core, and the Schwarzschild geometry is continuously recovered as $\ell_0\to0$. The horizon structure exhibits two horizons for $\ell_0<\ell_{0,\mathrm{crit}}$, a degenerate horizon at $\ell_{0,\mathrm{crit}}\simeq0.7698M$, and a regular horizonless configuration above this critical value. In the charged geometry, the central region remains regular, although its character depends on the combination of $\ell_0$ and $Q$, allowing both de Sitter-like and anti-de Sitter-like cores. At the critical value $32M\ell_0=3\pi Q^2$, the effective cosmological constant vanishes and the quadratic contribution to the metric function disappears. In this case, the central geometry becomes Minkowski-like, with $f(r)=1+\mathcal{O}(r^4)$ and vanishing curvature invariants at $r=0$. The combined analysis of the $(\ell_0/M,Q/M)$ parameter space further shows that these two parameters jointly determine the horizon structure of the charged geometry and the corresponding extremal boundary.

The geodesic analysis revealed that the regularization also produces significant modifications in the orbital structure in the strong-field regime. For null geodesics, increasing $\ell_0$ shifts the outer unstable photon orbit toward smaller radii. The electric charge produces a similar inward displacement of the outer photon orbit. Interestingly, the disappearance of the horizons does not necessarily coincide with the disappearance of circular null orbits. In both the neutral and charged geometries, there exists a finite interval in parameter space in which the configuration is already horizonless but still admits both stable and unstable photon orbits. For massive particles, both $\ell_0$ and $Q$ modify the extrema of the effective potential and, consequently, the location and stability of circular orbits. In particular, increasing the electric charge shifts the outer unstable circular orbit toward smaller radii, while moving the stable branch toward larger radii.

The tidal analysis reveals more directly the effects of the regularization in the strong-field region. In contrast to the divergent tidal behavior of the Schwarzschild geometry, the radial and angular components remain finite in the regular central region. The presence of the scale $\ell_0$ introduces extrema and sign changes in both components, giving rise to distinct stretching and compression regimes in the radial and angular directions. In particular, the neutral geometry exhibits an intermediate region in which both tidal components are simultaneously compressive, followed, at smaller radii, by a regular central regime in which both components become positive, corresponding to stretching in both directions. The inclusion of electric charge modifies the strength and location of these tidal features, reducing the magnitude of the main extrema of the radial and angular components and shifting the characteristic radii associated with their zeros and extrema.

The solutions of the geodesic deviation equations complement this local description by revealing the accumulated deformation of neighboring freely falling trajectories. For the two sets of initial conditions considered, the regularization scale $\ell_0$ modifies the radial evolution of the separation vector, while the inclusion of electric charge suppresses the accumulated radial stretching. On the other hand, the angular component is considerably less sensitive to the charge, revealing a distinct response of the radial and angular deformations to the presence of the electromagnetic field. Taken together, these results show that the minimum length scale associated with T-duality and the electric charge leave distinct signatures on both the orbital dynamics and the tidal response in the strong-field regime. At the same time, the appropriate Schwarzschild and Reissner--Nordstr\"om limits are smoothly recovered. 

As future perspectives, the present analysis could be extended to rotating T-duality-inspired geometries, in which rotation may give rise to a richer tidal structure, as observed in the Kerr spacetime and other stationary and axisymmetric spacetimes~\cite{Xin:2025ipu, LimaJunior:2020fhs, Gaete:2022ukm}. Another natural extension would be to consider more general trajectories, including observers in circular orbits, for which the tidal response may differ from that experienced during radial free fall~\cite{Vieira:2025vwe, Silva:2025hwl, Albacete:2024qja}. In addition, it would be interesting to investigate possible observational consequences of the modifications found in the orbital dynamics and tidal structure.

{\acknowledgments}

M.S.O. and G.A. would like to thank Conselho Nacional de Desenvolvimento Científico (CNPq) and Fundação Cearense de Apoio ao Desenvolvimento Científico e Tecnológico (FUNCAP) for financial support. M.S. thanks the CNPq, Brazil, for financial support through CNPq/PDE Grant No. 200218/2025-5.  We acknowledge TM Crispim for useful discussions and comments.  This work is also supported by the Spanish National Grant PID2024-157196NB-I00 funded by MICIU/AEI/10.13039/501100011033 and the Department of
Education, Junta de Castilla y Le\'on (Spain) and FEDER Funds, Ref. CLU-2023-1-05.


\bibliography{refs}



\end{document}